%% file: ms.tex
\documentclass[12pt,preprint]{aastex}

\newcommand\gtsim{\mathrel{\lower0.6ex\hbox{$\buildrel {\textstyle >}\over {\scriptstyle \sim}$}}}
\newcommand\ltsim{\mathrel{\lower0.6ex\hbox{$\buildrel {\textstyle <}\over {\scriptstyle \sim}$}}}
\newcommand\sersic{S\'{e}rsic }
\usepackage{aas_macros}
\usepackage[footnotesize]{subfigure}

\slugcomment{Submitted to ApJS.}
\shorttitle{3CR Hosts}
\shortauthors{Floyd et al.}

\begin{document}
\title{HST NIR Snapshot Survey of 3CR Radio Source Counterparts\altaffilmark{1}\\
II: An Atlas and Inventory of the Host Galaxies, Mergers and Companions}

\author{David J. E. Floyd\altaffilmark{2} \email{dfloyd@lco.cl}}
\author{David Axon\altaffilmark{3},
Stefi Baum\altaffilmark{3},
Alessandro Capetti\altaffilmark{4},
Marco Chiaberge\altaffilmark{2},
Duccio Macchetto\altaffilmark{2},
Juan Madrid\altaffilmark{2},
George Miley\altaffilmark{5},
Christopher P. O'Dea\altaffilmark{3},
Eric Perlman\altaffilmark{6},
Alice Quillen\altaffilmark{7},
William Sparks\altaffilmark{2},
Grant Tremblay\altaffilmark{2}}

\altaffiltext{1}{Based on observations with the NASA/ESA Hubble Space Telescope, obtained at the Space Telescope Science Institute, which is operated by the Assciation of Universities for Research in Astronomy, Inc. (AURA), under NASA contract NAS5-26555}
\altaffiltext{2}{Space Telescope Science Institute, 3700 San Martin Drive, Baltimore, MD 21218, U.S.A.}
\altaffiltext{3}{Department   of  Physics,   Rochester   Institute  of Technology, 85 Lomb Memorial Drive, Rochester, NY 14623}
\altaffiltext{4}{INAF--Osservatorio Astronomico di Torino, Strada Osservatorio 20, Pino Torinese I-10025, Italy}
\altaffiltext{5}{Leiden Observatory, P.O. Box 9513, NL-2300 RA Leiden, The Netherlands.}
\altaffiltext{6}{Florida Institute of Technology, Physics \& Space Sciences Department, 150 W. University Blvd., Melbourne, FL 32901 USA}
\altaffiltext{7}{Department  of Physics  and Astronomy,  University of Rochester, Bausch \& Lomb Hall, P.O. Box 270171, 600 Wilson Boulevard, Rochester, NY 14627}

 %%%%%%%%%%%%%%%%%%%%%%%%%%%%%%%%%%%%%%%%%%%%%%%%%
 
\begin{abstract}
We present the second part of an $H$-band (1.6~$\mu$m) ``atlas'' of $z<0.3$ 3CR radio galaxies, using the {\em Hubble Space Telescope Near Infrared Camera and Multi-Object Spectrometer} (HST NICMOS2). We present new imaging for 21 recently acquired sources, and host galaxy modeling for the full sample of 101 (including 11 archival) -- an 87\% completion rate.
Two different modeling techniques are applied, following those adopted by the galaxy morphology and the quasar host galaxy communities. Results are compared, and found to be in excellent agreement, although the former breaks down in the case of strongly nucleated sources.
Companion sources are tabulated, and the presence of mergers, tidal features, dust disks and jets are catalogued. The tables form a catalogue for those interested in the structural and morphological dust-free host galaxy properties of the 3CR sample, and for comparison with morphological studies of quiescent galaxies and quasar host galaxies.
Host galaxy masses are estimated, and found to typically lie at around $2\times10^{11}$~$M_{\odot}$.
In general, the population is found to be consistent with the local population of quiescent elliptical galaxies, but with a longer tail to low \sersic index, mainly consisting of low-redshift ($z<0.1$) and low-radio-power (FR~I) sources. A few unusually disky FR~II host galaxies are picked out for further discussion. Nearby external sources are identified in the majority of our images, many of which we argue are likely to be companion galaxies or merger remnants.
The reduced NICMOS data are now publicly available from our website~\footnote{http://archive.stsci.edu/prepds/3cr/}.
\end{abstract}

\keywords{galaxies: active --- galaxies: fundamental parameters}

 %%%%%%%%%%%%%%%%%%%%%%%%%%%%%%%%%%%%%%%%%%%%%%%%%
\section{Introduction}
\label{sec-intro}

Extra-galactic radio sources have long posed an enigma to our understanding of the universe. While they can occupy some of the most gravitationally dominant galaxies in clusters, their bolometric luminosity is dominated not by the starlight from their galactic hosts but by the (often physically much vaster) regions of low-density radio-emitting material that have an origin in a tiny region of space at their very core. We now understand that the physics of these objects is related to the super-massive ($\gtsim10^8 M_{\odot}$) black-holes residing at their cores and the fueling of these black holes by accretion of gas and dust. However it is still unclear why so many apparently optically similar galaxies exist in the universe with no sign of radio activity. What makes one black hole radio-active, and another not? The concept of a black-hole duty-cycle~\citep{richstone+98} as the fraction of black holes that are active at any time is a useful statistical characterization, but we still do not understand what physically drives it. 
It is known that, broadly, the comoving number density of black-holes evolves along with the cosmic star-formation rate, and this can be seen as a correlation between the evolution of the black-hole duty cycle, and that of the cosmic star formation rate~\citep{wang+06}.
Mergers have long been regarded as a strong candidate for triggering AGN activity, providing a mechanism for the delivering the necessary (gas and dust) fuel to the central regions of a galaxy and the black-hole therein~\citep{kauffmannhaehnelt00,dimatteo+05}. 
Studies have shown the presence of nuclear dust disks in the cores of many radio galaxies~\citep{sadlergerhard85,vandokkumfranx95}, but similar numbers are also found in quiescent galaxies~\citep{veroncettyveron88}. %De Koff, Lehnert, Martel, McCarthy
Early studies of quasars showed morphological disturbances or tidal features to be present in many of the host galaxies (e.g.~\citealt{smith+86, hutchingsneff92, bahcall+97}), but more recent (and better-selected) samples have shown that the proportion exhibiting major tidal features is indistinguishable from that in the quiescent elliptical galaxy population~\citep{dunlop+03,floyd+04}.
Thus today it remains unclear whether there are any optical-IR properties of radio galaxies that distinguish them as a class from the putative parent population of quiescent elliptical galaxies, rather than simply being inherited from them.
% gas and dust in many systems, but studies of quasar host galaxies~\citep{dunlop+03,floyd+04} show that the presence of tidal disruptions in quasars is statistically indistinguishable from that in quiescent ellipticals of comparable mass. This is difficult to reconcile with the ``unification'' of radio-loud quasars and radio galaxies. However, the two are necessarily studied in very different ways. In this paper we present the final installment of data from our near-infrared (NIR) snapshot survey, and proceed to study the entire sample using techniques frequently used in both the quasar host galaxy, and quiescent galaxy research fields.

%In quasars it is very difficult to obtain images of the host galaxies, let alone accurate measurements of their stellar populations, owing to the dominating glare of the active nucleus. Radio-loud quasars are understood to be the same population as the FR~II radio galaxies, but with their central engines aligned at different orientations with respect to the observer~\citep{barthel89}. The FR~I's are similarly unified with the BL~Lac sources~\citep{urrypadovani95}. If this is indeed correct, then a study of the detailed structural properties of radio galaxies can offer us far deeper understanding of the relationship between AGN and environment than a study of quasar host galaxies. But it is important to be sure that we are speaking the same language when discussing the properties of quasar host galaxies, and normal elliptical galaxies.

In recent years, much attention has been drawn to the correlation seen locally between bulge and black hole mass in normal quiescent elliptical galaxies (e.g.~\citealt{gebhardt+00,merrittferrarese01}. AGN provide two obvious possible feedback mechanisms under which such a relation might arise, through radiation pressure and through the action of radio jets emanating from the central engine. Theoretical developments, and recent numerical studies using semi-analytic models (e.g.~\citealt{croton+06}) to explore the evolution of galaxies have shown that AGN feedback can in principal enforce such a correlation by shutting off star-formation and ejecting gas and dust from the most massive galaxies. The same models offer a natural solution to the problems of over-production of very massive galaxy haloes, and simultaneously can explain the so-called ``red sequence'' of galaxies (e.g.~\citealt{cattaneo+06, dimatteo+05,springel+05}). \citet{best+06} showed that the radio-loud phase of AGN activity is able to suppress the cooling of the host galaxy halo gas sufficiently that the radio source can control the rate of growth of the elliptical galaxy. 
% The physical cause of the radio-loudness however, remains elusive. Numerous studies have compared the properties of different classes of AGN with those of normal quiescent galaxies, to test whether environmental effects play a role on AGN activity. However, observational evidence for any connection between host galaxy environment and AGN activity remains tantalizingly sparse. 

It is important to investigate whether any statistical differences can be detected between samples of radio-loud AGN and samples of quiescent galaxies or radio-quiet AGN. One difficulty is the way in which active and quiescent galaxies have traditionally been studied is rather different. The most detailed morphological studies of large samples of quiescent galaxies are not conducted in the same way as studies of radio galaxies, or the host galaxies of quasars and other AGN.
To investigate the causes of AGN activity and radio-loudness we need large samples of objects in which {\em both} AGN and environment can be explored at multiple wavelengths. The 3C makes an ideal such sample for studying the effect of radio-loud AGN activity on host galaxy environment, and vice-versa. Here we focus on the 3CR, a well-defined sub-sample of the most powerful northern hemisphere radio galaxies~\citep{bennett62, spinrad+85}. We aim to provide a complete inventory of the properties of the low-redshift ($z<0.3$) 3CR, including physical characterization of the host galaxies free of dust to determine the underlying galaxy structure and dominant stellar mass; the presence of companions, mergers, dust disks, and jets; and the clear detection of any unresolved optical-IR nuclear sources. This will enable us to explore whether anything distinguishes them from normal quiescent galaxies, and continue the ongoing study into the question of what makes radio-galaxies ``radio loud''.

\subsection{The 3CR snapshot program}
In HST cycle 13 (2004--5) we embarked on a Near-Infrared Camera and Multi-Object Spectrometer (NICMOS) $H$-band snapshot imaging campaign of $z<0.3$ 3CR sources (\citealt{madrid+06} - hereafter paper I) . The infrared is an important realm for our growing understanding of the relationship between AGN and galaxies, enabling us to study the underlying host galaxies free from the distorting effects of dust. The SNAP program is now complete, having continued through cycle 14 (2005--6) at reduced priority. Since publication of paper I (which presented images for the first 69 targets, observed during cycle 13) an additional 22 sources have been successfully observed, and images for these new targets are presented here. Finally, archival data for an additional 11 objects previously observed with NICMOS2 in F160W were obtained from the Multi-mission Archive at Space Telescope (MAST\footnote{http://archive.stsci.edu/}). 

In this paper we seek to characterize the local 3CR host galaxy population in such a way that they can easily be compared to similarly large samples of quiescent galaxies, mergers, radio-loud and radio-quiet quasar host galaxies, and brightest cluster members. 
We present modeling of all the data using elliptical isophote (1-dimensional radial profile fitting) and two-component 2-dimensional galaxy-modeling techniques. We present the sample, observations and data reduction in section~\ref{sec-sampDR}. We discuss the 1D and 2D modeling of the NIR ($H$-band) host galaxies in section~\ref{sec-mod}. In section~\ref{sec-newobj} we present notes on, and images of, the newly observed and archival targets that were not presented in paper I.
Basic host galaxy properties are presented, and the modeling techniques compared in section~\ref{sec-res}. In section~\ref{sec-comp} we present a census of the companion sources and merger environments of the sample. Section~\ref{sec-disc} presents a general discussion of the sample and study and suggestions for the future. Section~\ref{sec-conc} concludes with a summary of our main findings. Detailed comparison of the sample with control samples of ellipticals, quasars and mergers is left to a companion paper (Floyd et al. {\em in preparation}). We have made all of our reduced data publicly available on the internet\footnote{http://archive.stsci.edu/prepds/3cr/}.

 %%%%%%%%%%%%%%%%%%%%%%%%%%%%%%%%%%%%%%%%%%%%%%%%%
\section{Sample and data reduction}
\label{sec-sampDR}
\begin{figure}
\plotone{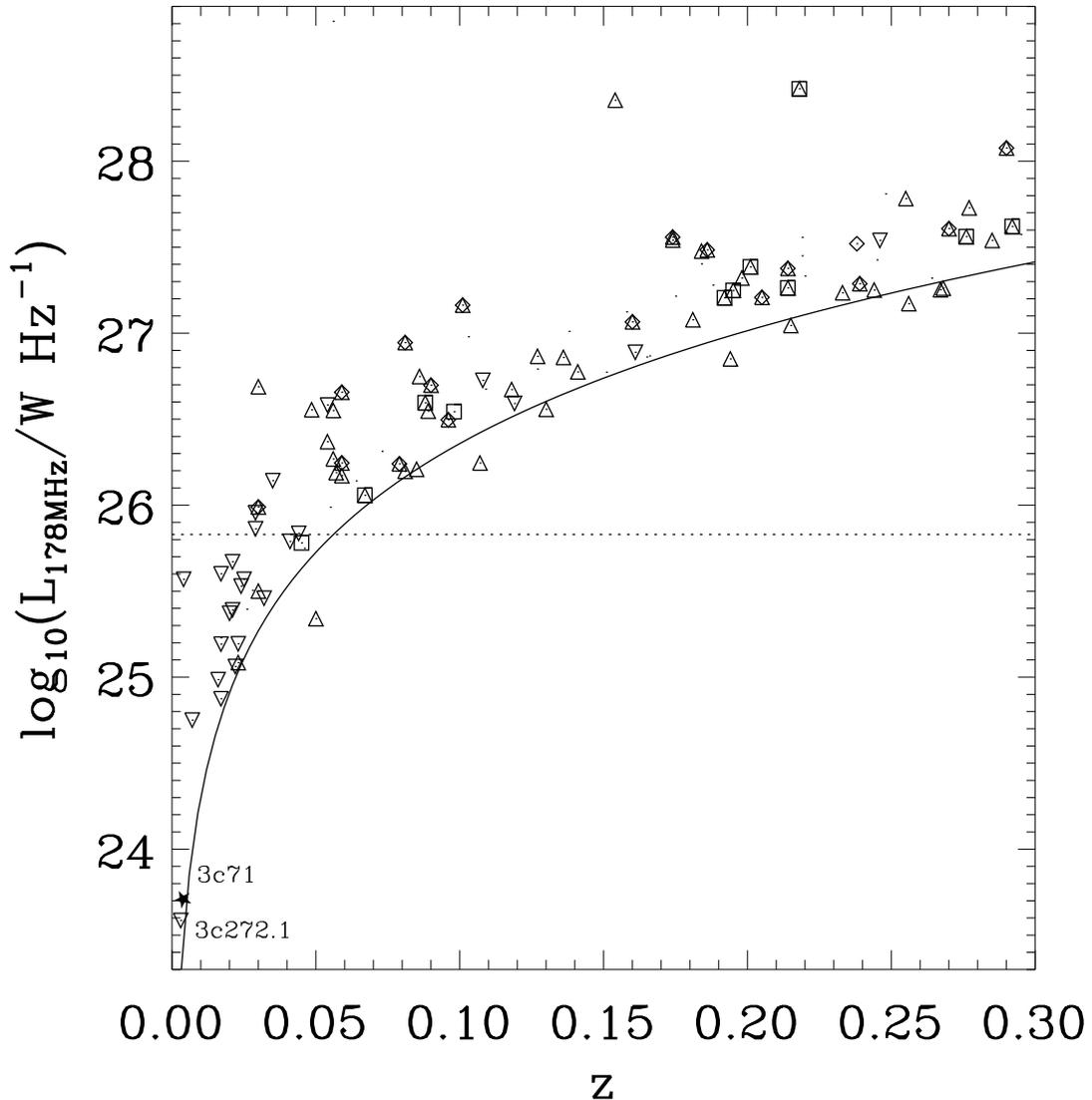}
\caption{\label{fig-samp178} 178~MHz radio luminosity of the sample. The dotted line indicates the conventional radio-loud / radio-quiet cut-off for quasars ($L_{5~{\mathrm GHz}}=10^{-24}~{\mathrm W~Hz^{-1}~sr}^{-1}$), converted to 178~MHz assuming a power-law index of 1 for the radio luminosity ($L_\nu\sim\nu^{-1}$). The solid line indicates the (9~Jy at 178~MHz) flux limit of the 3CR survey. Note that two of the most well-known low-redshift targets, 3C~71 (M~77, NGC~1068), and 3C~272.1 (M~84, NGC~4374) actually fall below the radio-loud quasar cut-off. Upward-pointing triangles represent FR~II's; Downward-pointing triangles are FR~I's; Diamonds are High Excitation Galaxies (HEG's); Squares Low-Excitation Galaxies (LEG's). The single solid star represents 3C~71.}
\end{figure}

The majority of the data that we use in this paper were taken during the HST Snapshot Program SNAP-10173 (PI: Sparks). The near-infrared images, fluxes and notes for the first 69 of our targets were presented in paper I. In this paper we present the imaging for 21 additional targets that were unobserved at the time of paper I's publication. An observing log for these newly observed sources is presented in table~\ref{tab-new}. In addition, we analyze archival NICMOS2 F160W data that exist for a further 11 objects -- see table~\ref{tab-arch}. 
Note that NIC1 and NIC3 lack the field of view and the resolution (respectively) for this morphological study, so observations of 3CR sources on those two chips in the archive have been omitted from study here. 
Finally, the NIC2 observations of 3C~273 were unsuitable for the present study as they provide only 30~s of on-galaxy integration, with the bulk of the observing time being dedicated to the famous jet.
Altogether this selection produces a near-complete sample of 101 powerful ($L_{5 {\mathrm GHz}}<10^{-24}$~W~Hz$^{-1}$~sr$^{-1}$) northern hemisphere radio sources at $z<0.3$ observed in a single infrared band. Radio luminosities (at 178~MHz) for the full sample are plotted in Fig.~\ref{fig-samp178}. Other basic sample data are presented in table~\ref{tab-props}.
Notes on, and images of the newly observed and archival targets are presented in section~\ref{sec-newobj}.  See paper I for images of, and notes on the other 69 sources.

We have re-reduced all of the original data presented in paper I, along with the new observations presented here, as described below. This new technique conserves the sky background throughout for all sources (allowing for better determination of the image statistics which is essential for accurate morphological characterization), and maintains the original pixel scale of 0\farcs075, which is near-critically resolved, and perfectly adequate for large-scale morphological studies such as this. On a large sample, the modeling of dithered images with double the spatial resolution results in an impractical increase in the computational time due to the convolution of each image with a point spread function.

We retrieved the data, pre-processed through the standard calibration pipeline, from the MAST.  We performed the data reduction described below using the Space Telescope Data Analysis System (STSDAS)  software running under PyRAF.
Two anomalies not  removed by the calibration pipeline are corrected during our data reduction: the pedestal effect, and the coronographic hole.
NIC2 has a quadrant  offset bias known as  the pedestal effect~\citep{noll+04}. We use the task {\sc pedsub} to remove this source of error. {\sc pedsub}  eliminates the pedestal effect but leaves the sky level untouched. We found that using {\sc pedsky} on nearby sources (as in paper I) led to some errors in the photometry due to inaccurate characterization and subtraction of the background flux.
The NICMOS coronographic spot appears as a small, well-defined region of erroneous flux on each individual dithered image, and is simply masked out before combining the images.

Each object was observed using a square dither pattern with four exposures of 288~s each (1152~s total integration time on each source). We use {\sc Multidrizzle}, a one-step task to perform image combination, dithering and cosmic ray rejection~\citep{koekemoer+02}.  We set {\sc Multidrizzle} to leave the  NIC2 native pixel size of 0\farcs075 and to not perform sky subtraction. Note that {\sc Multidrizzle} produces a final output image in units of electrons per second. We convert to DN before proceeding, by dividing the images by the Analogue-to-Digital Conversion (ADC) gain and multiplying by the integration time, obtained from the relevant FITS image headers (ADCGAIN and EXPTIME respectively).
Also note that the default behavior of {\sc multidrizzle} is to output weight maps that are exposure time maps. We adjusted this by setting {\sc final\_wht\_type = ``ERR''} to produce inverse-variance maps.

For each source we have $k$-corrected the apparent magnitudes to rest-frame H-band assuming a spectral index of $\alpha=1.5$ for the host galaxy and $\alpha=0.2$ for the nucleus where necessary ($f_\nu\propto\nu^{-\alpha}$). Galactic extinction corrections have been applied following~\citet{schlegel+98}.  We assume throughout a flat, $\Lambda$-dominated cosmology in which $\Omega_{\Lambda}=0.7$, and $h_0=0.7$.

%%%%%%%%%%%%%%%%%%%%%%%%%%%%%%%%%%%%%%%%%%%%%%%%%%

\section{Modeling Procedure}
\label{sec-mod}
We have used Ellipse~\citep{jedrzejewski87} and Galfit~\citep{galfit} to model each source in the sample. The steps required to implement these models properly are outlined below (section~\ref{sec-premod}).
To the elliptical isophotal values returned by Ellipse, we fit 1-dimensional \sersic-law quasi-radial profiles (section~\ref{sec-ellipse}). Properties of these profiles are compared to those obtained by direct 2-dimensional fitting to the image data using Galfit (section~\ref{sec-galfit}). We make use of the summary of photometric galaxy relationships by~\citet{caon+93} to convert between surface brightness, scale-length and total flux for different values of \sersic index, $n$.

\subsection{Pre-modeling}
\label{sec-premod}

\subsubsection{Masks}
\label{sec-mask}
We began by creating a mask and error frame for each target, using the output weight frame and hot pixel flags from {\sc Multidrizzle}. The input bad pixel masks are combined by {\sc Multidrizzle} to produce a final mask of all hot or bad pixels. To this we added an additional source mask to remove any external sources, companions, diffraction spikes, and the warm (ampglow-affected) corners of the image from the fit. These masks were applied to all the subsequent fits described below.

\subsubsection{Errors}
\label{sec-err}
The combined weight ($1/\sigma^2$) frames that are output by {\sc Multidrizzle} incorporate the Poisson noise and read noise on each pixel. We converted them to standard error ($\sigma$) frames.
For the nuclear-dominated sources (with nuclei providing $\gtsim50$\% of the total flux), the errors are dominated not by Poisson or read noise, but by the ``sampling error'' associated with the lack of knowledge of the point spread function (PSF). In these cases we follow the technique described in~\citet{floyd+04}, to provide accurate representation and weighting across the entire dynamic range, from the nucleus to the wings of the PSF. In practice, this means estimating the ``sampling error'' in concentric circular annuli, centered on the nuclei, from the deviation of the source from the PSF within each annulus. Where this drops to the Poissonian level (typically within a radius of $\sim0\farcs5$), we revert to the properly calculated individual pixel errors described above.

\subsubsection{Point spread functions}
\label{sec-psf}
Point spread functions (PSF's) were calculated for each target using {\sc Tinytim}~\citep{tinytim}, setting the position of the PSF to the average of the four central positions of the target. We generated a PSF at twice the resolution of the NIC2 chip, to provide oversampling, and accurate shifting and convolution.
Defocus was estimated for the time of each observation using the tables and figures provided on the NICMOS website~\footnote{http://www.stsci.edu/hst/observatory/focus/documents.html}, and this information was included in each  {\sc Tinytim} model. The  {\sc Tinytim} models were, last of all, rotated through the same angle as the target images, to give north up in each case.

\subsubsection{Sky background level}
\label{sec-sky}
For all targets, an initial measurement of the approximate background level in the image was made using {\sc imstatistics} to determine the photometric statistics on several empty regions of the chip. We attempted to measure the background in all four quadrants of the chip, as a check of the success of the quadrant bias removal. In practice, on all images where this was possible, the quadrants were found to have identical backgrounds, within the Poissonian errors. 
On objects that are too close for us to observe any background on the small $19 \times 19\arcsec$ NIC2 chip (in practice, almost all objects at $z < 0.1$ and some large objects at higher redshifts) we employed the technique developed by~\citet{sparksjorgensen93}: All of these targets have been observed previously using WFPC2 in either $R$ or $V$ band (or both). We obtained the reduced full WFPC2 mosaics from the 3CR database~\footnote{http://acs.pha.jhu.edu/$\sim$martel/}, and measured the background flux on these larger images. Profiles of the flux in each of the 2 bands were produced, and plotted against each other. In all cases there is a region of linearity in the 2 flux levels, and a linear fit was made in this region, extrapolating down to obtain the IR flux at the level of the optical (WFPC2) background. As a check of the robustness of this fitting technique, we also deployed it on images of five higher redshift sources, where a clear measurement of the infrared background was also possible from the NIC2 image. In these cases we produced a small cutout of the infrared image, such that no background light was visible, mimicking the situation for the lower redshift targets. In these tests, all five fits yielded infrared backgrounds that, again were within the Poissonian error of the true, measured background fluxes on the original infrared images.

\subsection{Ellipse fitting}
\label{sec-ellipse}
We used the {\sc Ellipse} modeling procedure in {\sc Iraf.stsdas} to fit elliptical isophotes to the sky-subtracted NICMOS images. 
We initialized each fit by feeding it the position of the centroid of the source, as measured using {\sc imcntr}, and estimated values of the position angle (PA) and ellipticity ($1-b/a$). We began the fit at 10 pixels (semi-major axis length) working outwards in geometric steps, increasing in size by a factor $\times1.1$. Once the fit has terminated due to insufficient pixels (less than half on a given annulus, {\sc fflag} = 0.5), or convergence (largest harmonic amplitude $<5\%$ of the residual RMS around the harmonic fit) it returns to fit the inner region with identical geometric steps, all the way to the centre of the target. The central position, ellipticity and PA of each ellipse were left free to explore the stability of the isophotal centering, and investigate structures in the galaxy deviating from the purely elliptical.
Deviation from a perfect ellipse at each radius is broken down with a Fourier series,  characterized by the 3rd and 4th order sine and cosine coefficients. We convert from the {\sc ellipse} parameters A3, A4, B3 and B4 to $a_{4}/a$, $b_{4}/b$ following the conversion described in~\citet{milvangjensenjorgensen99} -- see that paper for a useful discussion of all the ellipse fit parameters. Additional masking was performed, in software, on pixels deviating by more than 3$\sigma_{\mathrm annulus}$ from the mean value on a given fitted elliptical annulus.

Modeling with {\sc ellipse} allows us to track radial changes in the ellipticity and position angle of the isophotes of galaxies, providing a powerful analysis of the structure of the galaxy, and presence of any disturbances from a relaxed elliptical morphology. 
1-dimensional~\citet{sersic68} profile fits were made to the best-fitting ellipse-isophotal intensities, after convolution of the model with a point spread function (PSF), using a Levenberg-Marquardt non-linear least-squares minimization technique~\citep{numrec}. This technique has been widely used in the past for studying the properties of elliptical galaxies (e.g.~\citealt{BBF92,graham+96}) and mergers (e.g.~\citealt{rothberg+04}), but note that it does not provide a true ``radial profile'' since adjacent elliptical isophotes may have different position angles and ellipticities, thus overlapping, and contributing flux to one-another. 
For purposes of comparison with the 2D fits (see below) and with the literature, the median values of ellipticity, position angle, and diskiness were computed for the isophotal fits between a radius of 0\farcs5 (inside of which the nucleus may dominate) and where the signal drops to $3\sigma_{\mathrm sky}$, or the edge of the chip if that is reached first. The results of the elliptical isophote fitting are presented in table~\ref{tab-ellipse}.

\subsection{Galfit}
\label{sec-galfit}
Galfit~\citep{galfit} is a versatile fitting code which allows the user to fit multiple components to a galaxy in an iterative fashion, using a downhill gradient Levenberg-Marquardt technique. The model galaxy is convolved with a PSF at each iteration, and compared to the data through the $\chi^{2}$ statistic. We fitted each object using a single \sersic component convolved with a PSF, plus an additional unresolved nuclear component represented by the same PSF. Error frames and masks were produced as described in section~\ref{sec-err} above. The majority of objects are too distant to allow detection of a ``Nuker'' (e.g.~\citealt{faber+97}) or \citet{graham+03} type core, so we adopt just a single \sersic component for study throughout the sample for consistency, simplicity, and comparison with the bulk of the literature.

The complete (\sersic + PSF + sky) model has 11 free parameters: 
Centre of host $(x,y)$; Luminosity of the host galaxy ($L_H$); Scale length of the host galaxy ($R_{e}$); Position Angle of the host galaxy ($\theta$); Ellipticity of the host galaxy ($b/a$); \sersic parameter of the host galaxy ($n$); Luminosity of the nucleus ($L_N$); Position of nucleus $(x,y)$; Background flux.
We initially held fixed the values of the \sersic index ($n=4$ - i.e. a de Vaucouleurs elliptical), the ``diskiness'' parameter, ($c=0$), and the sky background flux (at a level determined as described in section~\ref{sec-sky} above). After a satisfactory fit was obtained, we freed these parameters one-by-one, in an iterative modeling procedure, in the order listed above. Last of all, the sky value was freed up to see if a better fit could be obtained, and to test the degeneracy between sky background and host galaxy flux. The models presented here include this fitted sky value, which was always consistent with the measured sky value. The results of the galfit \sersic model fitting are presented in table~\ref{tab-sers}.

As discussed above (section~\ref{sec-err}), we have adopted the error analysis technique of~\citet{floyd+04} for nuclear-dominated objects. Since this is not the standard error treatment used by Galfit, we performed extensive tests (discussed in the Appendix) of the effect of this assumption on the fits produced by Galfit using both real and synthetic sources. We found Galfit to produce identical results using each technique for sources with weaker nuclei. We argue that the \citet{floyd+04} technique is the formally correct analysis of the errors for sources with dominant nuclei since it properly accounts for all sources of error across the image. 

\subsection{Errors on the fitted parameters}
Errors on the fitted parameters in each case are calculated from the diagonal elements of covariance matrix for that set of parameters, following the assumption that each parameter is independent. The errors quoted in tables~\ref{tab-sers} and~\ref{tab-ellipse} are formal 1-$\sigma$ statistical errors. Note that in many cases, the true errors are dominated by the systematics, since the galaxies are not ideal. To investigate this, we performed simple tests of different galaxy fitting algorithms as described in the appendix. Including all systematics, we expect that all our morphological results are accurate to within 10\%, and host fluxes within 2\%

%%%%%%%%%%%%%%%%%%%%%%%%%%%%%%%%%%%%%%%%%%%%%%%%%

\section{Notes on individual objects}
\label{sec-newobj}
In this section we present notes and images for the newly observed objects in our SNAP program (i.e. those not included in paper I) -- Fig.~\ref{fig-newobjects}, and those taken from the archive -- Fig.~\ref{fig-archobjects}.

\subsection{Newly observed sources}
\label{sec-newnewobj}
\subsubsection{3C~15; $z=0.073$}
This source exhibits a prominent optical-IR jet~\citep{martel+98} in an apparently undisturbed round elliptical that almost fills the NICMOS2 field-of-view. No bright nuclear point source is detectable in H-band.

\subsubsection{3C~17; $z=0.219$}
This very peculiar radio source~\citep{morganti+99} and BLRG has a quite round elliptical galaxy, and a prominent IR nucleus. There are several faint companions in the immediate surroundings of the host galaxy.

\subsubsection{3C~33; $z=0.059$}
An FR~II HEG source with a prominent dust disk aligned $\sim45^\circ$ to the radio axis. The host galaxy is elliptical, appearing slightly disturbed due to the dust disk.

\subsubsection{3C~98; z=0.030}
A NLRG FR~I, hosted by an undisturbed elliptical with smaller companion elliptical. No internal structure is visible in the NICMOS2 image. The radio source is double lobed, running almost north-south~\citep{miller+85}.

\subsubsection{3C~132; $z=0.214$}
3C~132 appears near the western corner of the NIC2 image, and has a somewhat elongated, smooth elliptical host galaxy. Several companions are visible, including one bright foreground star at the image center. The radio axis of this FR~II source is almost perpendicular to the major axis of the host galaxy.

\subsubsection{3C~153; $z=0.277$}
3C~153 is a radio galaxy (angular size 6 arcsec=25~kpc) oriented at PA$\sim50^\circ$ with an  FRII radio morphology and a large arm-length ratio $\sim1.9$~\citep{laing81}. It has been classified as a CSS source (e.g.~\citealt{akujor+91}) but it would be at the upper end of the size distribution for such sources. 
The [OII] emission line nebula is slightly extended along the radio axis~\citep{mccarthy+95}. ~\citet{dekoff+96} note that the galaxy has an elliptical nucleus and strong emission lines. 
The NICMOS image shows a fairly round host galaxy. There are a couple of unresolved sources nearby but they don't correlate with knots or hotspots in the radio source. Two other features are caused by the intersection of diffraction spikes.

\subsubsection{3C~166; $z=0.245$}
An unusual radio source featuring two lobes with very different morphologies~\citep{spanglerbridle82}. The radio axis runs north-south, with Spangler \& Bridle's southern features C and D detected on the IR image as infrared hotspots. The source has a clearly-detected, unresolved infrared nucleus.

\subsubsection{3C~234; $z=0.185$}
3C~234 exhibits a bright quasar-like nucleus in the NICMOS image, and is known to be a broad-line radio galaxy~\citep{mccarthy+95} with classical double FR~II radio source. In the optical~\citep{dekoff+96,mclure+99} there are features emanating from the nucleus in the east and west direction. The western feature, described as a tidal arm by McLure et al. is detectable on the new NICMOS image. There are two additional sources to the west in the same direction, the furthest of which is also detectable in F555W, and F702W. Both are detectable in F675W.
This galaxy is known to have strong extended emission lines which likely explain these features. Pa~$\beta$ (1.28~$\mu$m) is in the passband of F160W at the redshift of 3C~234.

\subsubsection{3C~258; $z=0.165$}
\label{sec-3c258}
This galaxy is a well-known compact steep spectrum (CSS) radio source, with a compact, double radio structure seen on scales of a few tenths of an arcsecond along PA$\approx33^\circ$~\cite{spencer+89,akujor+91,ludke+98}, although its arcminute scale structure is aligned along a nearly north-south direction \cite{strom+90}. Its redshift has been quoted as z=0.165 \cite{Sm76}, but this value is in dispute as \cite{dey94} note that spectra taken at Lick show a likely background quasar, based on the observation of a single emission line at 7111\AA which would give $z=1.54$ if due to Mg~II 2798\AA. However, due to poor seeing at the time the spectra were taken, no firm claims can be made as to the origin of this line. Our NICMOS data are consistent with a significantly higher redshift, as they show a very compact host galaxy that is much fainter than one would expect for an object at z=0.165.  The nuclear point source in our image is also quite bright, with an arc-like object 3" ESE of the AGN which may be an irregular foreground ($z=0.165$) galaxy causing the confusion.

\subsubsection{3C~284; $z=0.239$}
An FR~II HEG radio source, hosted by a disturbed elliptical with a SE tidal tail toward its most prominent companion 4\arcsec SE. Several fainter companions are also visible on the NICMOS chip. In the optical the source has a disky appearance and double nucleus, owing to this same tidal tail and associated dust lanes.

\subsubsection{3C~296; $z=0.025$}
3C~296 is hosted by an extremely boxy elliptical galaxy. It has a double lobed jet with clear radio emission along the jet axis (FR~I). The WFPC2 image~\citep{martel+99} shows a truncated edge-on disk embedded in the galaxy, similar to that seen in NGC~4261. The disk is aligned with the major axis of the elliptical galaxy, with the jet perpendicular to the disk.
The nucleus is seen in both WFPC2 image (though faint) and in the NICMOS image (brighter).  X-ray and radio observations of this galaxy are discussed in~\citet{hardcastle+05}.

\subsubsection{3C~300; $z=0.270$}
A double-sided FR~II HEG, hosted by a compact, elongated $\sim 3$~kpc elliptical host galaxy with a faint companion 8\arcsec East, and a very faint tidal distortion. The radio axis runs roughly SE-NW, and there is a faint thread of emission along this axis on the NW side, which is a candidate jet. However, it is almost lost in the noise of the NICMOS chip edge.

\subsubsection{3C~323.1; $z=0.264$}
This QSO has a bright nucleus that produces diffraction spikes in the near-infrared  image. It has a close  companion to the northwest that does not show clear signs of interaction, but is consistent with being the remnant nucleus of a galaxy in the final stages of merging~\citep{miller+03,canalizostockton97}.
The Keck spectroscopy presented by Miller et al. demonstrate that the host is not a normal elliptical galaxy, and from our $H$-band image there is some evidence of disturbance, which may explain the small scale length fit to the data. It is also possible that the radio emission from this source is Doppler boosted, making it an artificially radio-loud object.

\subsubsection{3C~326; $z=0.090$}
A double source - the new NICMOS image shows only the southern, which is the stronger of the two in the radio~\citep{rawlings+91}. A larger companion elliptical sits 25\arcsec (43~kpc) to the north. A dust disk bisects the nucleus of the small elliptical host galaxy, giving the appearance of a disk. 3 very faint small companion sources are visible within, or close to the main galaxy halo. Unusually the radio axis is aligned close to the axis of the disk.

\subsubsection{3C~332; $z=0.270$} 
A powerful FR~II radio galaxy with a prominent quasar-like infrared nucleus marking the center of an elliptical host galaxy. There is a smaller companion 3\arcsec to the SW.

\subsubsection{3C~357; $z=0.166$}
3C~357 is an FR~II radio galaxy with angular size 90\arcsec (250~kpc) at a PA~110 degrees~\citep{fanti+87,harvanekhardcastle98}. In both lobes the hotspots are off to one side suggesting there has been a change in the jet direction. There is emission line gas extending a few arcsec from the nucleus along the radio axis~\citep{mccarthy+95}. \citet{dekoff+96} and ~\citet{capetti+00b} note that there are filamentary dust lanes southwest of the nucleus. The galaxy isophotes are very elliptical with no remarkable features. There is a single faint candidate companion source 8\arcsec East, with a larger companion just on the eastern edge of the chip. No unresolved nucleus is detected, but the core of the galaxy appears unusually bright, and the host has a high \sersic index ($n=5.34$).

\subsubsection{3C~403.1; $z=0.055$}
This double radio source lies at low galactic latitude ($b\approx -14^\circ$), resulting in numerous stars on the image. In spite of this one can see that the undisturbed-looking elliptical host galaxy is located in a fairly dense environment, with one large companion seen on the NICMOS chip, and 3 others similar in brightness visible on the WFPC2~\citep{martel+99} image. 

\subsubsection{3C~410; $z=0.248$}
Our observations of 3C~410 suffered a pointing error, resulting in only two usable images of the source -- hence the rather noisy image. A strong central nuclear point source is visible centered on a round elliptical host galaxy. Numerous unresolved sources are visible in the field.

\subsubsection{3C~424; $z=0.127$}
Faint tidal distortions are visible to the north on our NICMOS image, and this source lies in a dense environment, with numerous large companions visible on the WFPC2~\citep{dekoff+96} image. The radio source apparently lies on the edge of the cluster. Two bright unresolved sources are visible to the north, with a resolved object NW, on the edge of the NICMOS chip, barely detected at $R$.

\subsubsection{3C~442; $z=0.026$}
Only two exposures can be used of the four, due to telescope pointing errors. The host galaxy is a smooth elliptical, with 2 companions visible on the WFPC2 image~\citep{martel+99}. The point source on the edge of the NICMOS chip to the WSW is in line with the radio axis, and is a candidate hotspot - also clearly visible on the WFPC2 images.

\subsubsection{3C~459; $z=0.219$}
A strong IR nucleus, and heavily distorted off-center host galaxy distinguish this FR~II radio galaxy. The host is dominated by young stellar population~\citep{tadhunter+00}, and must have undergone a
recent merger. The radio source is small (10\arcsec) and also very asymmetric~\citep{morganti+99}.

\begin{figure*}[ht]
\plottwo{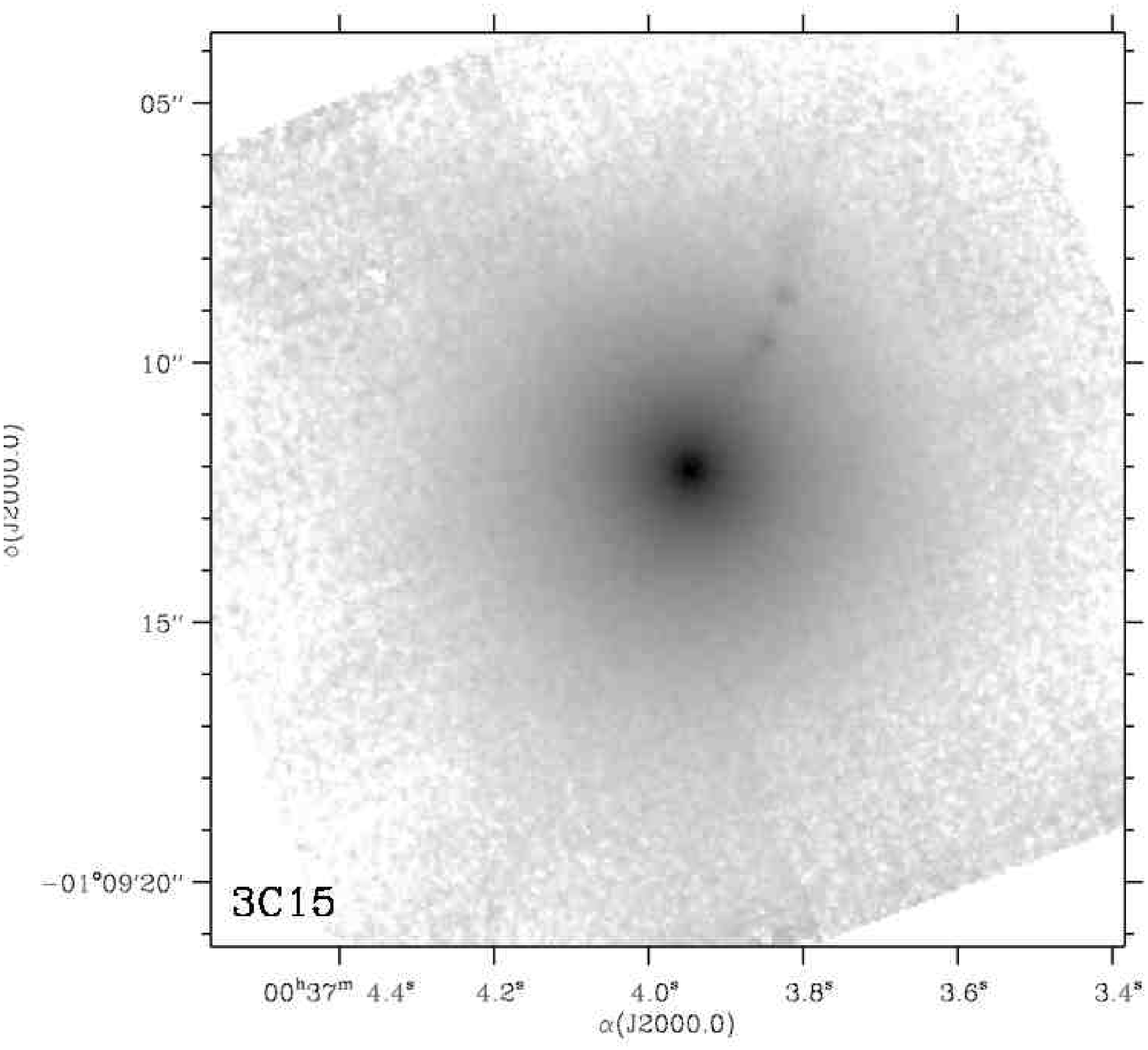}{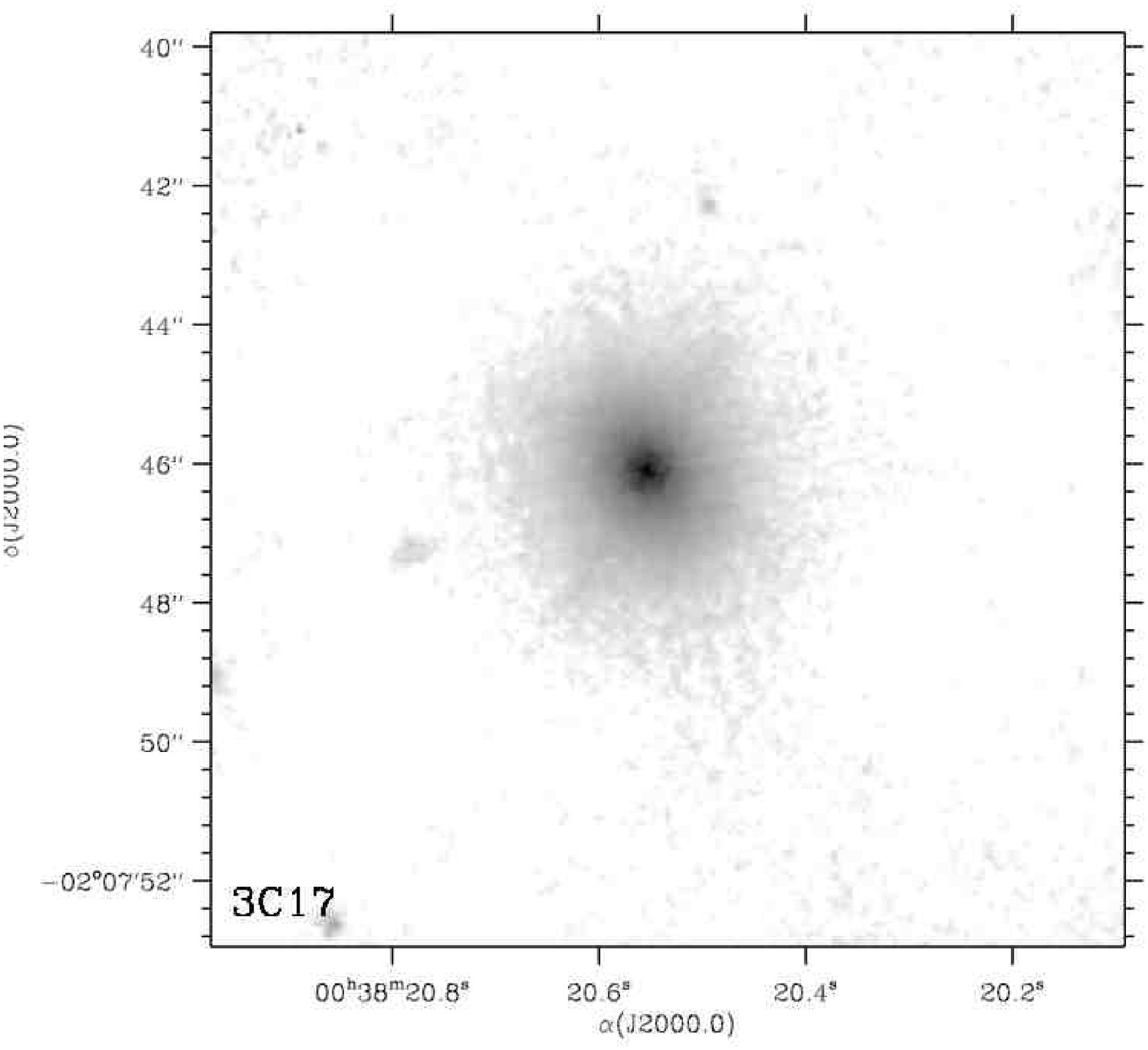}
\plottwo{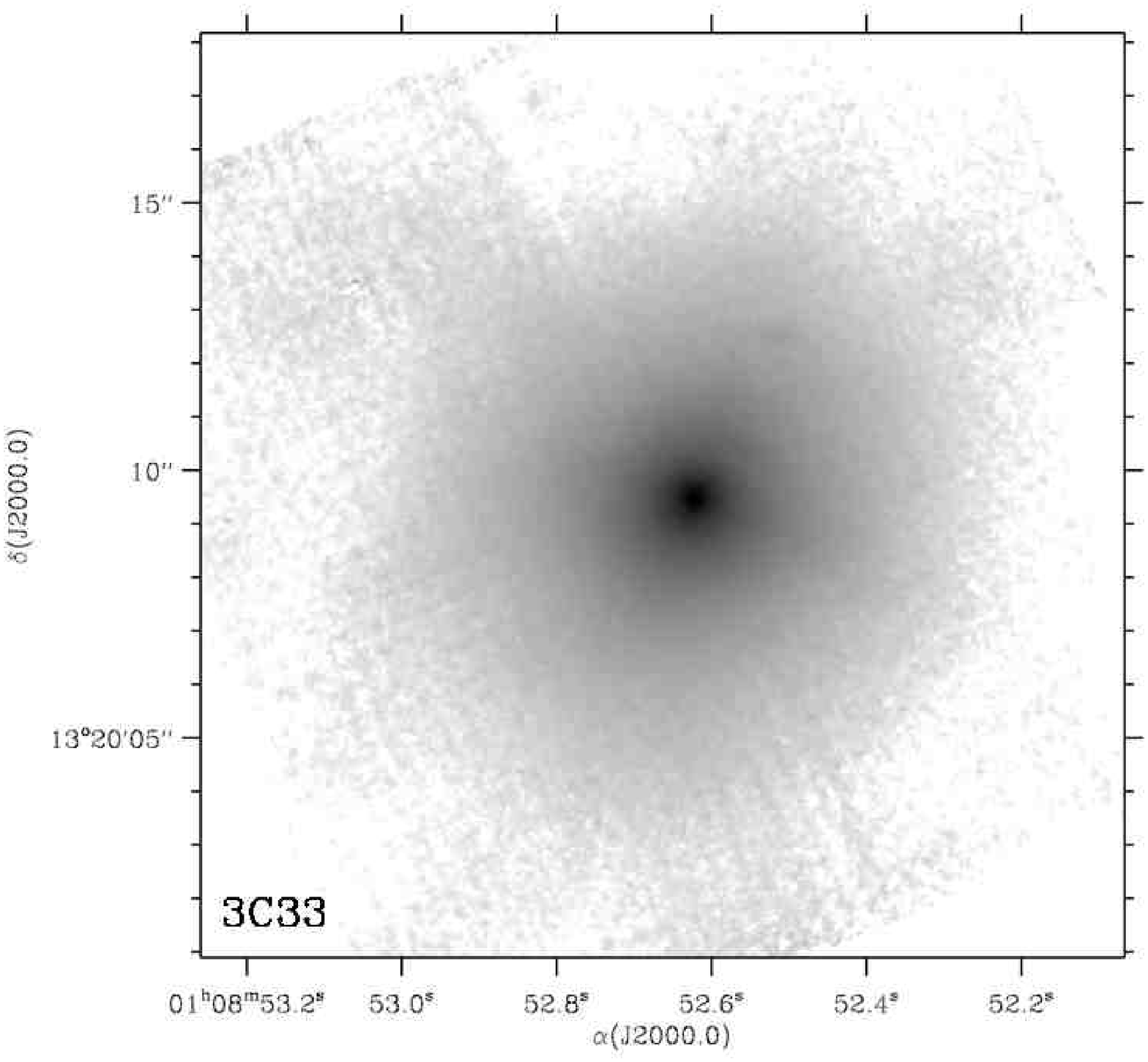}{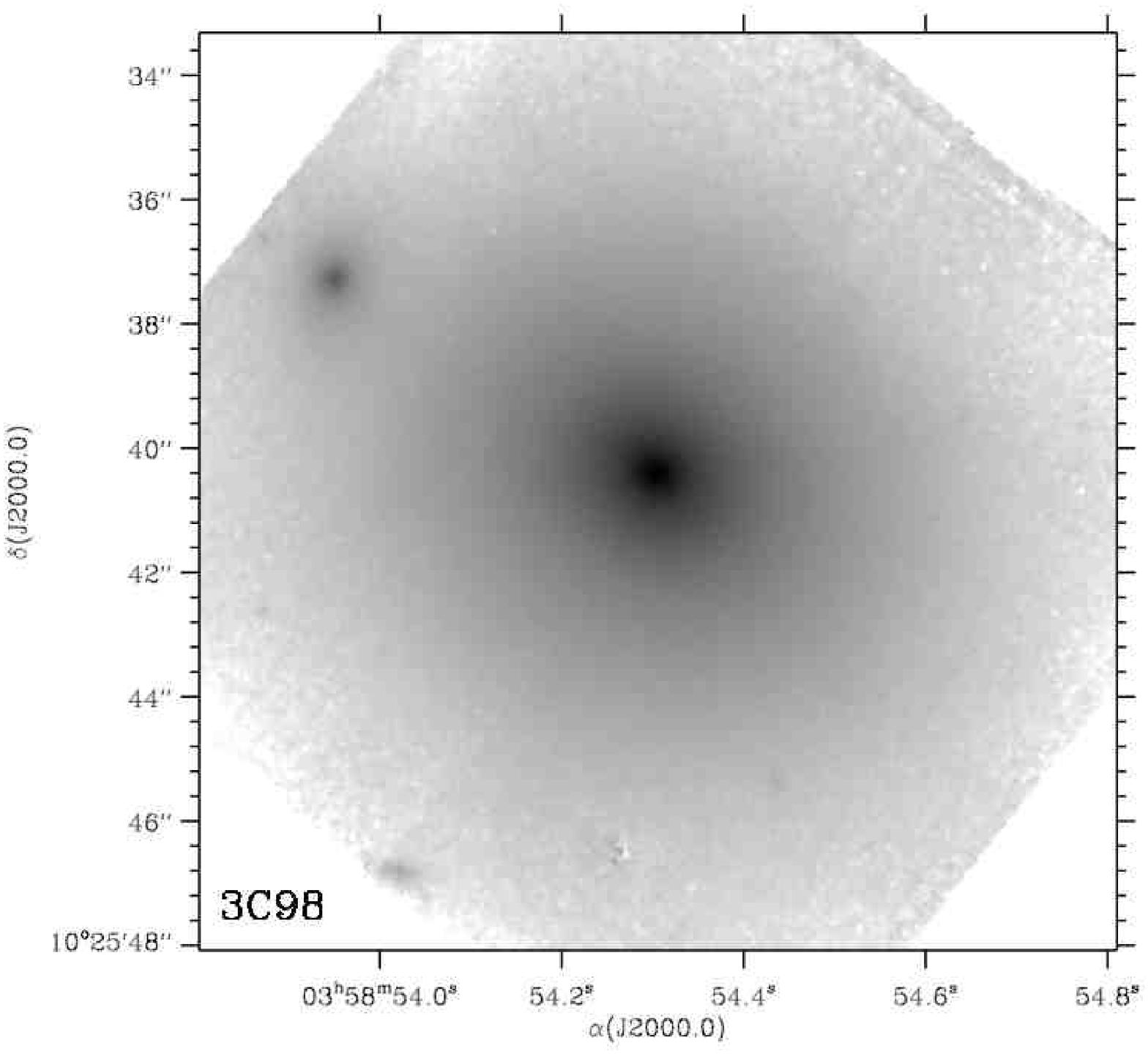}
\plottwo{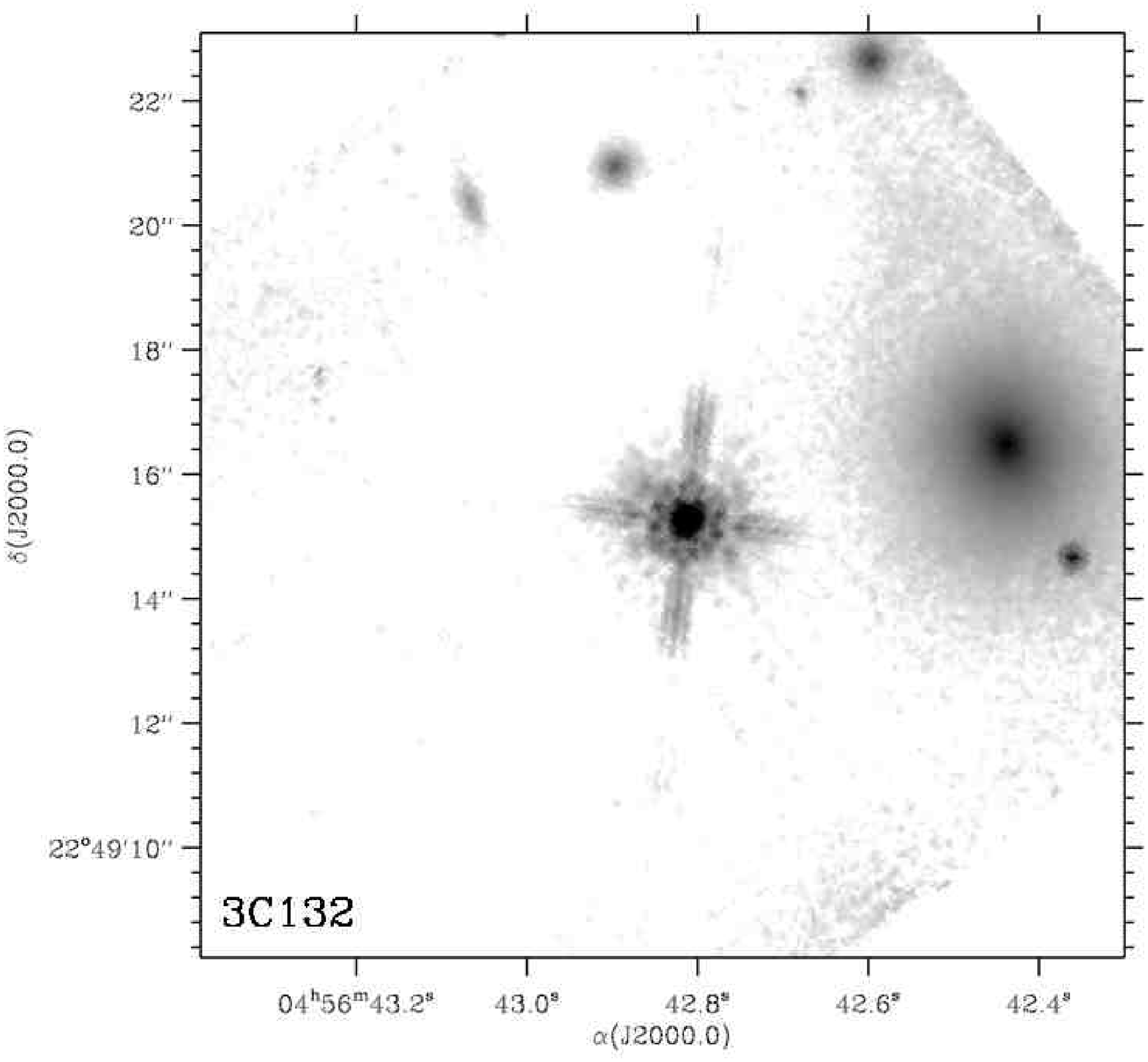}{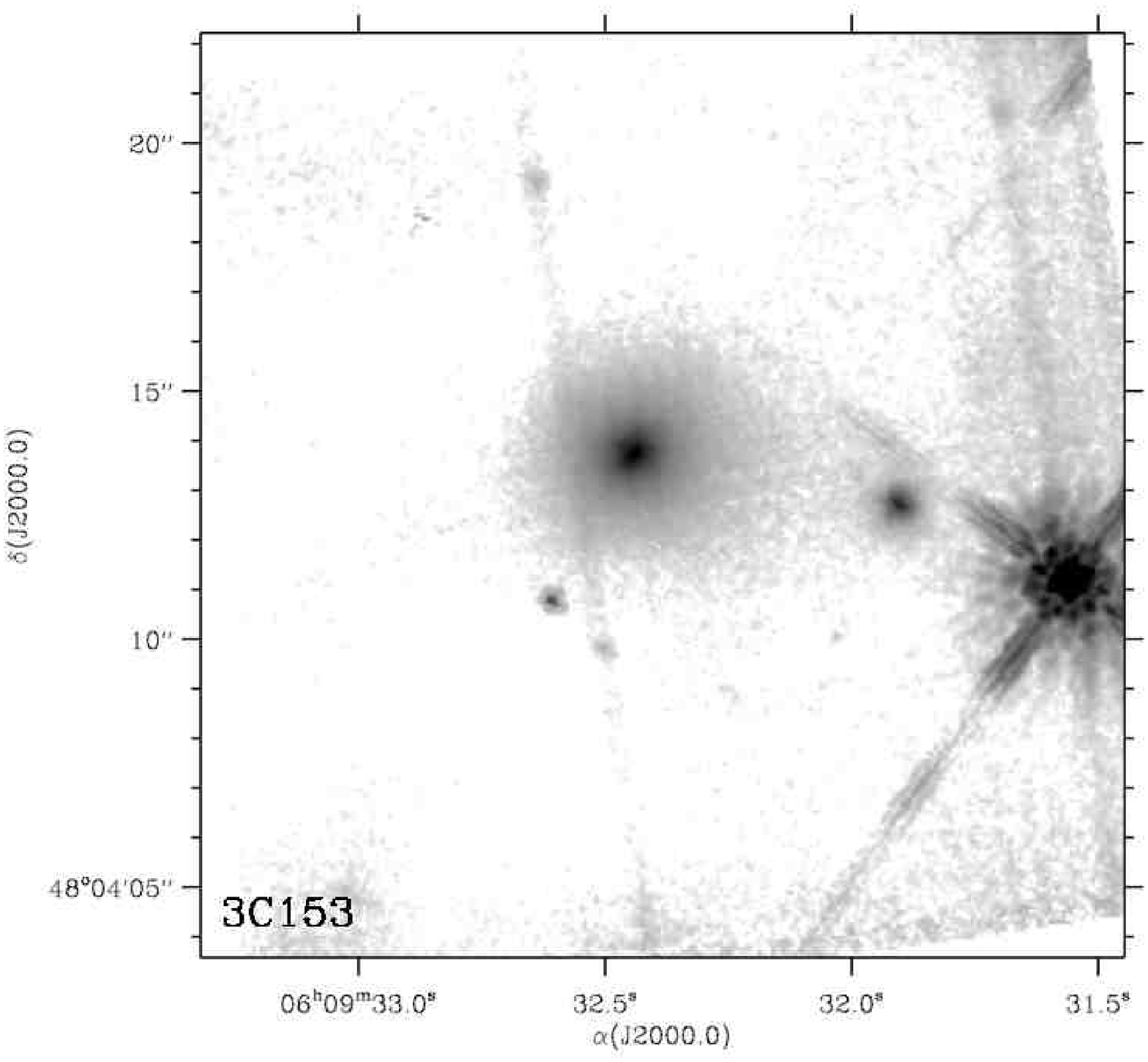}
\caption{\label{fig-newobjects} NICMOS2 f160w ($H$ band) images of the newly observed objects aligned north up -- see table~\ref{tab-new}.}
\end{figure*}

\addtocounter{figure}{-1}

\begin{figure*}[ht]
\plottwo{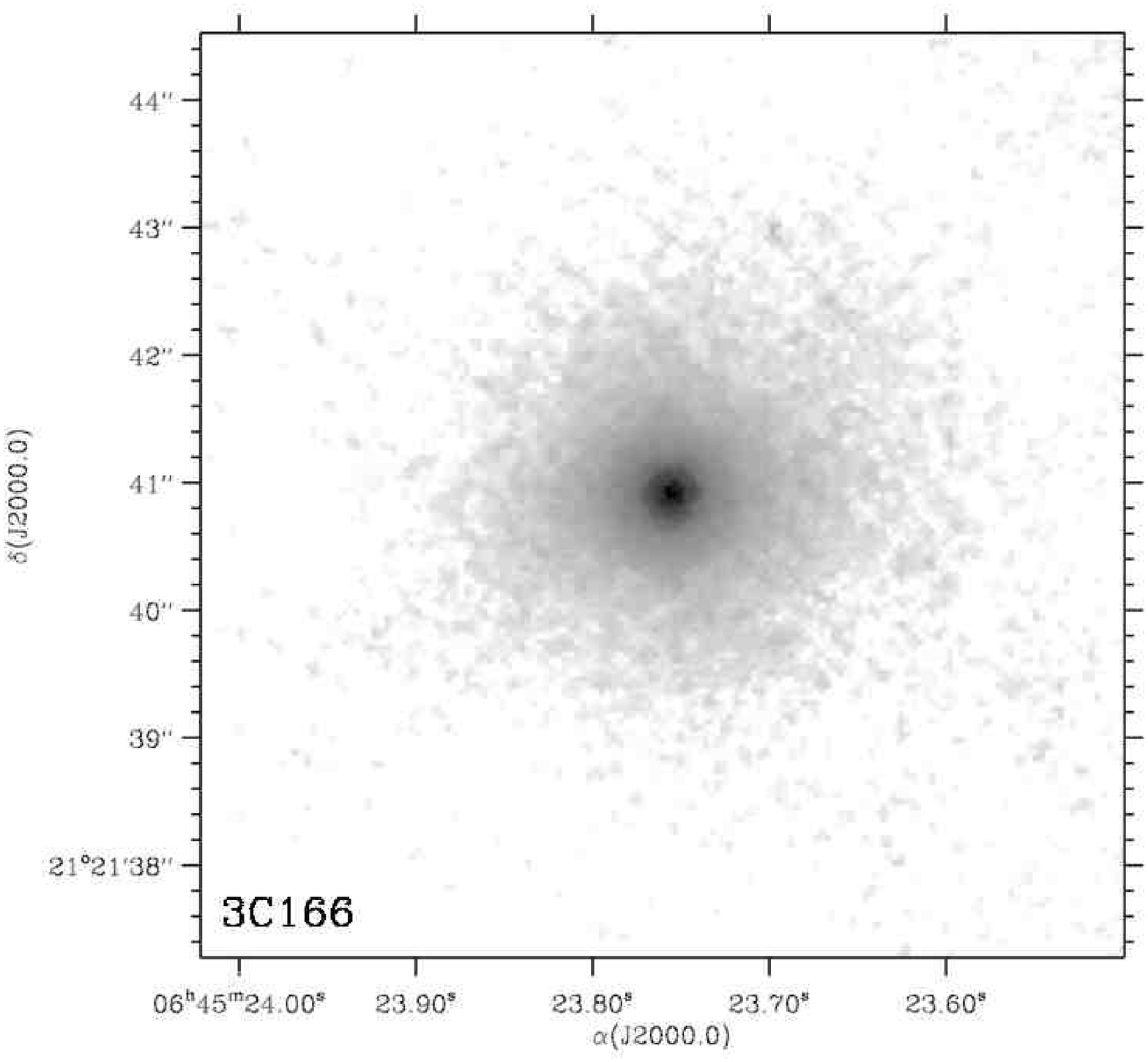}{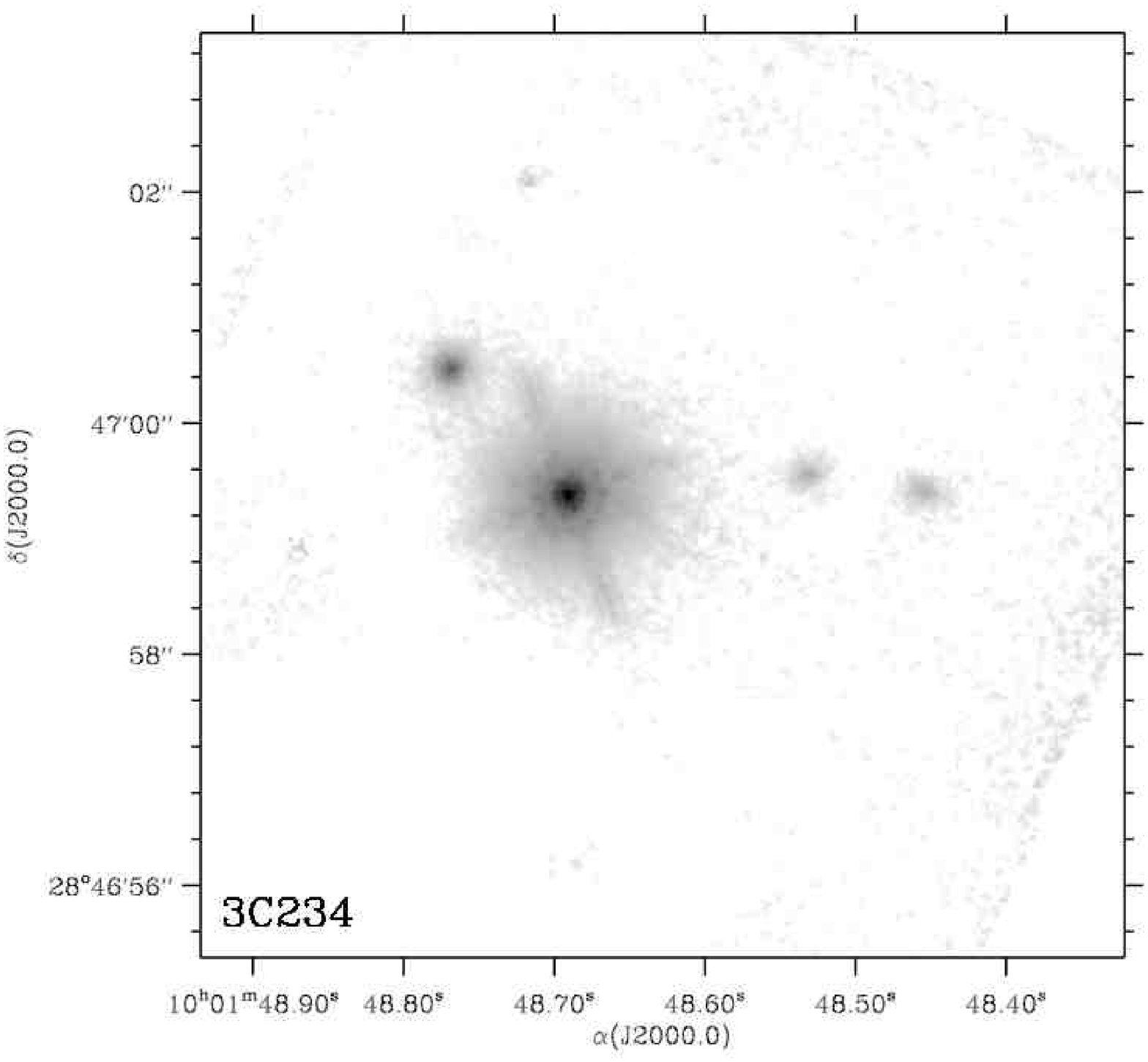}
\plottwo{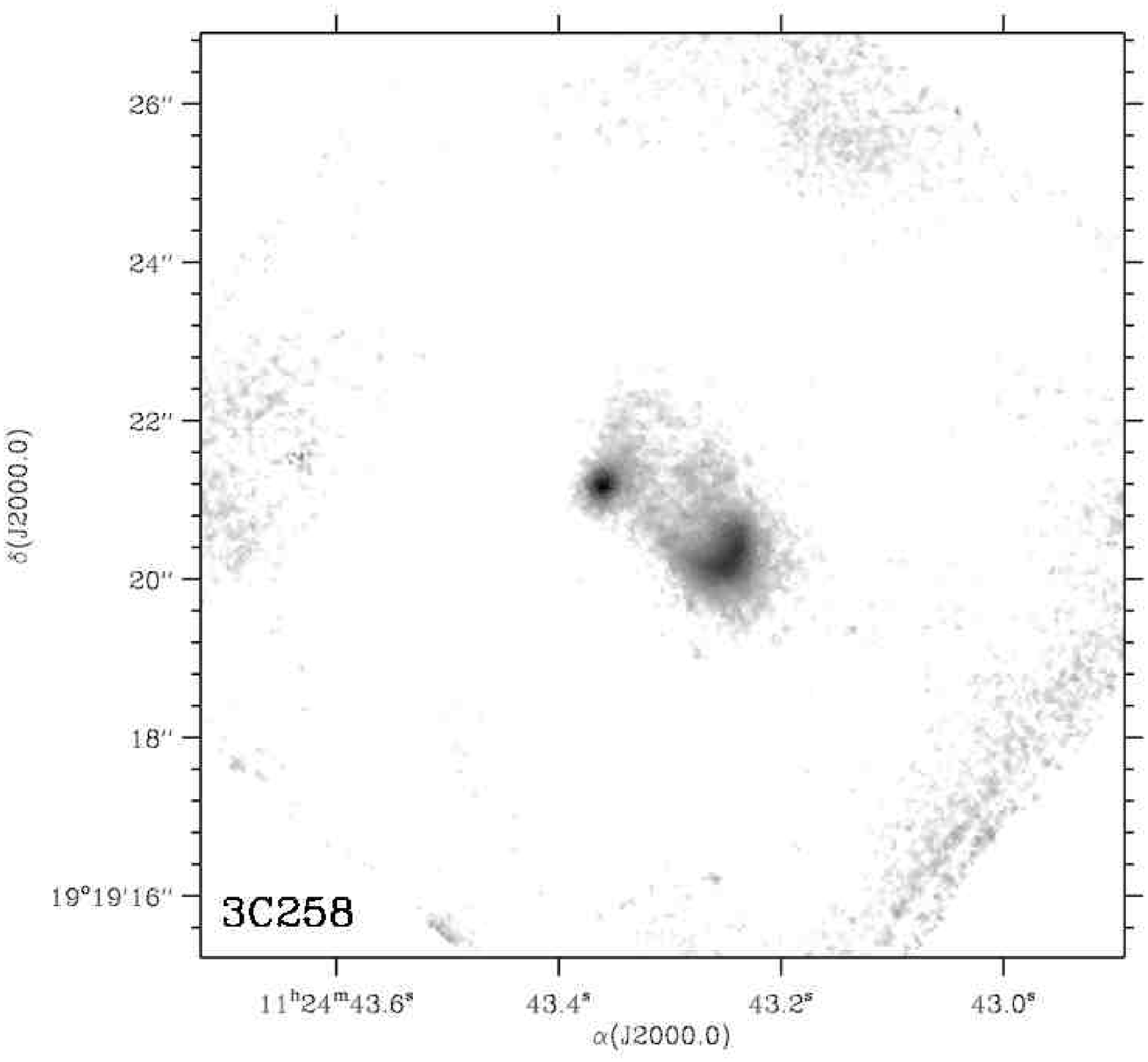}{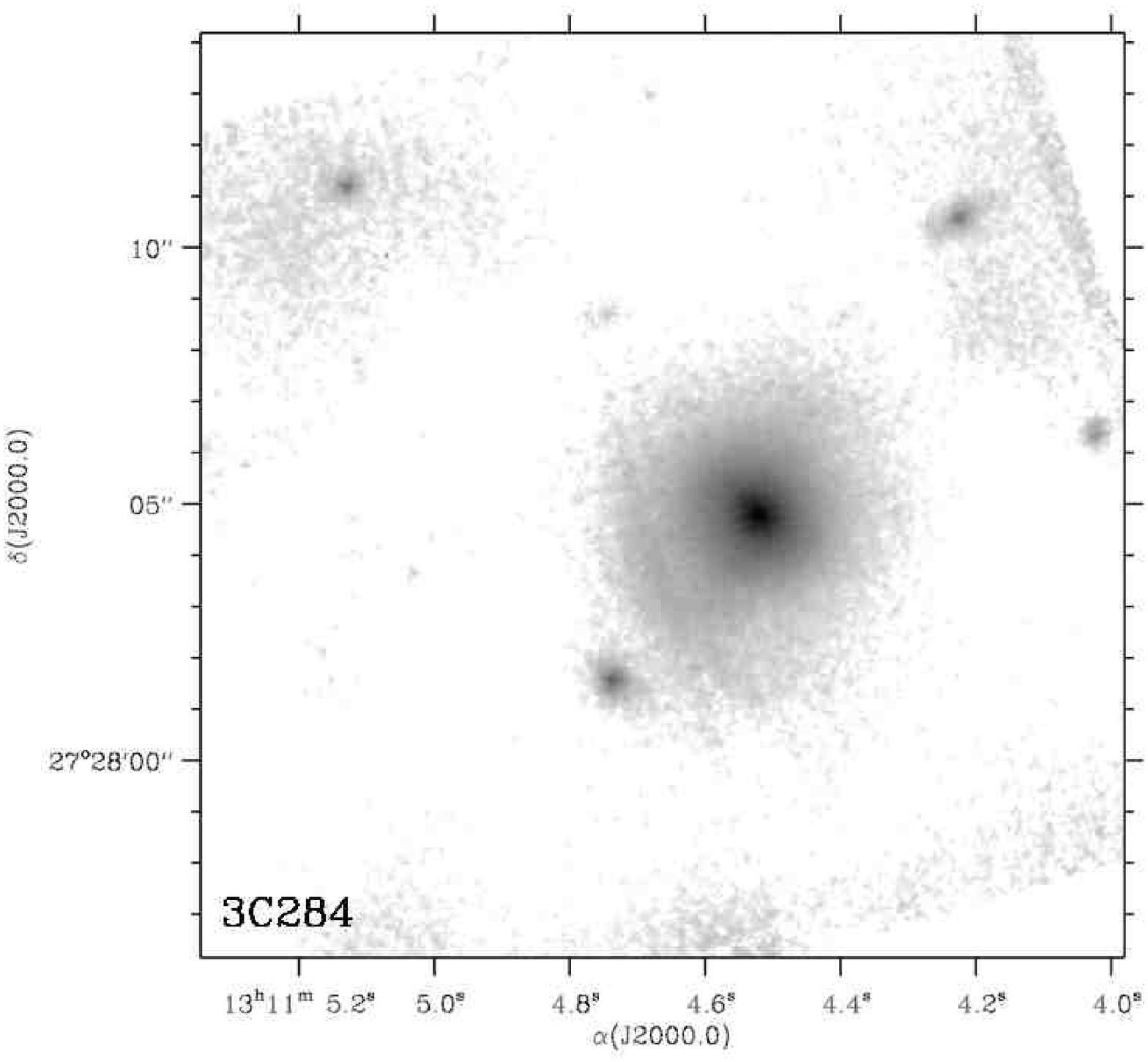}
\plottwo{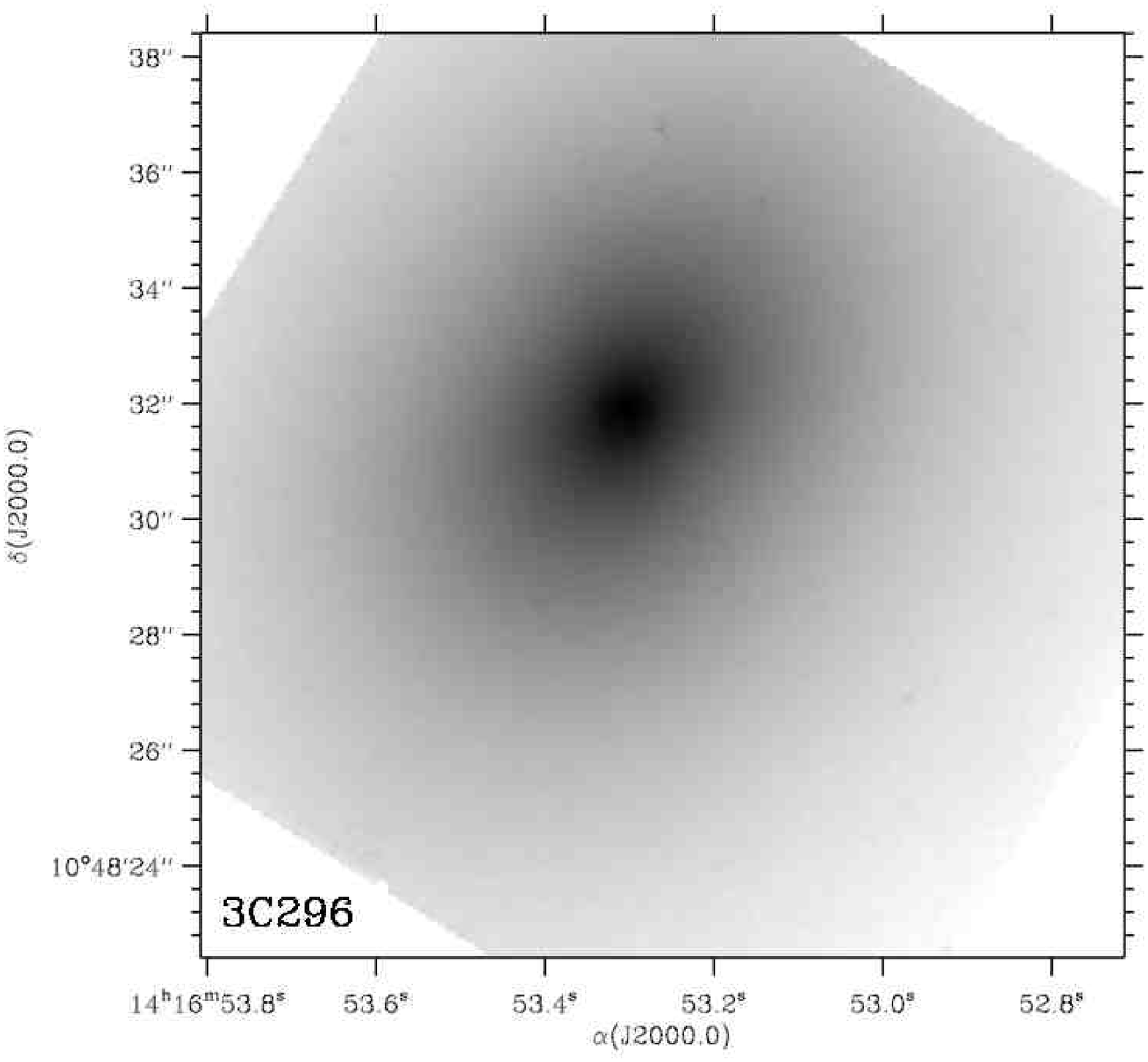}{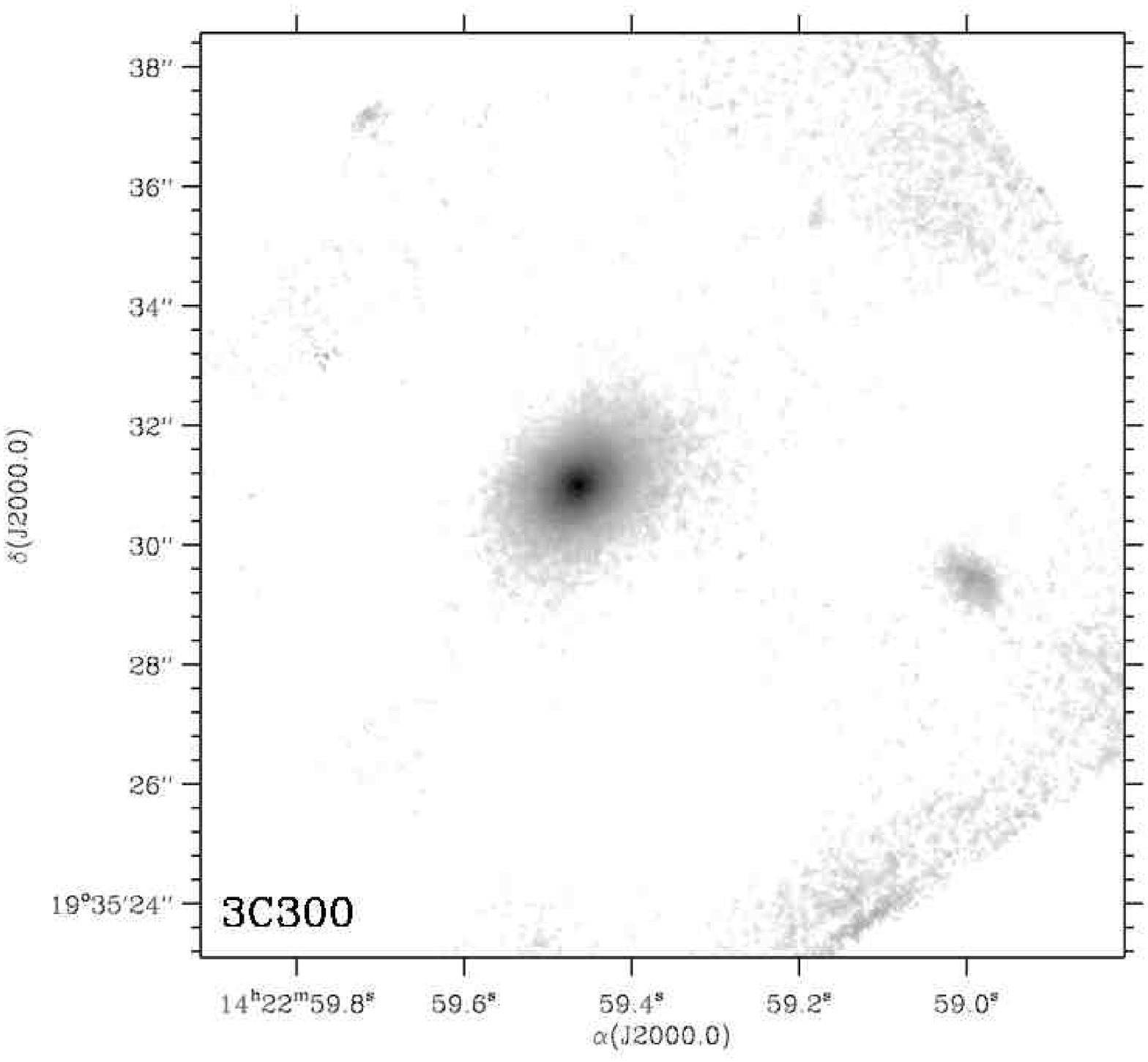}
\caption{(continued). NICMOS2 f160w ($H$ band) images of the newly observed objects aligned north up -- see table~\ref{tab-new}.}
\end{figure*}

\addtocounter{figure}{-1}

\begin{figure*}[ht]
\plottwo{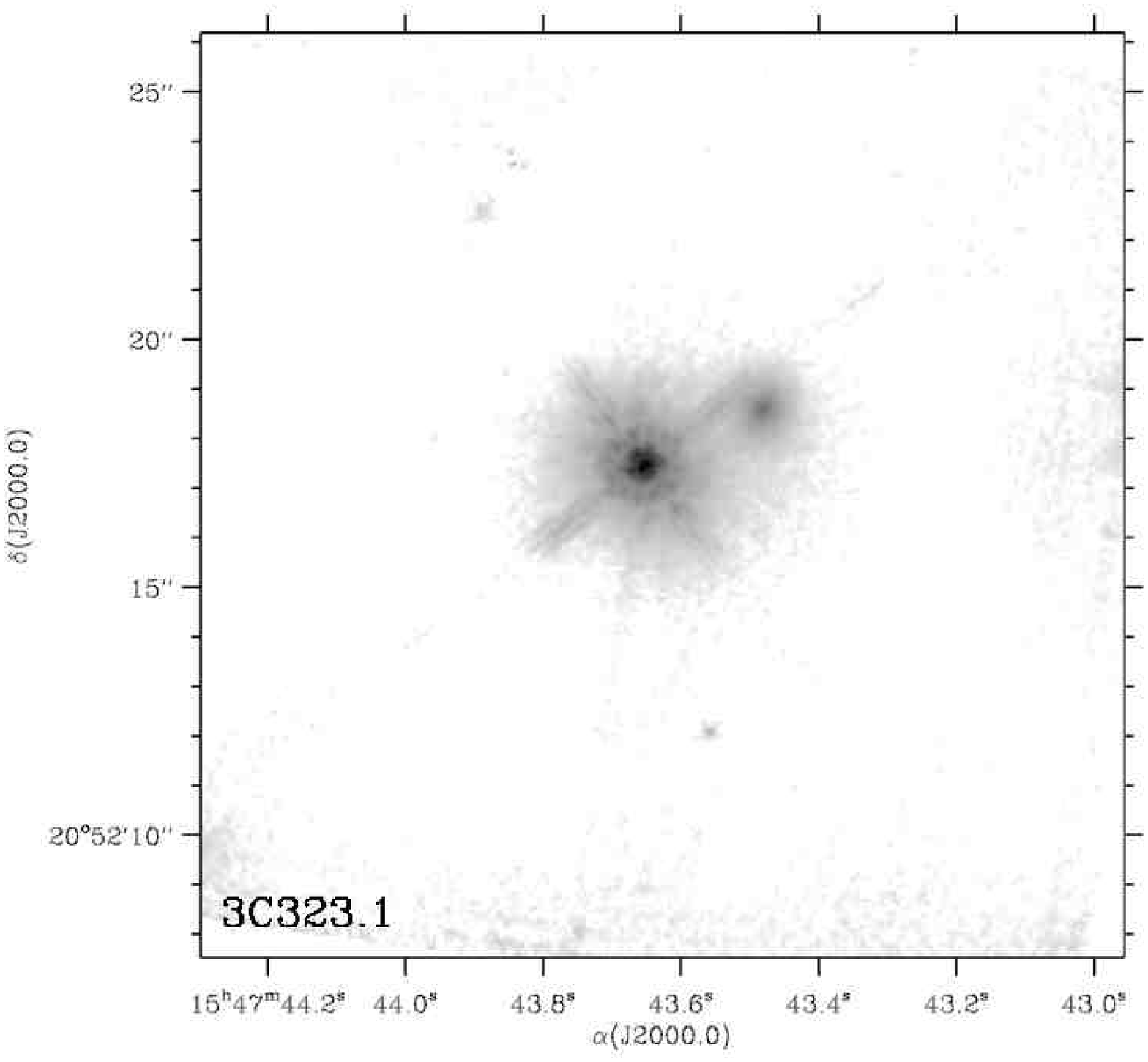}{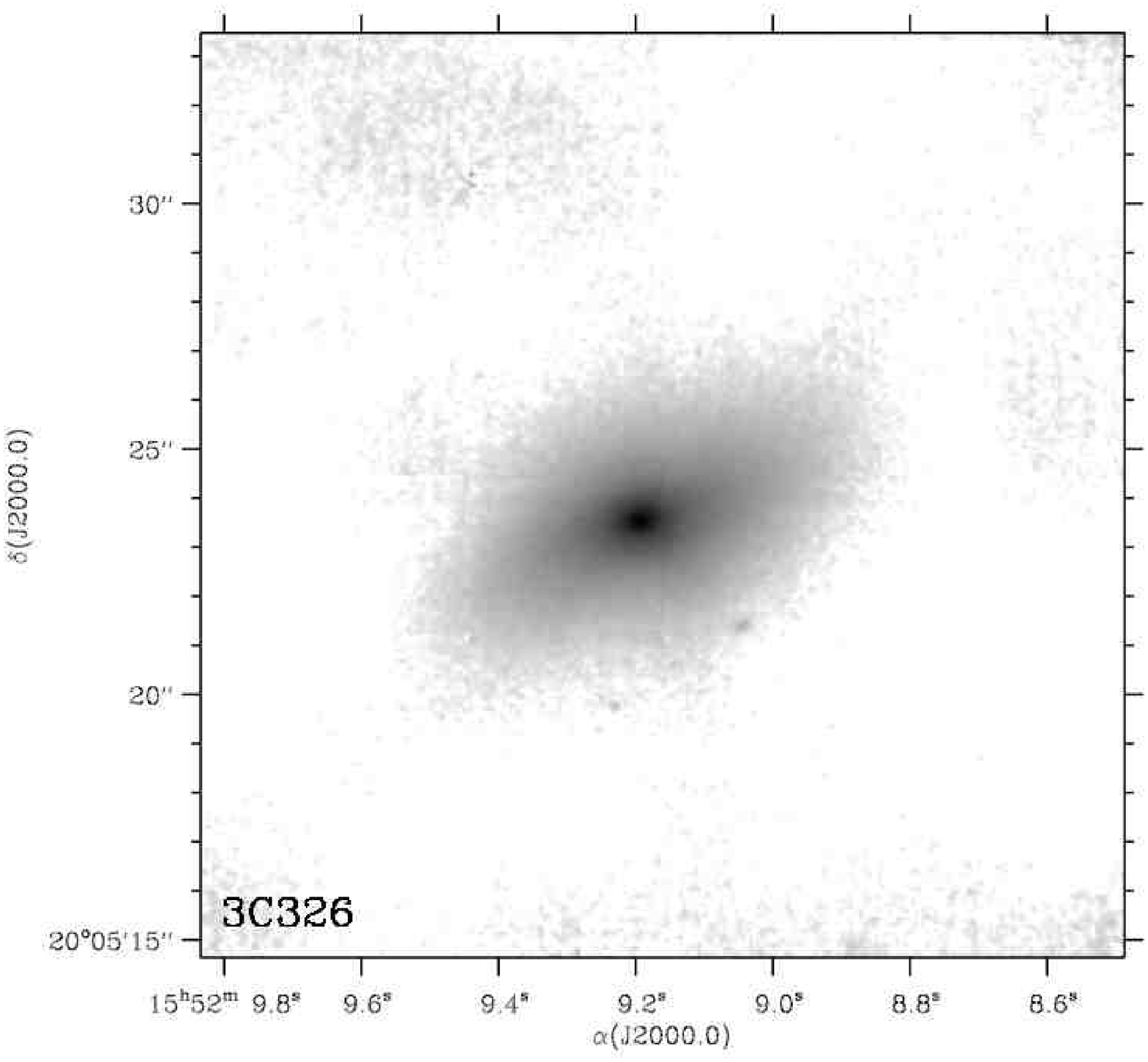}
\plottwo{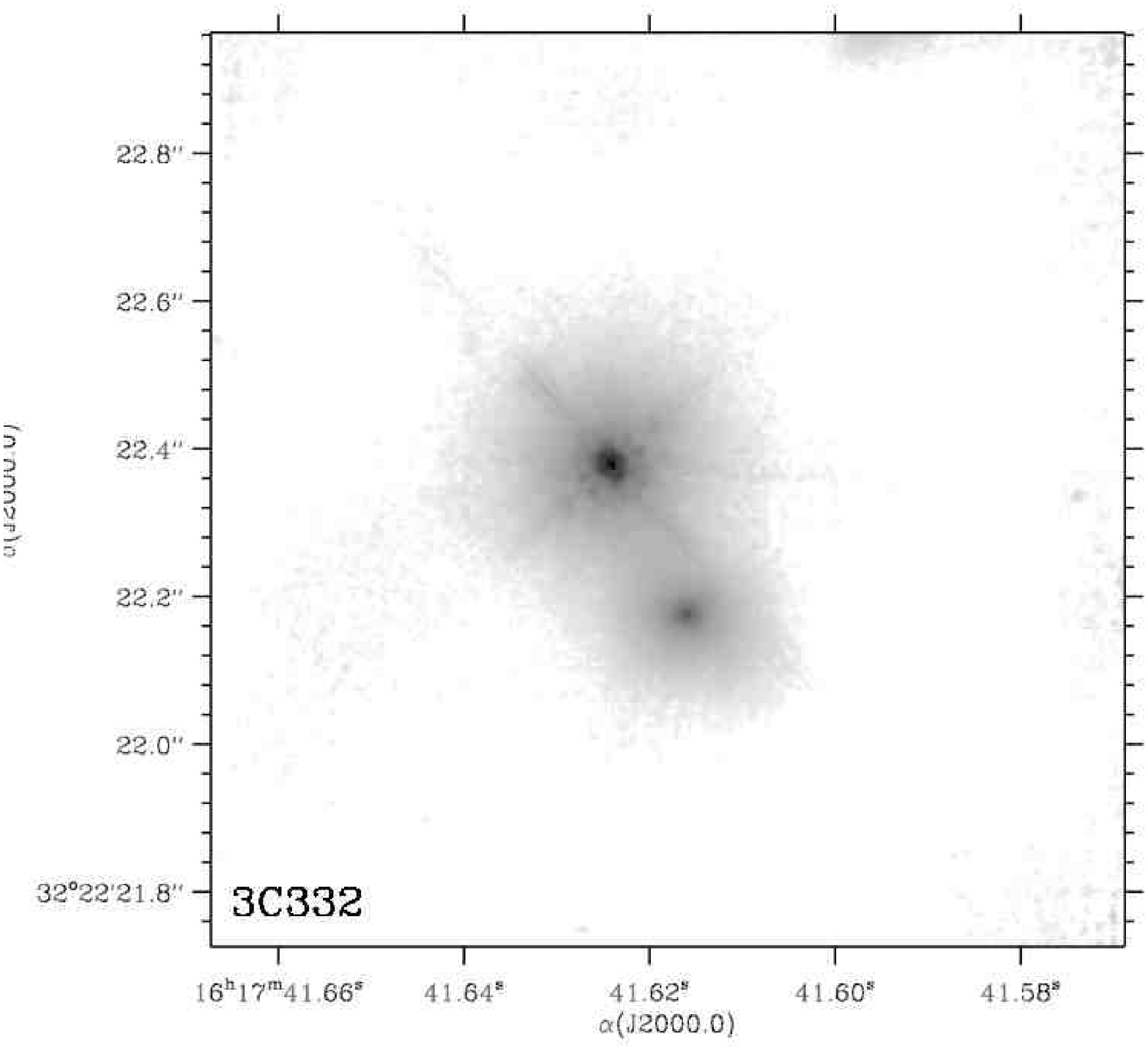}{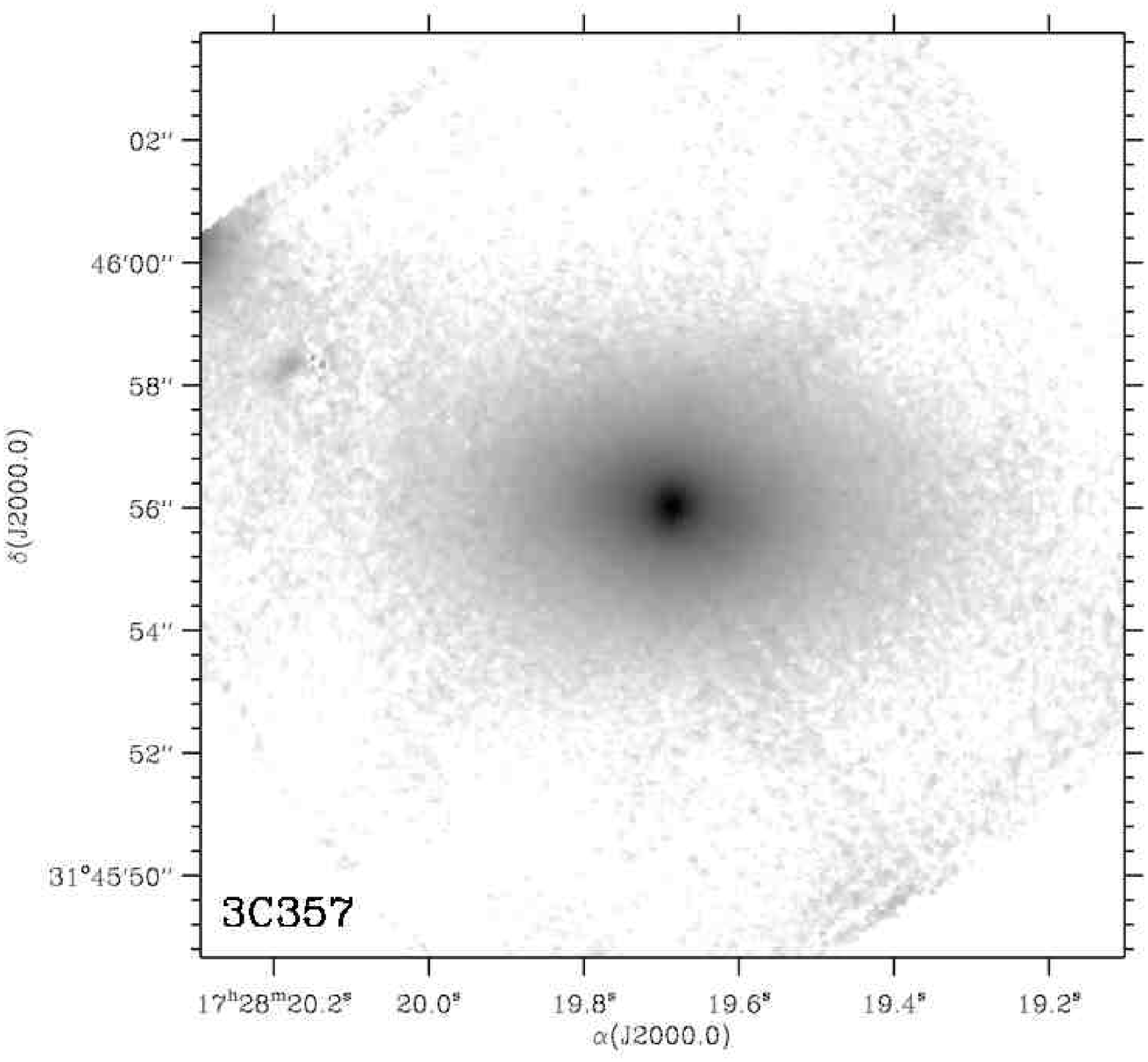}
\plottwo{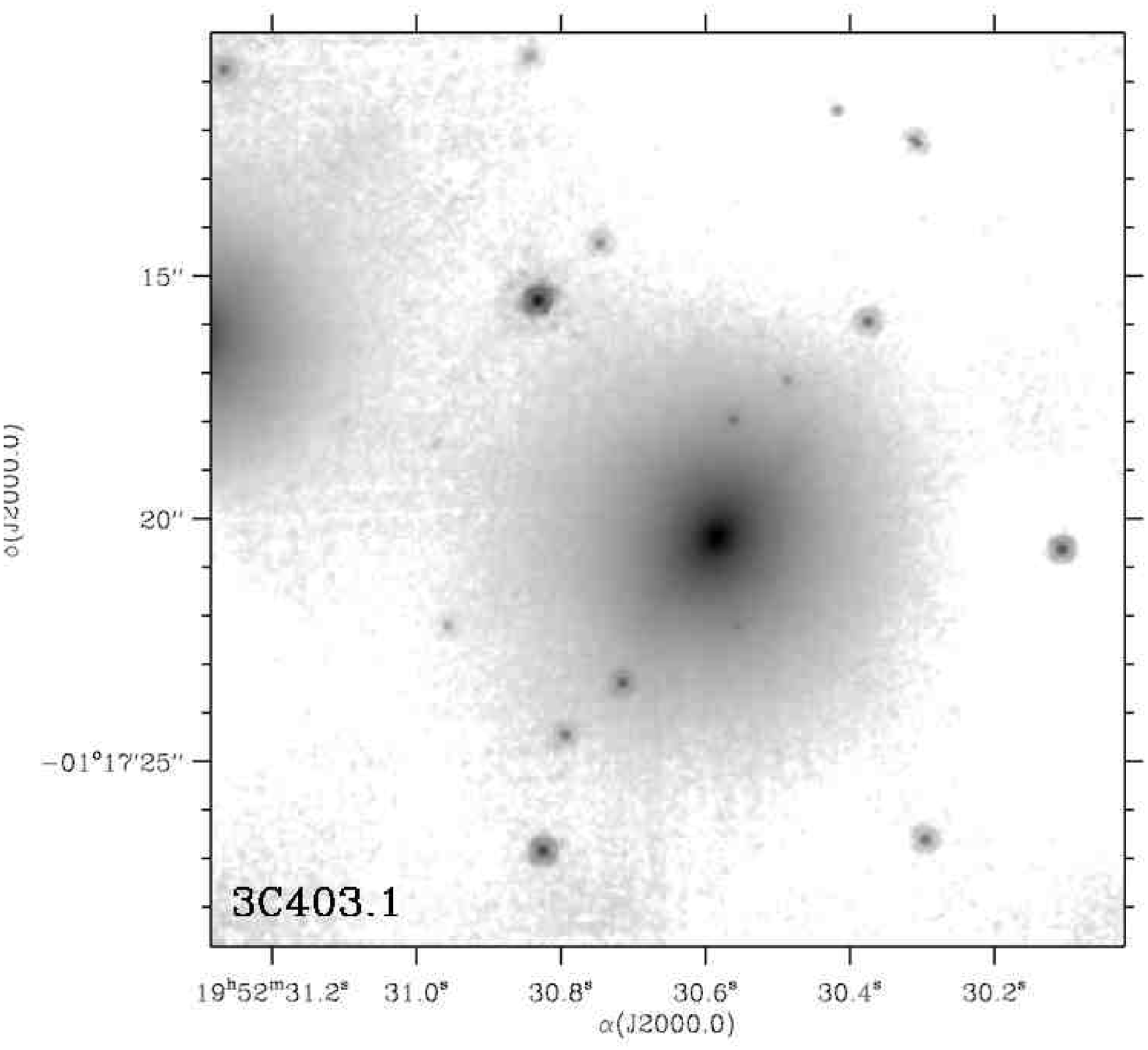}{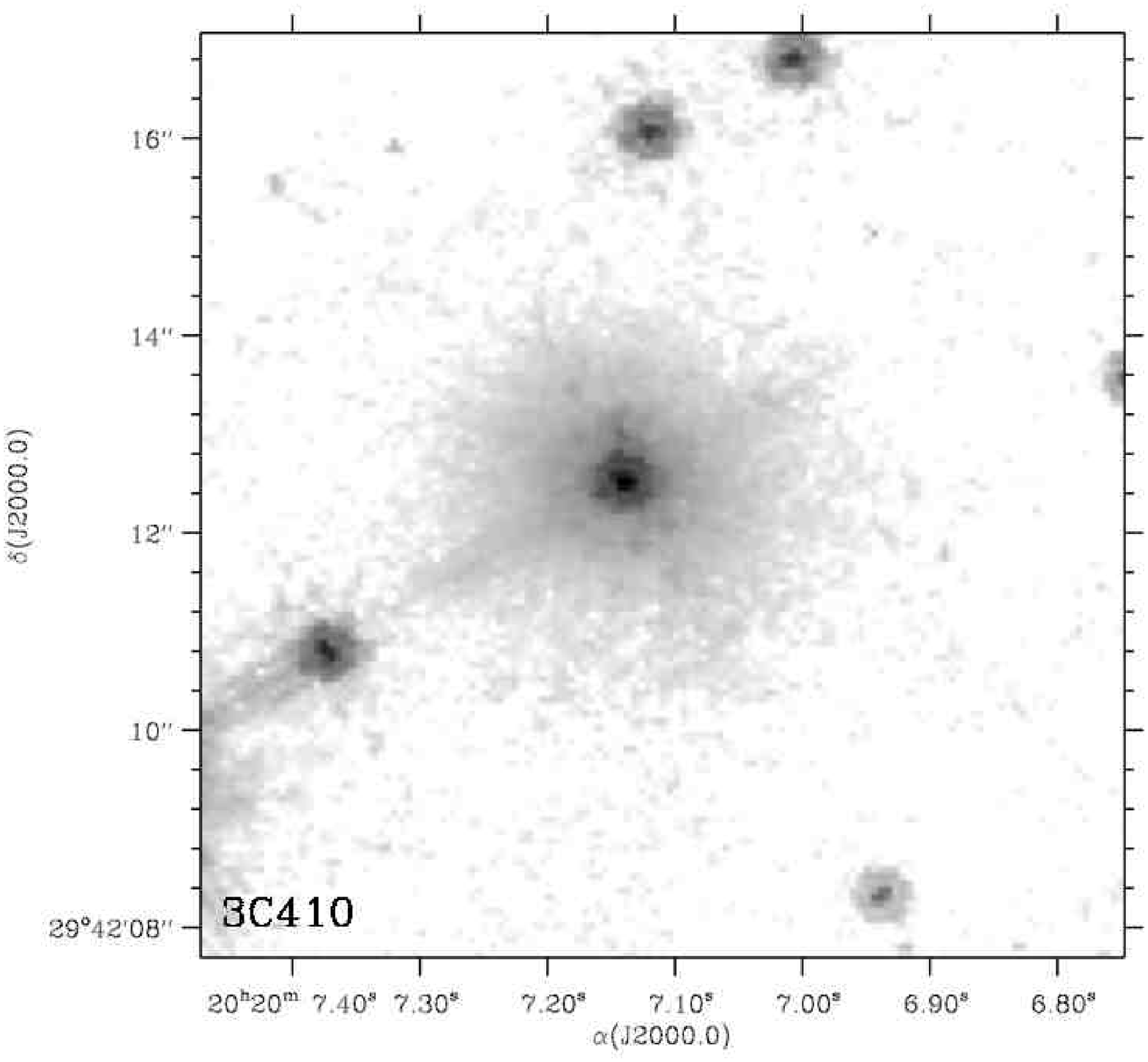}

\caption{(continued). NICMOS2 f160w ($H$ band) images of the newly observed objects aligned north up -- see table~\ref{tab-new}.}
\end{figure*}

\addtocounter{figure}{-1}

\begin{figure*}[ht]
\plottwo{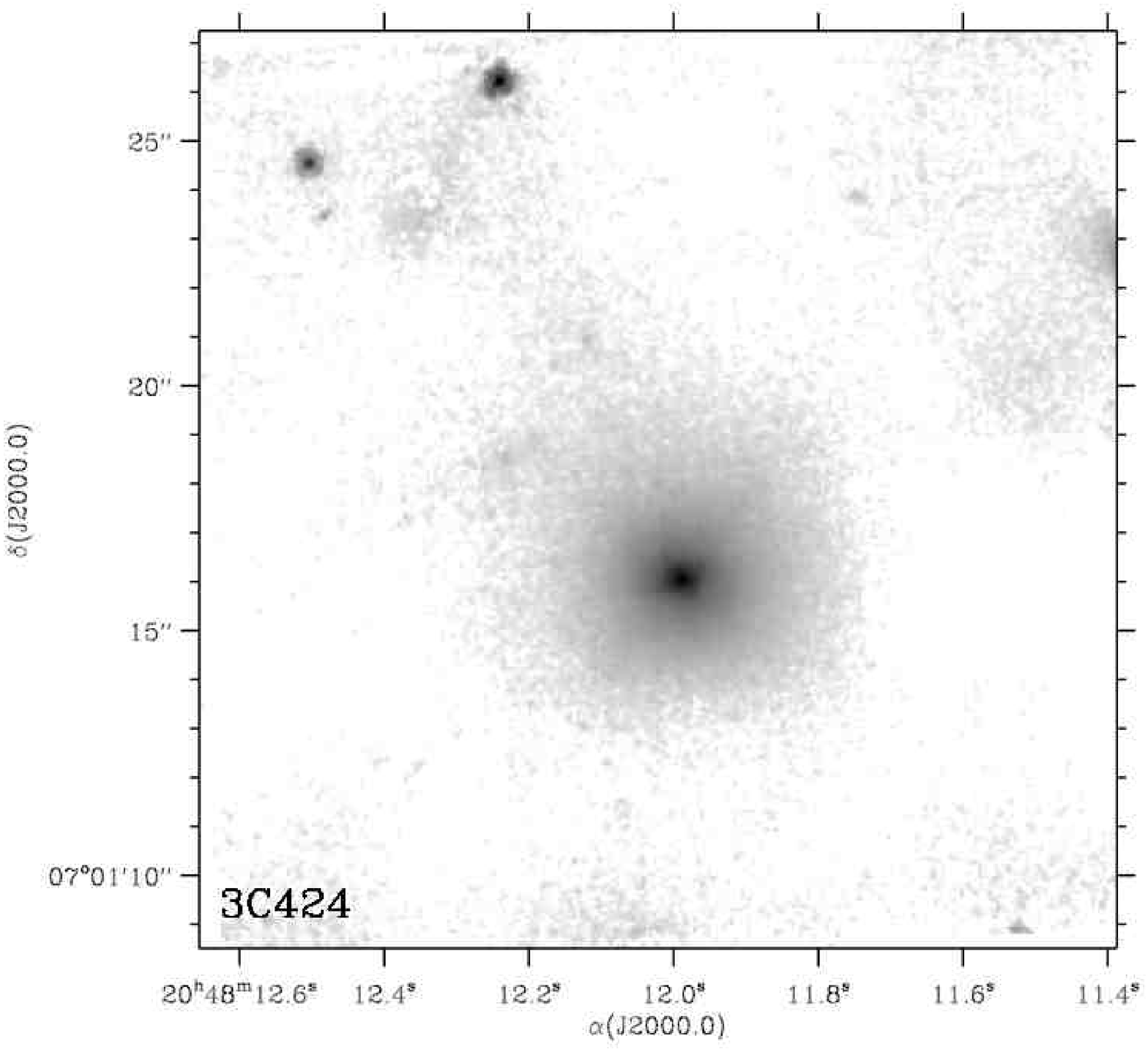}{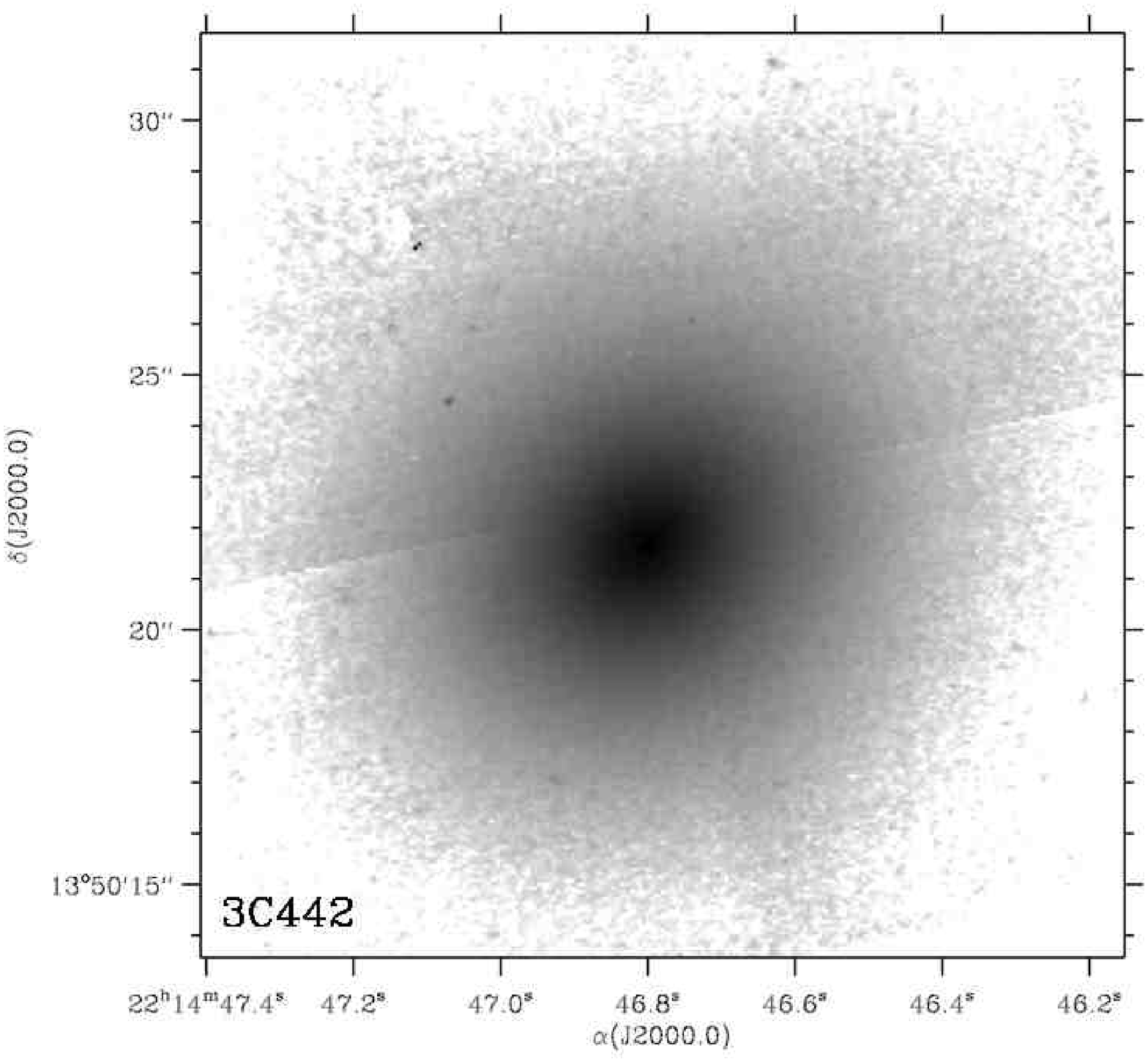}
\epsscale{0.45}
\plotone{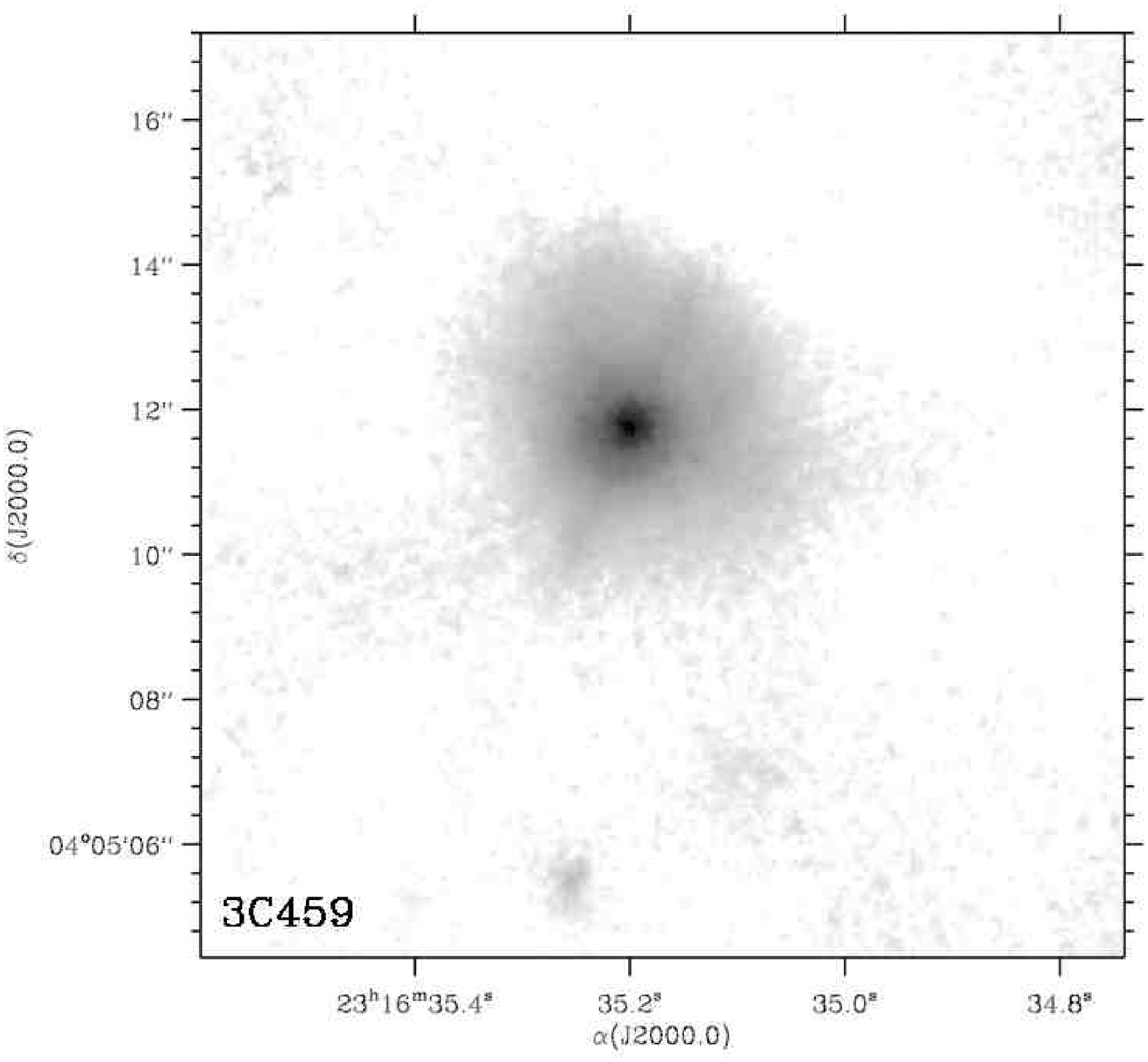}
\epsscale{1.0}
\caption{(continued). NICMOS2 f160w ($H$ band) images of the newly observed objects aligned north up -- see table~\ref{tab-new}.}
\end{figure*}

\clearpage

%%%%%%%%%%%%%%%%%%%%%%%%%%%%%%%%%%%%%%%%%%%%%%%%%

\subsection{Archival data}
\label{sec-archobj}

\subsubsection{3C~71 (NGC~1068, M~77); $z=0.003793$}
\label{sec-3c71}
A well-known spiral galaxy, and the archetypal Seyfert 2, NGC~1068 is not truly ``radio-loud'' in terms of its radio luminosity. It has an obscured Sy~1 spectrum~\citep{antonuccimiller85} and a compact (few arcsecond) radio jet aligned NNE-SSW. The NICMOS data were obtained as part of the NICMOS GTO time by~\citet{thompsoncorbin99}. The galaxy dwarfs the small NIC2 chip, and even the WFPC2 mosaic (PROPOSID's 5479, 5754). We were unable to produce an accurate photometric fit to this object, since we could not estimate the background flux. The spiral structure and active nucleus are readily apparent on the NICMOS image, as are numerous globular clusters. There is a vast literature on this source.

\subsubsection{3C~84 (NGC~1275, Per~A); $z=0.017559$}
The central galaxy of the Perseus cluster contains an FR~I radio source and Seyfert 2 nucleus at the centre of a cooling flow. It shows a large number of companion sources in the infrared, and a complicated dust morphology in the optical-IR. The NICMOS data was first published by ~\citet{martini+03}. Only the central region is visible in the NIC2 image. We used the F702W image from~\citet{martel+99} to estimate the background. However, due to the immense size of this cD galaxy, even the full WFPC2 mosaic (PROPOSID 6228) does not detect the edge of the galaxy, and the results are photometrically uncertain.

\subsubsection{3C~264; $z=0.021718$}
3C~264 is a tailed radio source in the cluster Abell~1367\footnote{http://www.jb.man.ac.uk/atlas/dragns.html}. On the arcsec scale there is a bright one-sided jet which is detected in the radio~\citep{baum+97,lara+99}, optical~\citep{crane+93}, and X-ray~\citep{tilak06,padgett+05}. ~\citet{baum+97} note the existence of an inner dusty disk which causes an apparent ring in the galaxy surface brightness where the disk ends. The NICMOS image shows a bright nucleus, the jet and the ring noted by Baum et al. The data was previously published by~\citet{capetti+00}.

\subsubsection{3C~270 (NGC~4261); $z=0.007465$}
A well-studied nearby FR~I with a boxy elliptical host galaxy and a circumnuclear dust disk. \cite{ferrarese+96} used imaging and spectroscopy of this source to determine its black-hole mass. A faint unresolved nuclear point source is visible in the IR. Data first published by~\citet{quillen+00}.

\subsubsection{3C~272.1 (NGC~4374, M~84); $z=0.003536$}
A large well-studied elliptical in the Virgo cluster with a prominent dust lane visible in the optical, and faintly detectable in the NICMOS image~\citet{bower+00}. The radio source is a double-sided FR~I jet and core. However, its radio power is somewhat lower than the fiducial cutoff for radio-loud quasars. It is described as a ``weak radio galaxy'' by~\citet{laingbridle87}.

\subsubsection{3C~274 (NGC~4486, M~87, Virgo A); $z=0.004360$}
An FR~I with celebrated optical jet. The NICMOS data was originally presented by~\citep{corbin+02}, where the source is studied in detail.

\subsubsection{3C~293 (UGC~08782); $z=0.045034$}
A merger-remnant with significant distortion at optical and IR wavelengths. The prominent optical dust lanes partially cover an optical jet that is clearly seen in the infrared. This object was studied in detail by~\citet{floyd+06a}.

\subsubsection{3C~305 (IC~1065; UGC 09553); $z=0.041639$}
A nearby FR~I and merger remnant, with prominent dust lane across the nucleus in the optical. Studied in detail by~\citet{jackson+03}. The host galaxy is a large, elongated elliptical, with clear signs of disturbance in the optical and the IR. See~\citet{jackson+03} for original publication of the data, and a detailed study of the object.

\subsubsection{3C~317 (Abell~2052; UGC~09799); $z=0.034457$}
3C~317 is  a slightly elongated elliptical with  an unresolved nucleus. Two small elliptical galaxies are present in the field of view, one is located $7\arcsec$ to the northeast,  and the second one $9\arcsec$ to the  northwest.  The  image  of this  galaxy  taken by~\citet{martel+02} shows a peculiar UV filament  $\sim$ 4~kpc south of the nucleus. This filament is described as a region of active star formation likely triggered by a recent merger by~\citet{martel+02}. The data was first published by~\citet{quillen+00}.

\subsubsection{3C~338 (Abell~2199; NGC~6166); $z=0.030354$}
An FR~I radio source in a dense environment. Two companions are prominent on the NICMOS image. The elliptical host galaxy is disturbed, with a dust lane crossing the nucleus, faintly visible in the IR. First published by~\citet{jensen+01}. See also~\citet{ravindranath+01}.

\subsubsection{3C~405 (Cyg~A); $z=0.056075$}
3C~405 is the archetypal FR~II radio galaxy~\citep{carillibarthel96}. There is a buried quasar detected in the host galaxy~\citep{djorgovski+91}. The hot spots are detected in the X-ray (e.g.~\citealt{wilson+00}. The galaxy envelope fills the $19\times19$\arcsec NIC2 chip, with the hot spots lying outside the NICMOS image. The inner galaxy contains interesting patchy and filamentary structure. There is strong dust absorption running roughly E-W below the nucleus. There is a strong patch of dust just NE of the nucleus. There is also what looks like part of an ionization cone pointing to the north-west. At the highest brightness levels, we see a bright point source centered on a rough ``X'' shape which may define the edges of the ionization cone. Similar results have been described by ~\citet{tadhunter+99}.

\begin{figure*}[ht]
\plottwo{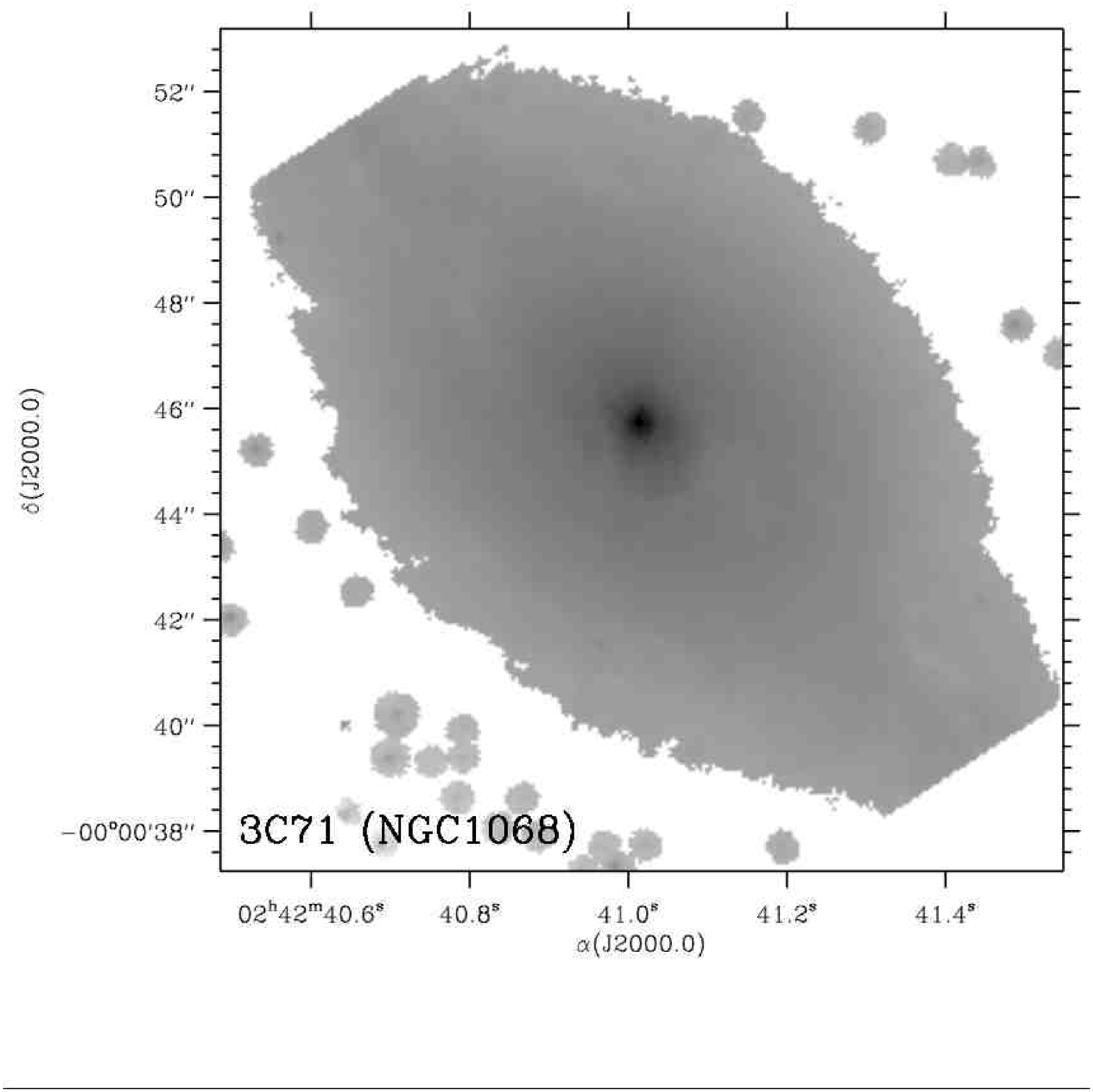}{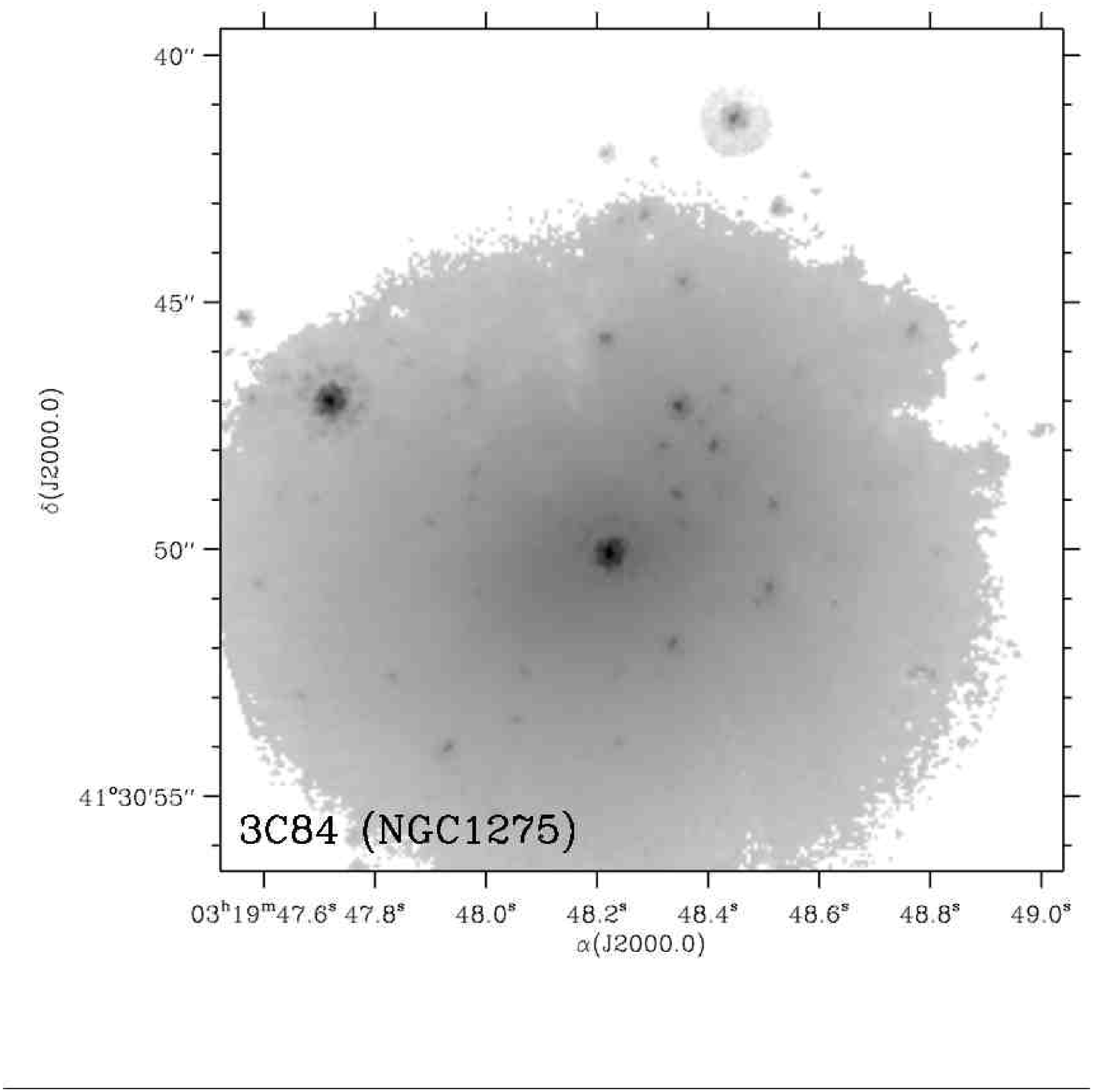}
\plottwo{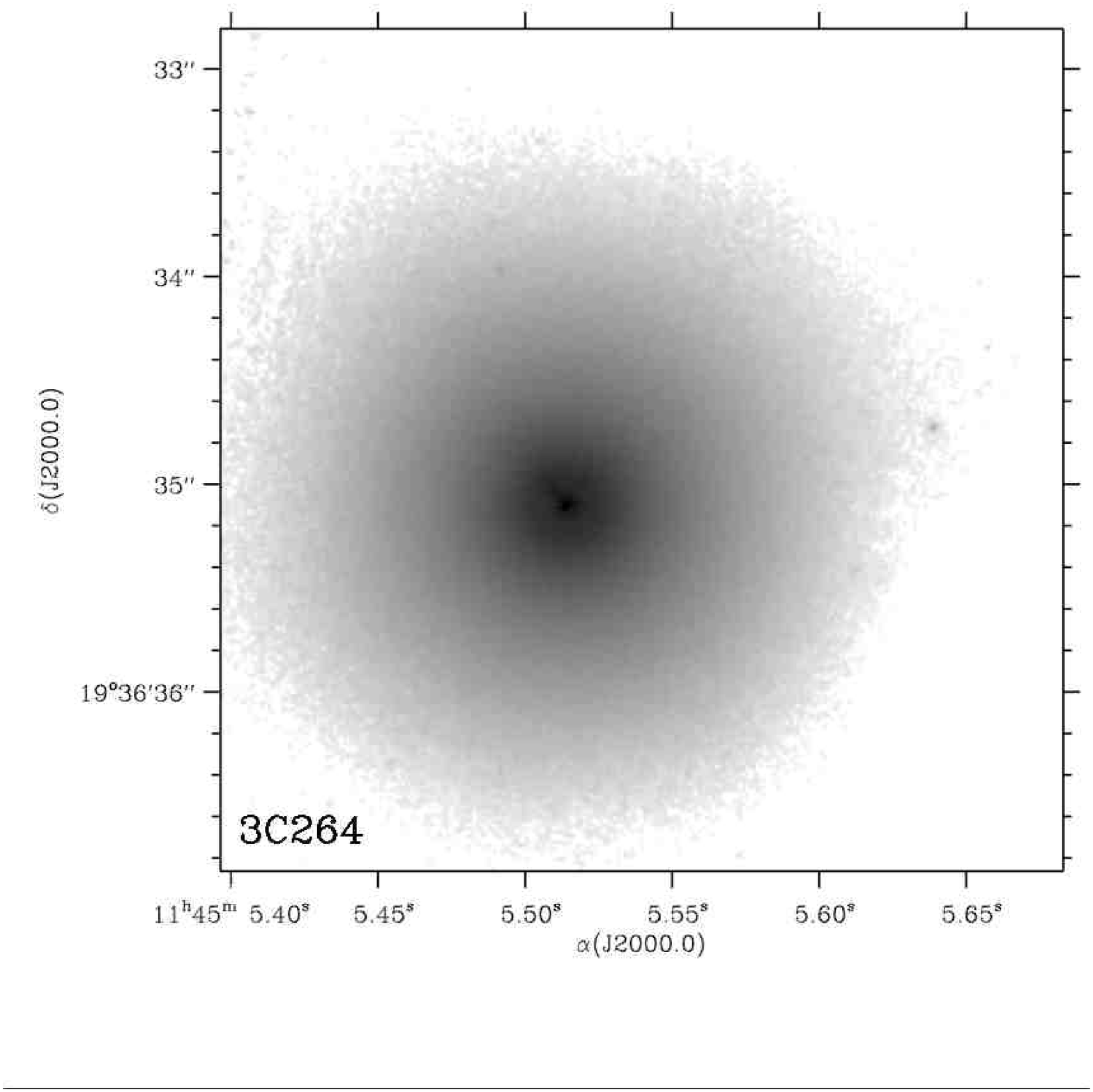}{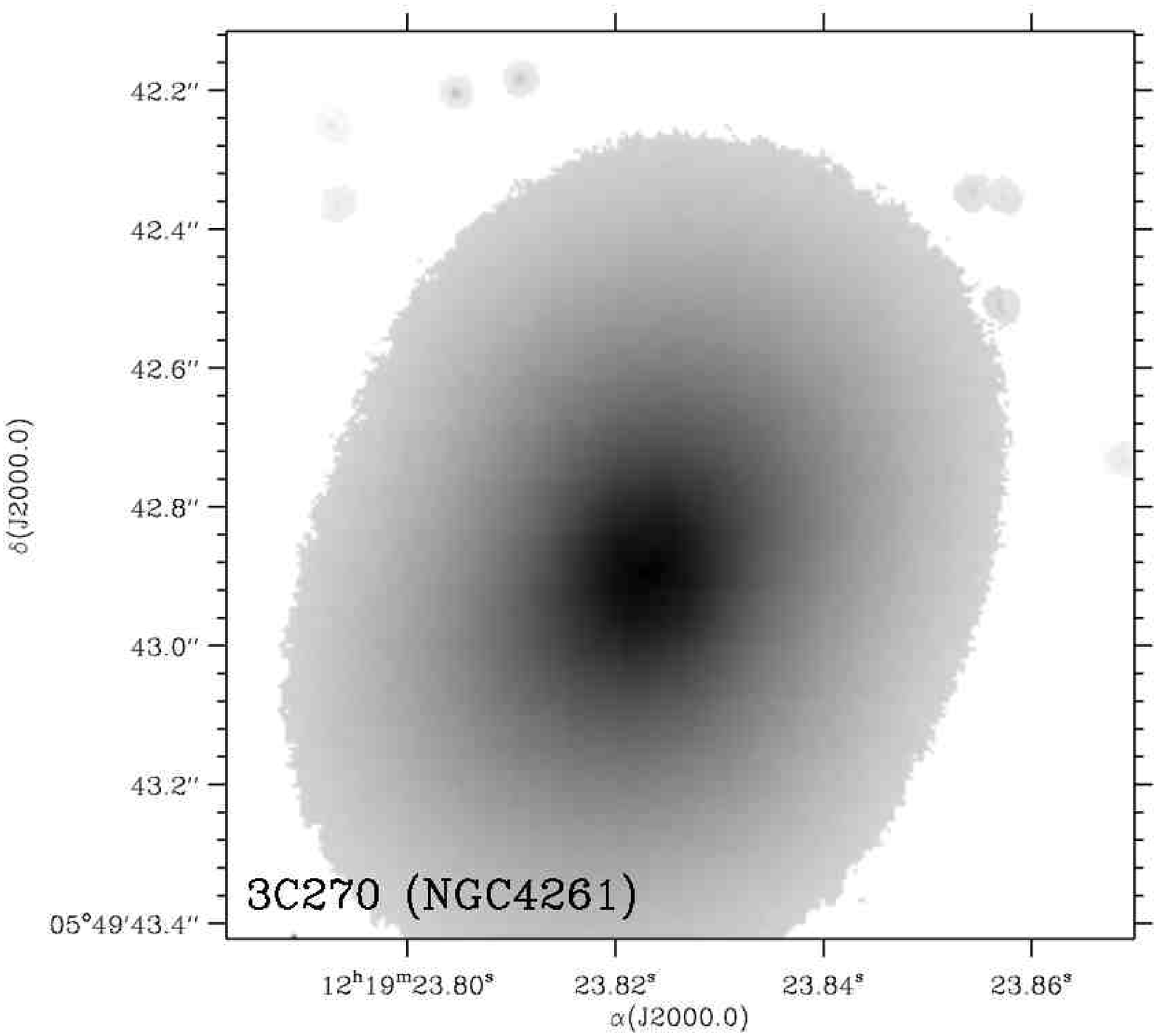}
\plottwo{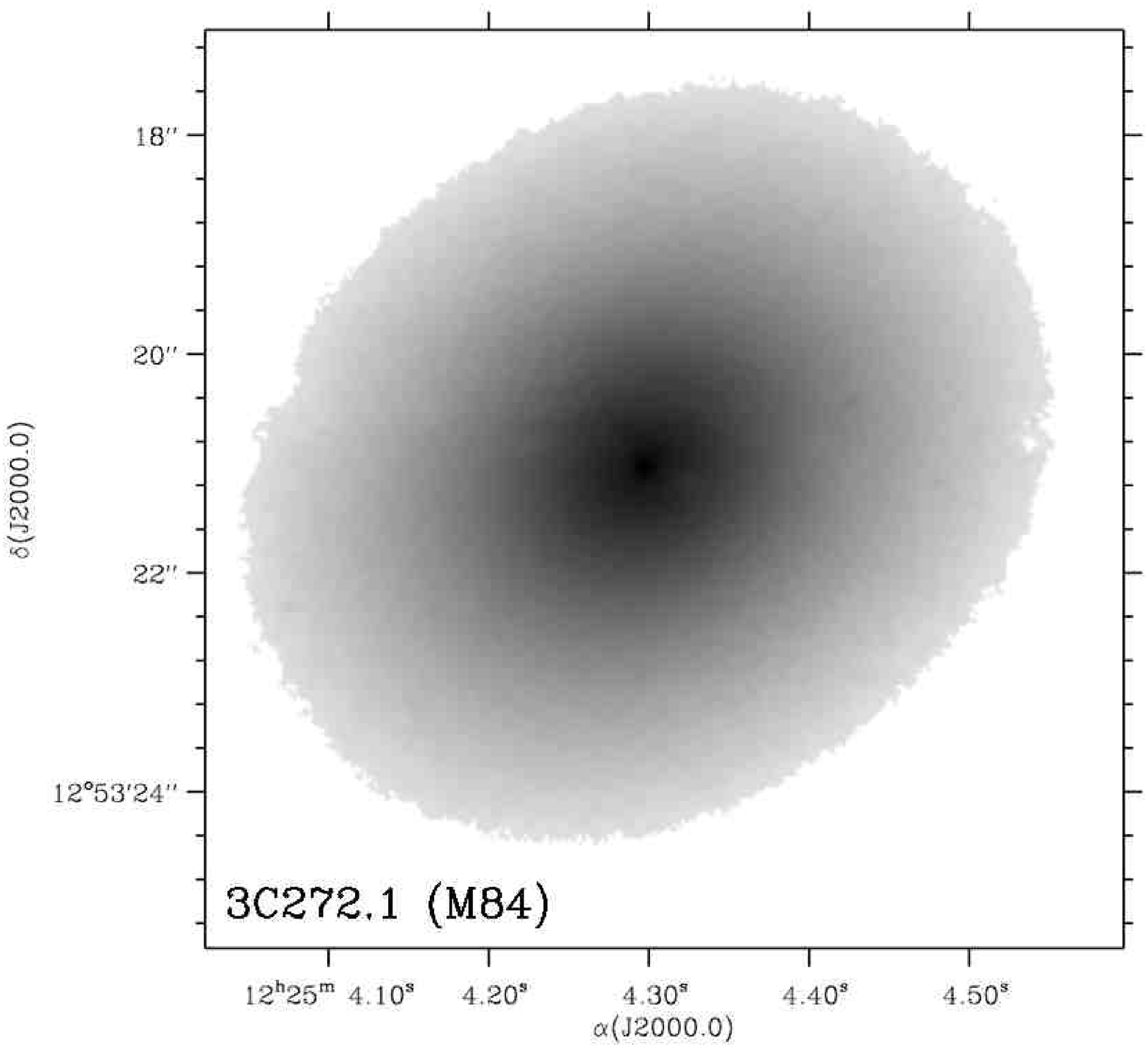}{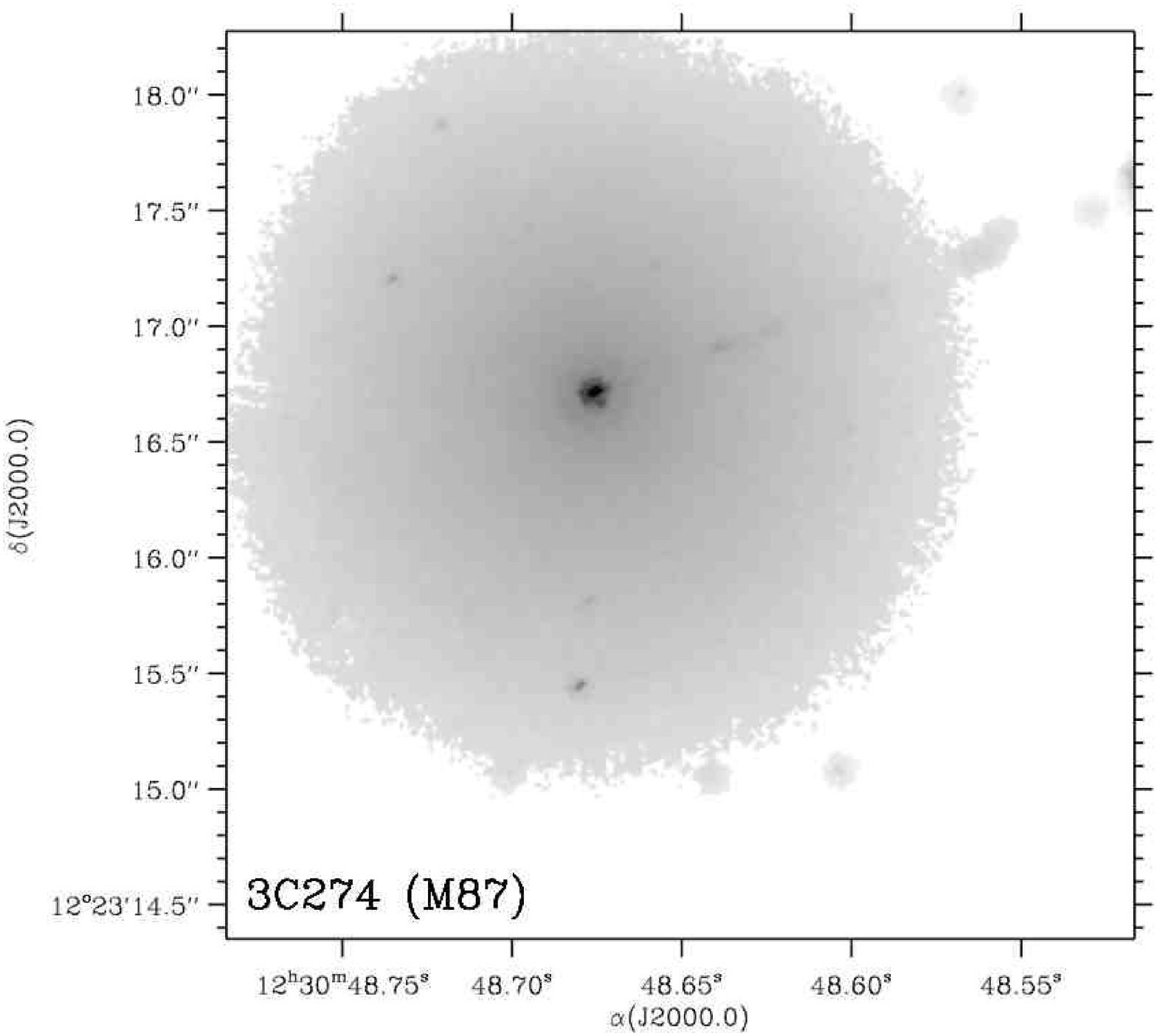}
\caption{\label{fig-archobjects} NICMOS2 f160w ($H$ band) images of the archival objects -- see table~\ref{tab-arch}.}
\end{figure*}

\addtocounter{figure}{-1}

\begin{figure*}[ht]
\plottwo{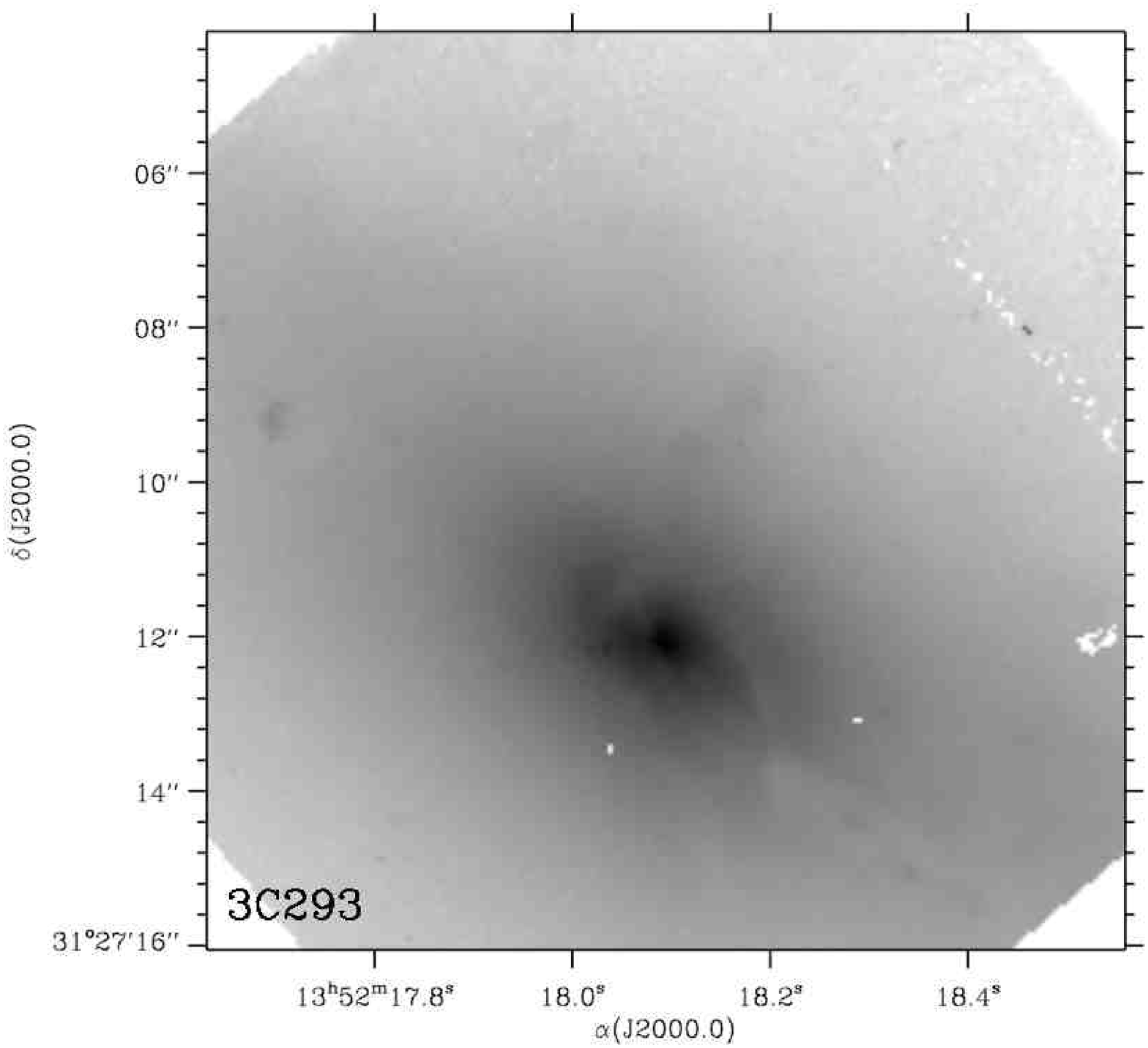}{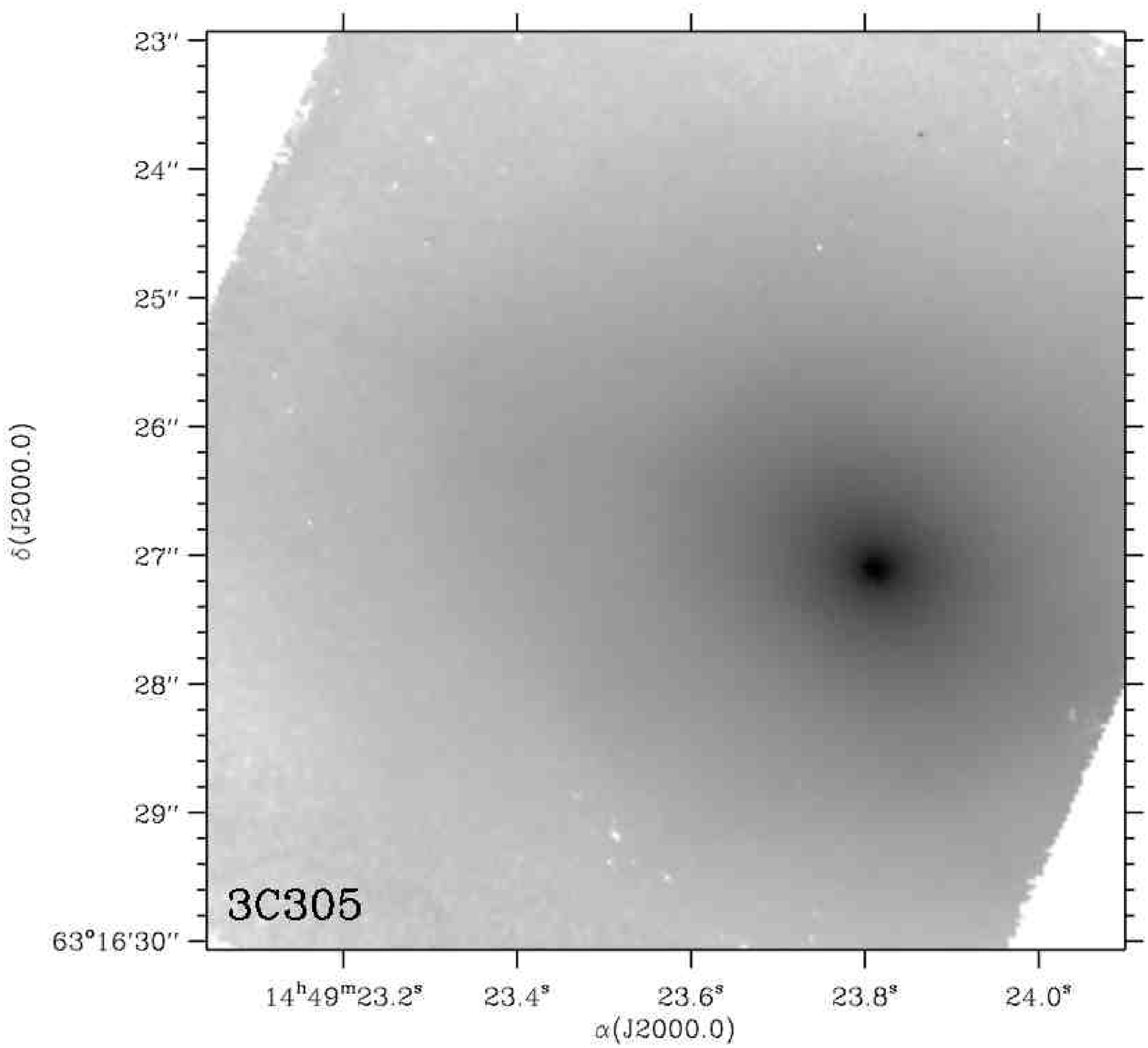}
\plottwo{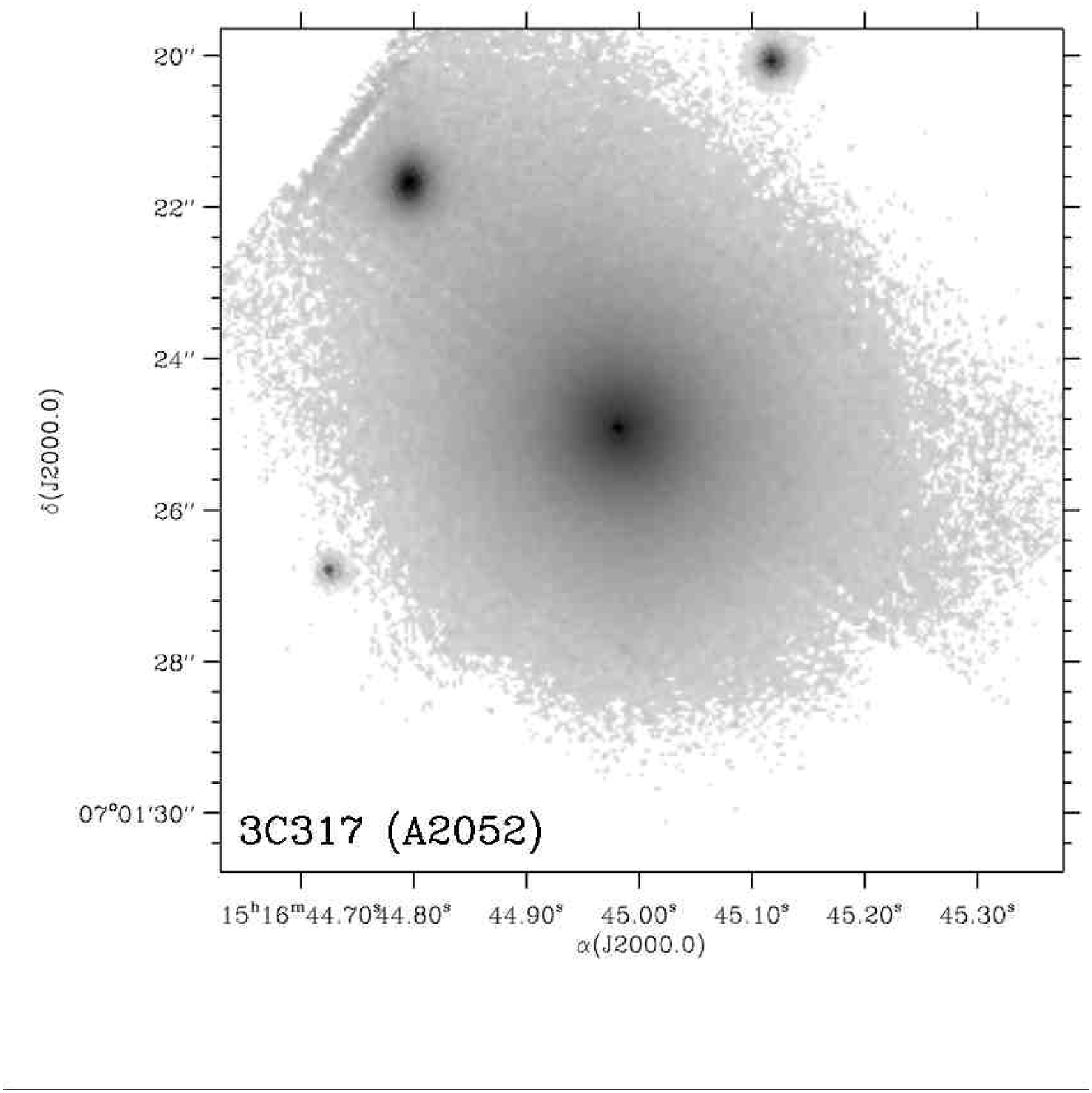}{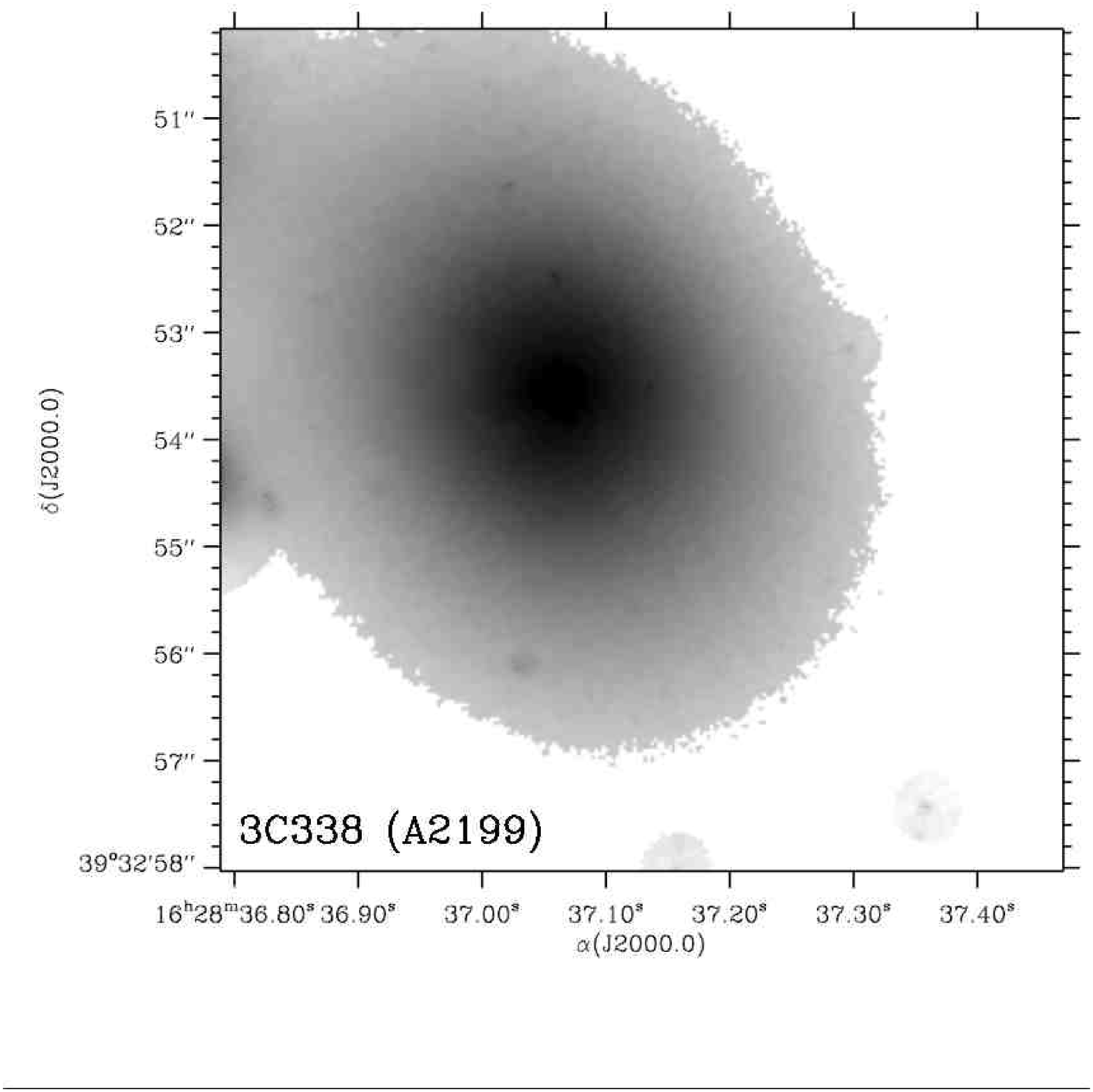}
\epsscale{0.45}
\plotone{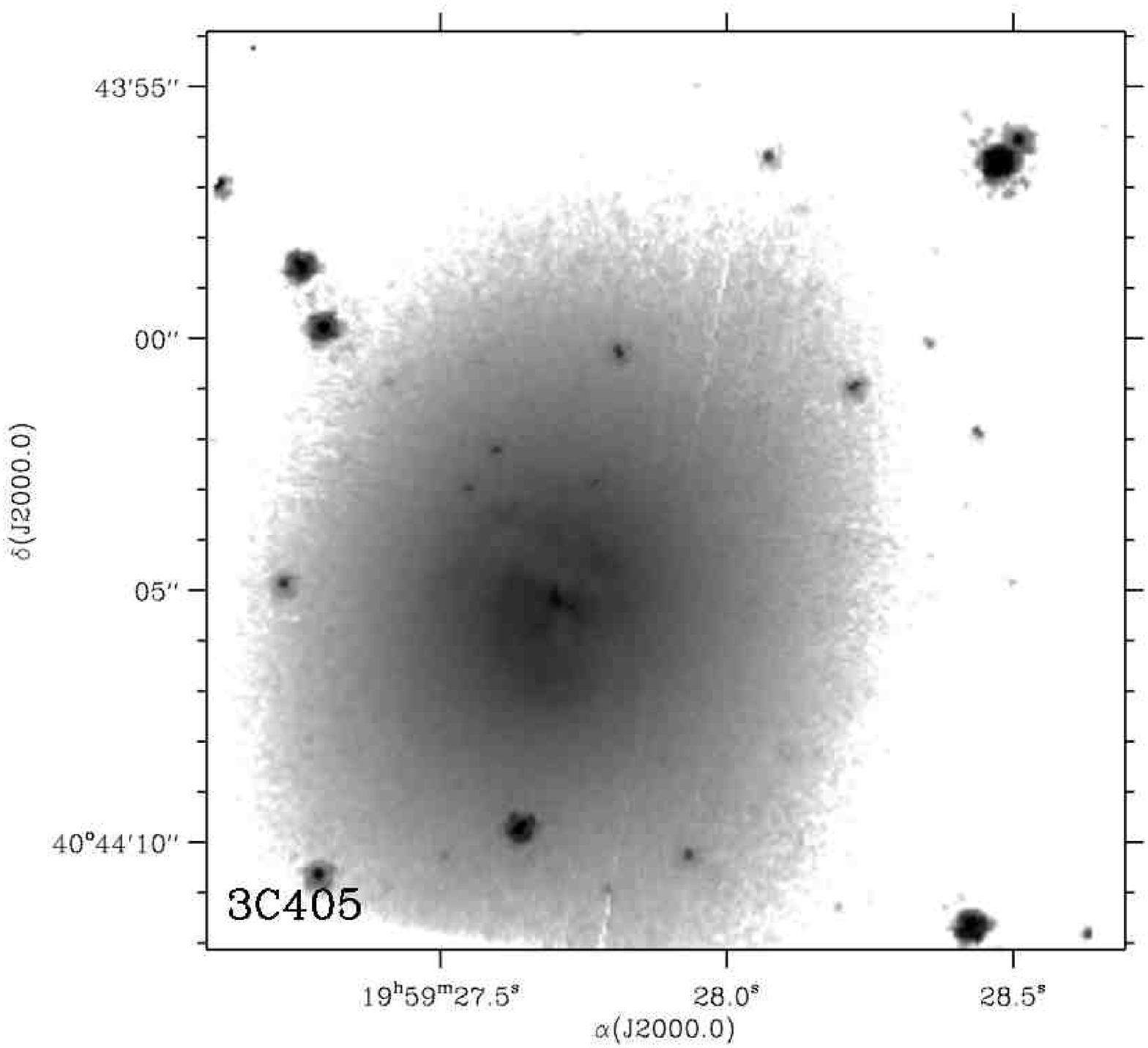}
\epsscale{1.0}
\caption{(continued). NICMOS2 f160w ($H$ band) images of the archival objects -- see table~\ref{tab-arch}.}
\end{figure*}

\clearpage

 %%%%%%%%%%%%%%%%%%%%%%%%%%%%%%%%%%%%%%%%%%%%%%%%%

\section{Host galaxy modeling results}
\label{sec-res}
In this section we present results for the 1-D and 2-D fits, a basic comparison of the results from the two techniques, and statistics describing the modelled properties  of the host galaxies, for comparison to other galaxy samples.

\subsection{Ellipse quasi-radial profile fits}
Properties of the ellipse fits are presented in table~\ref{tab-ellipse}.
We note that the ellipse-fitting technique is only reliable in cases where there is no strong nuclear point source. In the presence of a strong nuclear point source, much of the galaxy structure is obscured by the asymmetric PSF, the central region of the galaxy is entirely obscured, and thus the galaxy cannot be accurately fit. In these cases, the 2-dimensional fit is able to perform far better as it can take account of the asymmetrical 2-D structure of the PSF. We also note once again, that the 1-D approach does not provide a true ``radial profile'' since adjacent elliptical isophotes will in general have different position angles and ellipticities, thus overlapping and contributing flux to each other.

\subsection{Parametric galaxy fits}
The direct 2-D \sersic fits are summarized in table~\ref{tab-sers}. 
The technique is unable to follow all of the isophotal twists in a disturbed galaxy's profile, but provides an effective tool for modeling the bulk of a galaxy's flux. In addition, due to the convolution with a sub-pixel sampled PSF, and using the weighting scheme discussed above, we can reliably determine basic morphological parameters for even galaxies with strong quasar-like nuclei.

\subsection{Consistency}
\begin{figure*}
{\includegraphics[width=5.0cm]{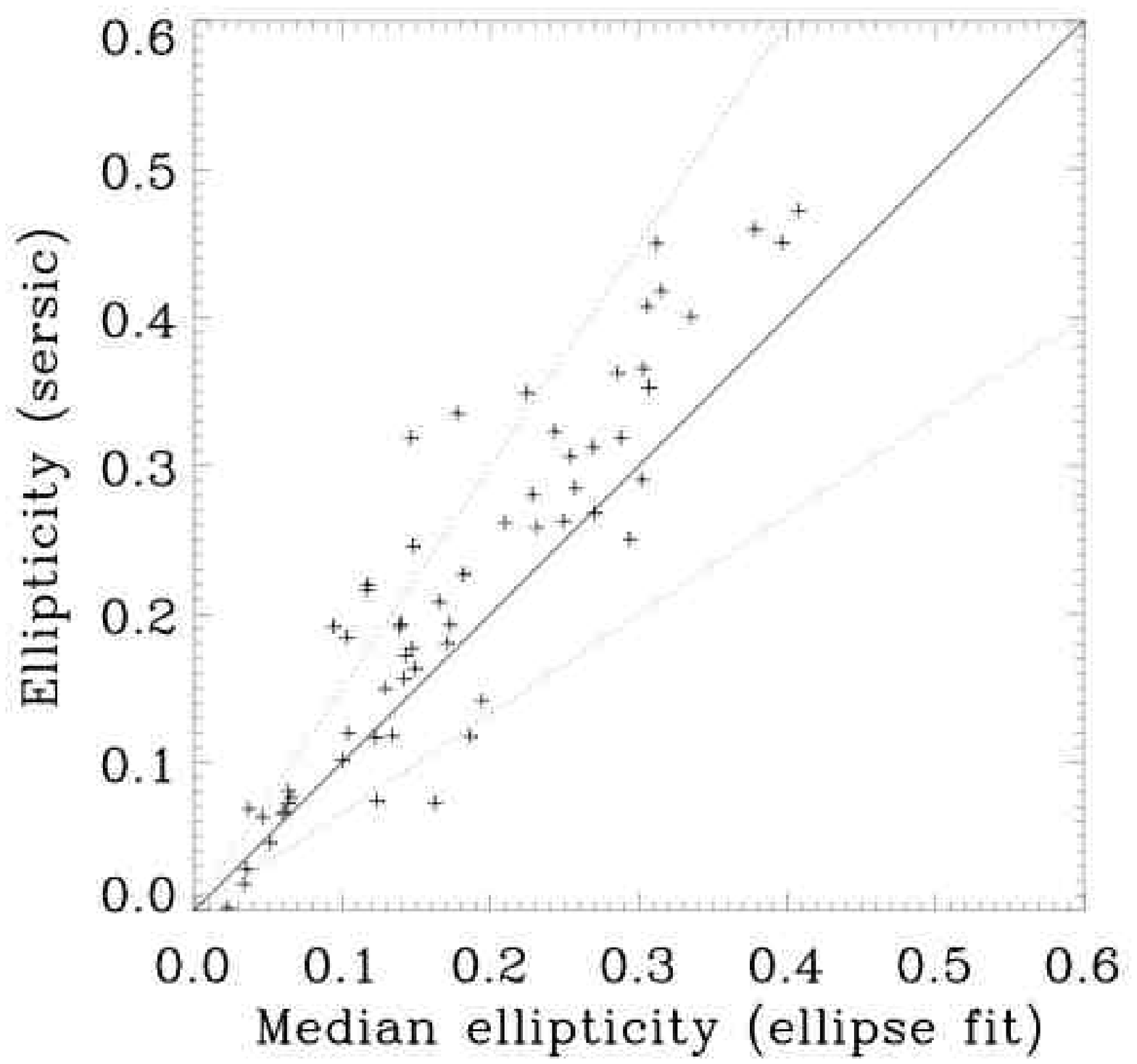}}%/Users/floyd/3C/Work/plots/ellip.eps}}
{\includegraphics[width=5.0cm]{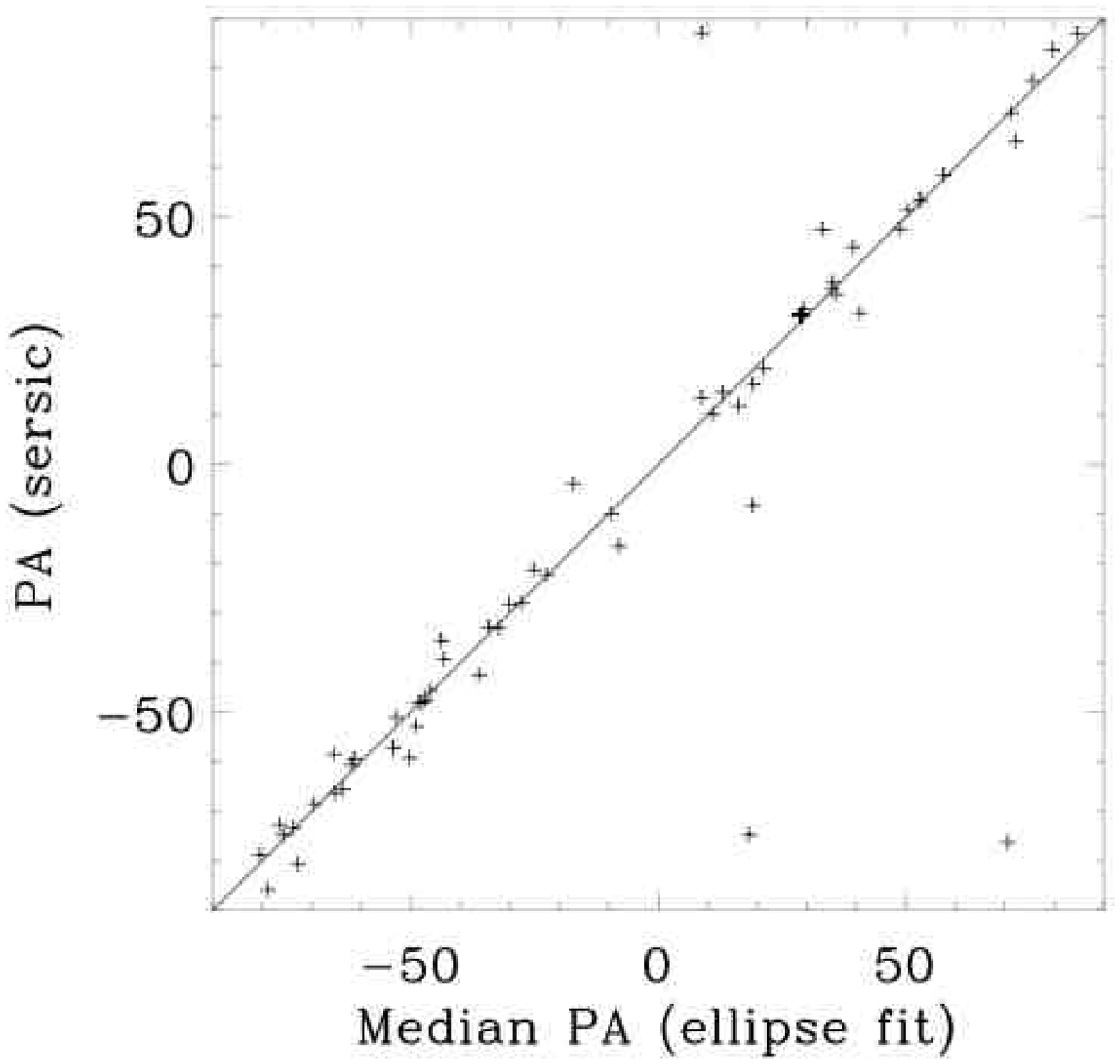}}%/Users/floyd/3C/Work/plots/PA.eps}}
{\includegraphics[width=5.0cm]{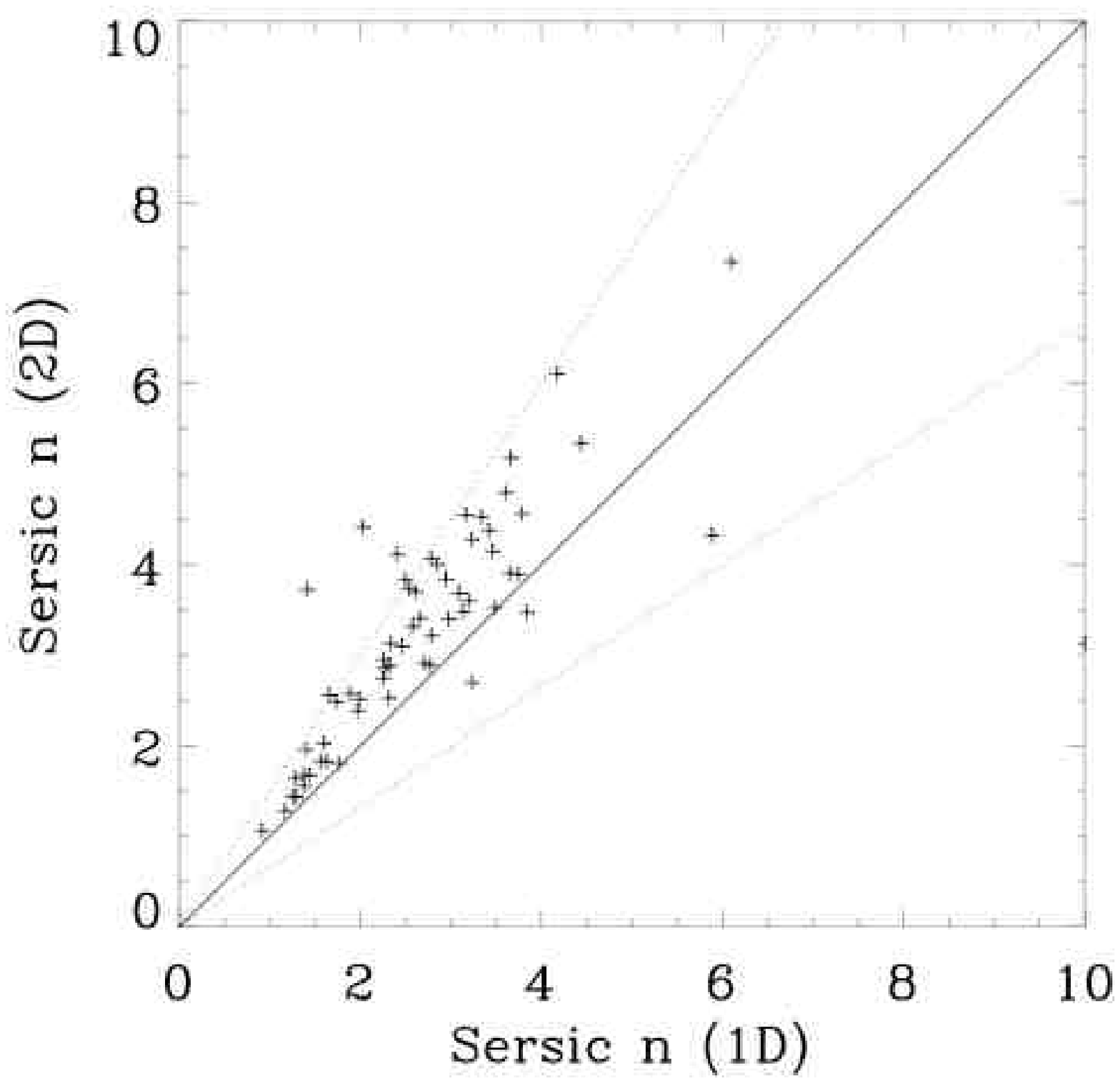}}%/Users/floyd/3C/Work/plots/sers.eps}}
\caption{\label{fig-comp} Results of the Ellipse (1D) and Galfit (2D) fits compared for targets without strong nuclear point sources (i.e. in which the nucleus contributes $<10$\% of the total luminosity). Left-to-right: Ellipticity, position angle (PA), and \sersic index. We plot the median ellipticity and PA of the ellipse fits, which are found to be consistent with the 2D parameters. The \sersic profile indices, $n$, fitted to the elliptical isophotal intensities differ from those fitted directly to the data through the 2-dimensional parametric model by up to a factor of 1.5 (indicated by the dotted lines). The 3 outliers are 3C~88, 3C~296 and 3C~338.}
\end{figure*}

We checked the consistency of the 1-D and 2-D fits by comparing the median ellipticity and position angle determined for the ellipse fits with the best-fit ellipticity and position angle in the parametric fits. For this comparison we exclude objects that have a significant nuclear point source, defined here as being one that contributes $\geq10$\% of the total flux of the source. 
The median position angles of each source can clearly be seen to be consistent with the fitted parametric values (Fig.~\ref{fig-comp}), with three outliers: 3C~129, 3C~192 and 3C~321. 
The ellipticities show somewhat more variation between the two modeling techniques, with fourteen sources differing by $>50\%$: 3C~79, 3C~88, 3C~105, 3C~130, 3C~135, 3C~192, 3C~264, 3C~284, 3C~319, 3C~321, 3C~338, 3C~401, 3C~438, 3C~465 
However, the \sersic parameters obtained from the 2-dimensional parametric fit, and the 1-dimensional fit to the elliptical isophote intensities exhibit a correlation, but are seen to differ significantly in many individual cases. The vast majority of objects have \sersic indices measured in each way falling within a factor of 1.5 of each other. Three significant outliers have higher 2-D \sersic indices than 1-D: 3C~88, 3C~296 and 3C~338.

Ellipse fits allow a more detailed tracing of the structure, and thus have significantly better residuals than the 2-D \sersic model fits, which cannot cope with the twists in isophotes seen in many sources. Nevertheless, both techniques do a good job of modeling the bulk of the source flux in the majority of cases. 
We consider the 2-D radial profile fits to be the ``correct'' approach to obtaining a meaningful radial profile. The 1-D fit to elliptical isophotal data does not constitute a true radial profile due to overlapping adjacent elliptical annuli, and thus the form of the \sersic law determined from this technique can only be an approximation. 
Fitting with two or three Sersic components was attempted in several cases, and while this results in significant improvement in the fit, it is at the cost of physical interpretation of the resulting parameters. See~\citet{donzelli+07} for 2-component fits to the elliptical isophotes for a large number of objects in the present sample. Note that the introduction of a second fitted component can drastically alter the form (\sersic index, scale length, luminosity) of the primary modeled component. Donzelli et al. likely better model the true  luminosity of the galaxy as their model provides a better overall fit  to its form. However, they do not provide fits to the nuclei, and thus the quality of the Donzelli fit to the inner 1.7\arcsec is generally poor. The present technique provides a reasonable estimate of the luminosity and form of the bulge and of the brightest nuclei, and allows meaningful comparison with existing samples of quasar host galaxies and quiescent ellipticals in the literature.

If we exclude all the nuclear-dominated sources (for which one cannot trust the 1D fit results due to the asymmetric diffraction spikes dominating the ellipse-fitted isophotes) we find that the 1D and 2D fitting results are essentially identical. Indeed, in simulations where we began with simulated galaxies and progressively added stronger and stronger nuclear components and random noise, we found that the 2D approach was always able to return the original parameters of the underlying galaxy, within the errors, even once the 1D approach became unworkable due to the strength of the diffraction spikes, and resulting asymmetry in the isophotes. Furthermore, adopting only the non-nucleated sample which one can study using {\em both} techniques, we find no significant change in our overall results. 

%This implies that 3CR sources with optical-IR nuclei are essentially identical in terms of host galaxy properties as those without. -- see section on nuclei.
%, and confirms that radio-loud quasars are indeed hosted by identical objects to FR~II radio galaxies.

 %%%%%%%%%%%%%%%%%%%%%%%%%%%%%%%%%%%%%%%%%%%%%%%%%
\subsection{Host galaxy properties}
\begin{figure*}
\plottwo{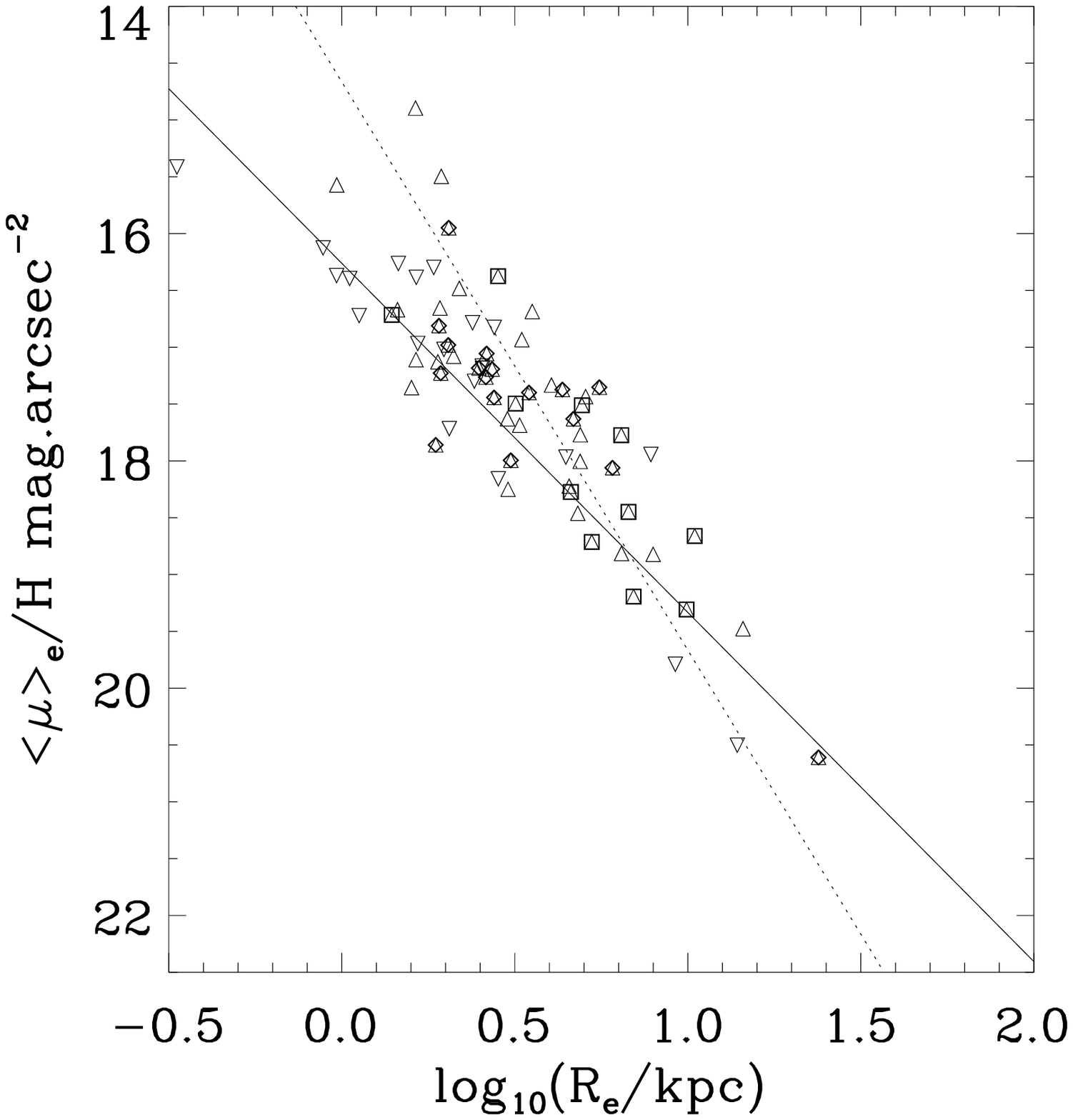}{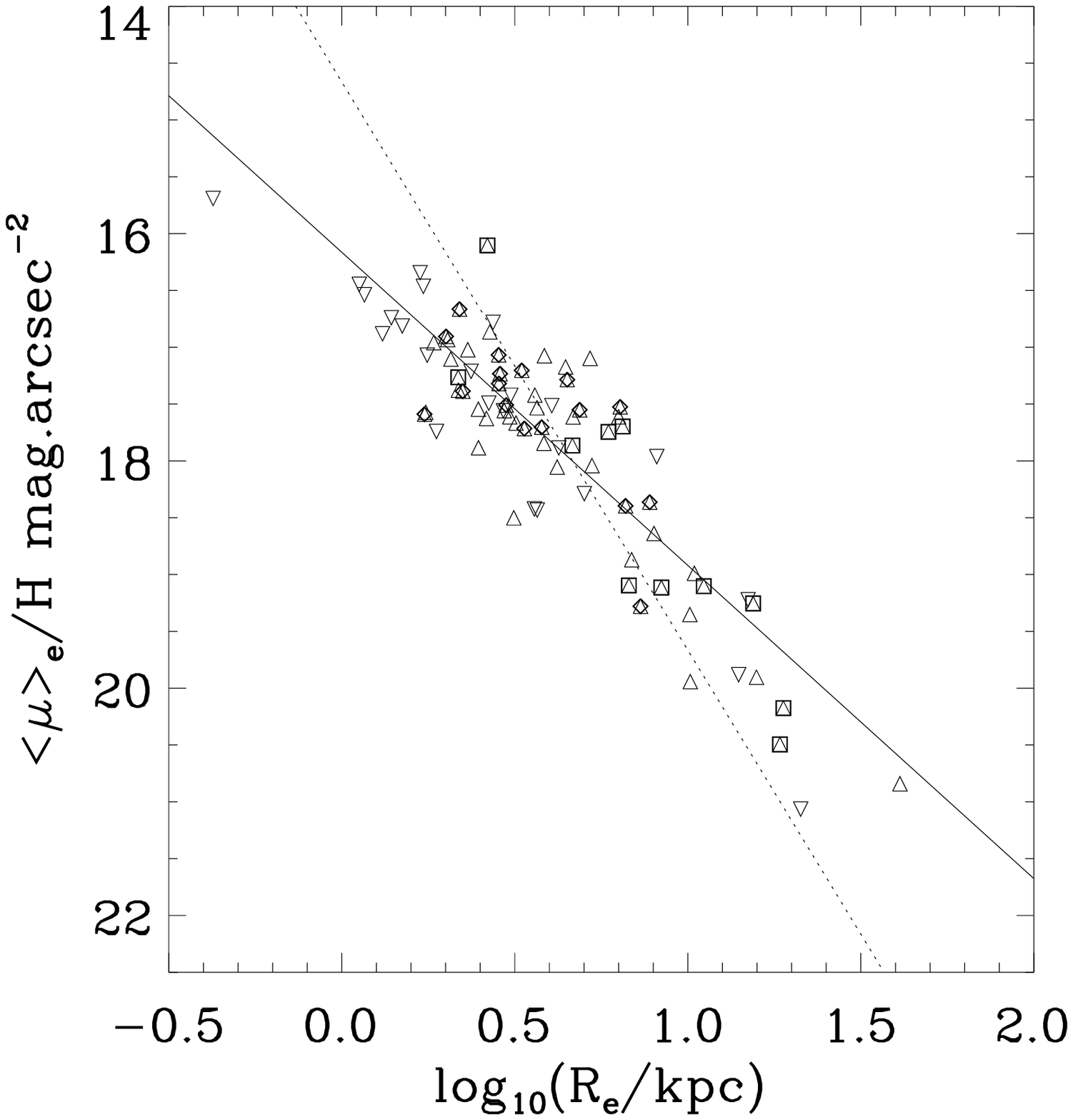}
\caption{\label{fig-korm1} Scale-length -- Surface-brightness relation for the 1D (left) and 2D (right) model fits to the full sample, excluding the outliers 3C~71 and 3C~258 (not plotted -- see text). Symbols are the same as in Fig.~1.}
\end{figure*}

Fig.~\ref{fig-korm1} shows the $R_{e}-\mu_{e}$ distribution resulting from each modeling technique, excluding two outliers, 3C~71 and 3C~258.
3C~71 has an unreliable fit due to its extension across both the NIC2 and WFPC2 fields (section~\ref{sec-3c71}).
We believe the highly peculiar source 3C~258 to be a higher redshift quasar behind a $z=0.165$ irregular galaxy (see section~\ref{sec-3c258}). 
Using the 2D results for the reasons discussed above, the mean effective radius, or scale-length (defined throughout as the half light radius of the fitted models) for the entire sample is found to be $\langle R_{e} \rangle =  \langle R_{1/2} \rangle = 7.46\pm 2.00$~kpc (median= 3.63~kpc), with the mean surface within the half light radius, $\langle\mu\rangle_{e} = 18.12\pm0.25$~mag.arcsec$^{-2}$ (median= 17.59~mag.arcsec$^{-2}$). Quoted errors are standard errors on the mean throughout.
The mean host luminosity for the sample is $\langle M_H \rangle = -23.53\pm0.27$~mag (median=-23.75).
The mean (2D) \sersic index for the sample is $\langle n \rangle = 3.68\pm0.19$ (median=3.41), and the mean ellipticity using the same technique is $\langle 1-b/a \rangle = 0.20\pm0.02$ (median=0.18).

%At this point we exclude from further statistical analysis the single source that dominates the scatter, lying $\geq3\sigma$ from the mean in terms of scale length or surface brightness, the highly peculiar source 3C~258, which we believe to be a higher redshift quasar behind a $z=0.165$ irregular galaxy (see section~\ref{sec-3c258}). We also exclude 3C~71 due to its extension and resulting unreliable fit (section~\ref{sec-3c71}).

Excluding 3C~258 and 3C~71 we obtain the following 3-$\sigma$ rejected means and standard errors for the sample: 

$\langle R_{e} \rangle = 7.59\pm 2.00$~kpc (median = 3.66~kpc).

$\langle\mu\rangle_{e} =  17.89\pm 0.12$~mag.arcsec$^{-2}$ (median = 17.57).

$\langle M_H \rangle = -23.81\pm 0.10$~mag (median = $-23.77$).

$\langle n \rangle =  3.69\pm 0.19$ (median = 3.47).

$\langle 1-b/a \rangle =     0.20\pm 0.02$ (median = 0.18).

We now briefly consider the main subdivisions of the sample into FR~I and FR~II types.

\subsubsection{FR~I's}
For the FR~I subsample, the 3-sigma rejected statistics are as follows:

$\langle R_{e} \rangle = 12.03\pm 7.66$~kpc (median = 2.74).

$\langle\mu\rangle_{e} = 17.75\pm     0.30$~mag.arcsec$^{-2}$ (median = 17.49).

$\langle M_H \rangle = -23.47\pm 0.27$~mag (median = $-23.29$).

$\langle n \rangle = 2.77\pm 0.31$ (median = 2.39).

$\langle 1-b/a \rangle = 0.18\pm 0.02$ (median = 0.18).

The FR~I's as a class tend to have lower $n$ (``diskier'') host galaxies than the combined sample, or any other individual subsample. They account for most of the lowest redshift ($z<0.05$) sources, with few at higher redshifts. They exhibit a large spread in scale length, being found both in normal elliptical galaxies, and in cD galaxies like 3C~84.

\subsubsection{FR~II's}
For the FR~II subsample the 3-sigma rejected statistics are as follows:

$\langle R_{e} \rangle = 5.94\pm 0.80$~kpc (median = 3.84).

$\langle\mu\rangle_{e} = 17.95\pm 0.13$~mag.arcsec$^{-2}$ (median = 17.61)

$\langle M_H \rangle = -23.88\pm 0.10$~mag (median= $-23.84$).

$\langle n \rangle = 3.88\pm 0.21$ (median = 3.70)

$\langle 1-b/a \rangle =     0.21\pm 0.02$ (median = 0.19).

The FR~II's represent a slightly more homogeneous population of giant ``true'' elliptical galaxies, with \sersic indices close to 4. 
The median size is larger than that of the FR~I's, and the scale-length spread is much smaller. However, a two-sided Kolmogorov-Smirnov (K-S) test reveals little significant difference between the distributions in host luminosity or scale length. There is an 18\% chance of the FR~I and FR~II \sersic indices being drawn from the same distribution ($D=0.26$).

%A simple 2-sided Kolmogorov-Smirnov (K-S) test reveals the significance of this difference ???
%It thus appears to take a ``normal'' giant elliptical to host an FR~II, with few being located in the dominant central galaxies of clusters.

%%%%%%%%%%%%%%%%%%%%%%%%%%%%%%%%%%%%%%%%%%%%%%%%%%
\section{Companions, jets, mergers and interactions}
\label{sec-comp}

We used the Source Extractor software~\citep{SEx} to detect and measure each candidate companion source from the {\sc ellipse} galaxy modeling residual images. 
We also obtained WFPC2 optical images of all of our sources from Andr\'{e} Martel's 3CR database\footnote{http://acs.pha.jhu.edu/$\sim$martel/}, or directly from the MAST\footnote{http://archive.stsci.edu/}, most of which were previously published by~\citet{dekoff+96,martel+99}. Table~\ref{tab-opt} summarizes the origins of the optical images used in this study.
We present basic companion astrometry and $H$-band photometry in table~\ref{tab-companions}, along with an inventory of the number of companions to each 3CR host galaxy visible on the NIC2 chip. 
We also manually investigated each object in the optical and the infrared, to classify each according to the presence of the following features:
Shells~\citep{malincarter83}; 
tidal tails;
mergers (clearly undergoing merger with a companion from distortion or overlapping isophotes);
pre-mergers (appear to be close to merging with a companion source);
major companion sources (with infrared luminosity $<1$~$H$~mag. dimmer than the primary host galaxy assuming the same redshift);
minor companion sources (a resolved companion source $>1$~$H$~mag. dimmer than the primary host galaxy assuming the same redshift);
unresolved point-like companion source (a candidate companion source that is unresolved in the NICMOS image);
jets, 
and dust disks.
The table lists the presence of each type of artifact. Below we discuss the main types encountered.

\subsection{Globular Clusters}
In the very nearest sources, the large number of detected faint, unresolved sources are likely to be Globular Clusters within the halo of the host galaxy. This is really only an issue in the most well-studied, nearby objects, i.e. those from archival data. In particular, 3C~71 (better known as NGC~1068) is well known for its globular cluster population. Other objects with candidate globular cluster detections are: 3C~29 ($z=0.05$); 3C~31 ($z=0.02$); 3C~84 (Per A / NGC~1275; $z=0.02$); 3C~88 ($z=0.03$); 3C~129 and 3C~129.1 (both at $z=0.02$).
At present the sample is too small, but it would be interesting from the point-of-view of AGN feedback models to explore whether radio jets affect the formation of globular clusters, by examining the color distributions of the globular cluster populations in these objects compared with those in quiescent ellipticals.
In the merger scenario for the formation of elliptical galaxies, an older stellar population combining with a younger one produces a spread in metallicity and a bimodal distribution in the metallicity and hence color of the globular clusters~\citep{vandenbergh00}. It is important and interesting to determine whether the globular cluster populations of radio galaxies differ from those of normal quiescent elliptical  galaxies, but this lies outside the scope of the present study.

\subsection{Jets and Hotspots}
Two new optical-IR jets have been discovered by the present NICMOS imaging survey; Those of 3C~133~\citep{floyd+06b} and 3C~401~\citep{chiaberge+05}.
3C~130 exhibits numerous point sources, but is too distant (at $z=0.109$) for them to be explained by Globular Clusters. Many of these may be hotspots, corresponding to the unusual radio morphology, and numerous hotspots observed in the radio by~\citep{hardcastle98}.
We identify a further six sources with candidate IR-synchrotron hotspots, based on their distance, the number of unresolved sources, unusual radio morphologies and/or spatial coincidence of the unresolved sources with features on the radio map:
3C~52; 
3C~61.1 (single bright unresolved source to north of galaxy); 
3C~66B; 
3C~219; 
3C219.1 and 3C~452.

Jets are identifiable in a further ten sources:
3C~15~\citep{martel+98}; 
3C~66B~\citep{butcher+80}; 
3C~133~\citep{floyd+06b};
3C~264~\citep{crane+93}; 
3C~274~\citep{curtis17}; 
3C~277.3~\citep{miley+81}; 
3C~293~\citep{floyd+06a}; 
3C~346~\citep{dey+94}; 
3C~371~\citep{nilsson+97}; 
and 3C~401~\citep{chiaberge+05}
Extended emission line regions, with the appearance of jets are clearly visible in 3C~171 and 3C~234.

\subsection{Dust disks}
31 objects in the sample are seen to possess $\sim 100$~pc scale distributions of dust and gas in their nuclear regions, often in settled, disk-like structures surrounding the AGN. See~\citet{tremblay+07} for further details. 
%3C~28; 3C~31; 3C~33; 3C~52; 3C~61.1; 3C~76.1; 3C~83.1; 3C~84; 3C~173.1; 3C~180; 3C~196.1; 3C~223.1; 3C~236; 3C~264; 3C~270; 3C~285; 3C~293; 3C~296; 3C~305; 3C~321; 3C~326; 3C~338; 3C~346; 3C~348; 3C~357; 3C~403; 3C~405; 3C~430; 3C~433; 3C~436; 3C~438; 3C~449~\citep{tremblay+06}.
11 of these are FR~I sources; 16 are identified as FR~II's.

\subsection{Mergers and pre-merger candidates}
Six sources are clearly identifiable as mergers through their overlapping isophotes and disturbed appearance: 3C~79; 3C~293~\citep{floyd+06b}; 3C~321; 3C~346; 3C~405 and 3C~433. There are an additional twenty-eight sources that qualify as ``pre-mergers'', or candidate mergers that cannot be confirmed by the overlapping of isophotes of the two partners.
``Major'' secondary sources are identifiable in twenty sources (20\%). Minor secondaries are identified in fifty-five sources, or 56\% of the sample.
Fifty-nine sources have unresolved companions. A number of these may turn out to be foreground stars, especially for sources at low galactic latitudes, but it seems unlikely that all such sources are so explained. For a small number we also obtain detections in the optical, and these sources typically have colors of $R-H=2-3$. The exact nature, and redshifts of these sources must await a detailed spectroscopic study, which could confirm whether the candidate companions indeed lie at the same redshift as the radio source, and further identify the stellar makeup of the sources. 
However, based on the sheer number of small companions, we conclude that these are of significant importance to the 3CR sample and to the radio source phenomenon itself. The broad-band colors of many of these sources are consistent with an old stellar population, perhaps the stripped core of a late-type spiral, or compact early-type galaxy that has undergone merger with the main galaxy, losing its gas and dust to the more massive system. Such a mechanism would provide a natural means to fuel the AGN activity.
%\citet{???} predicted that large numbers of such sources would be found in elliptical galaxies by HST. 
As far as we are aware, this is the first such detection in large numbers of sources across a range in redshifts. Examples of dwarf ellipticals and of globular clusters have been observed in a number of nearby sources, e.g. Fornax.

 %%%%%%%%%%%%%%%%%%%%%%%%%%%%%%%%%%%%%%%%%%%%%%%%%
\section{Discussion}
\label{sec-disc}

\subsection {Galaxy morphologies}
\begin{figure}
\plotone{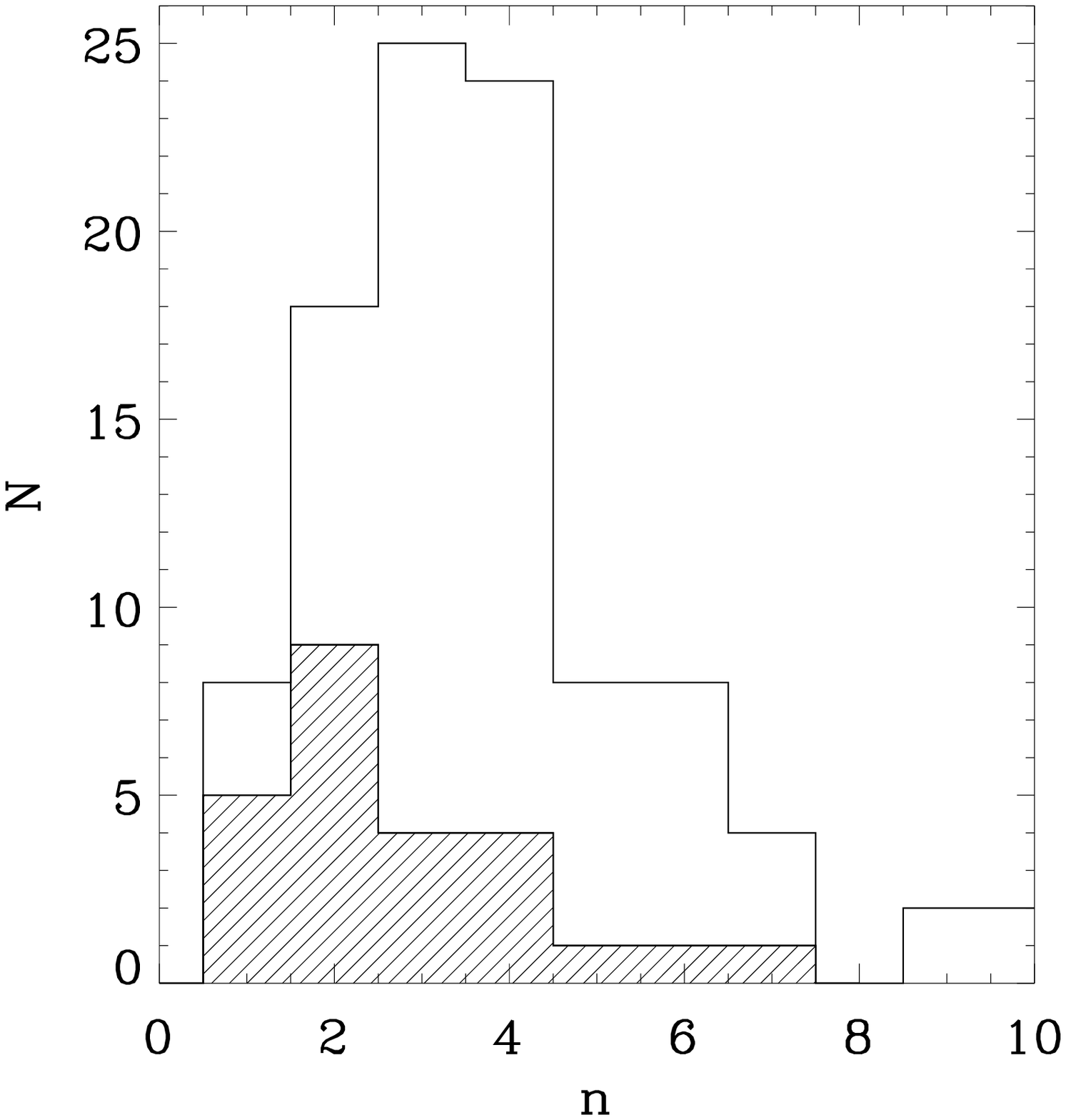}
\caption{\label{fig-n-hist} Host galaxy morphology distribution in terms of \sersic index, $n$: FR~I sources are shown shaded, with the solid line indicating the entire sample. The host galaxies exhibit a broader range of \sersic $n$ than a population of pure ellipticals, with a tail of low-$n$ galaxies (nineteen at $n<2$). These ``disky'' sources are listed and briefly discussed in the main text.}
\end{figure}

\begin{figure*}
\plottwo{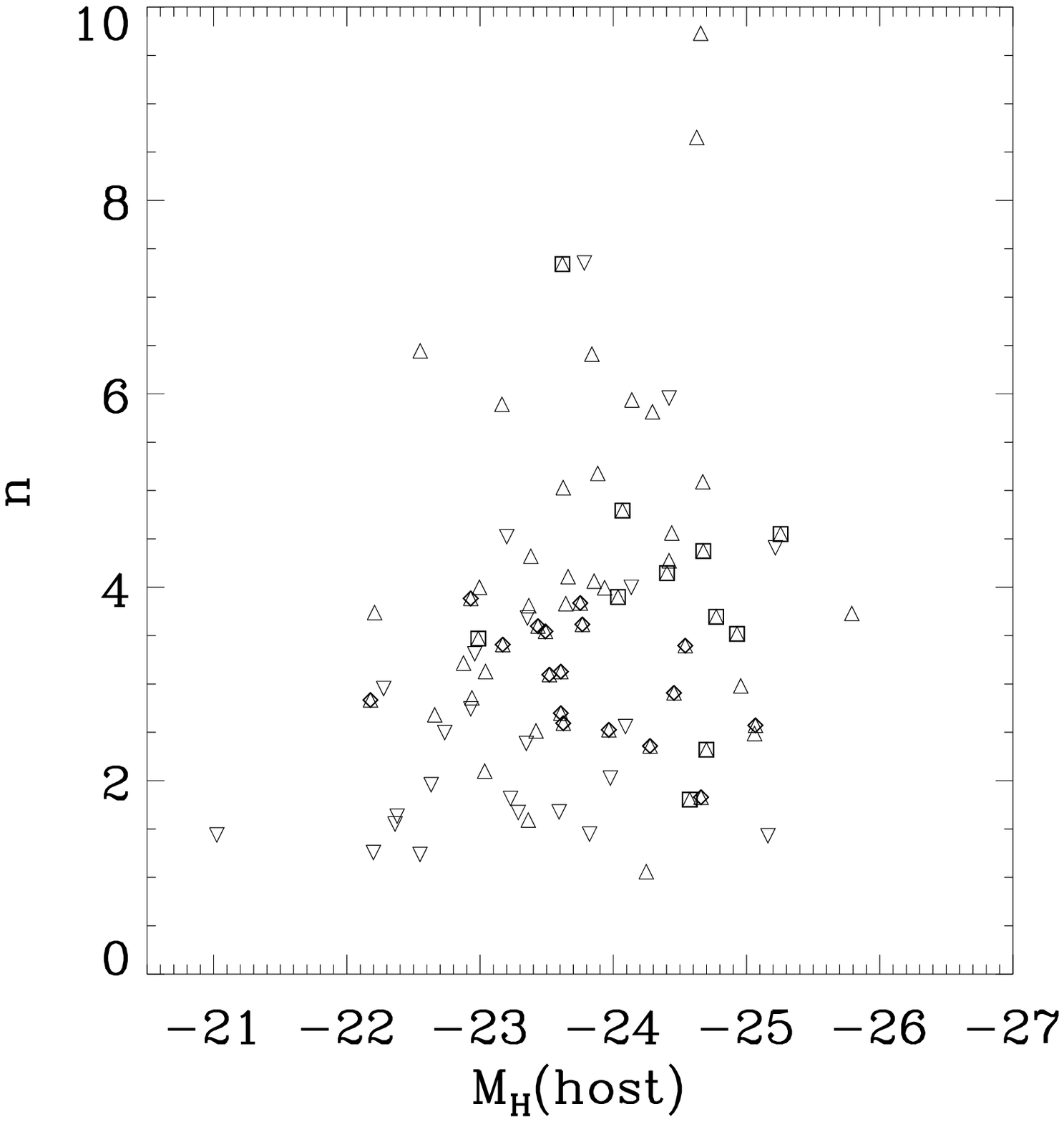}{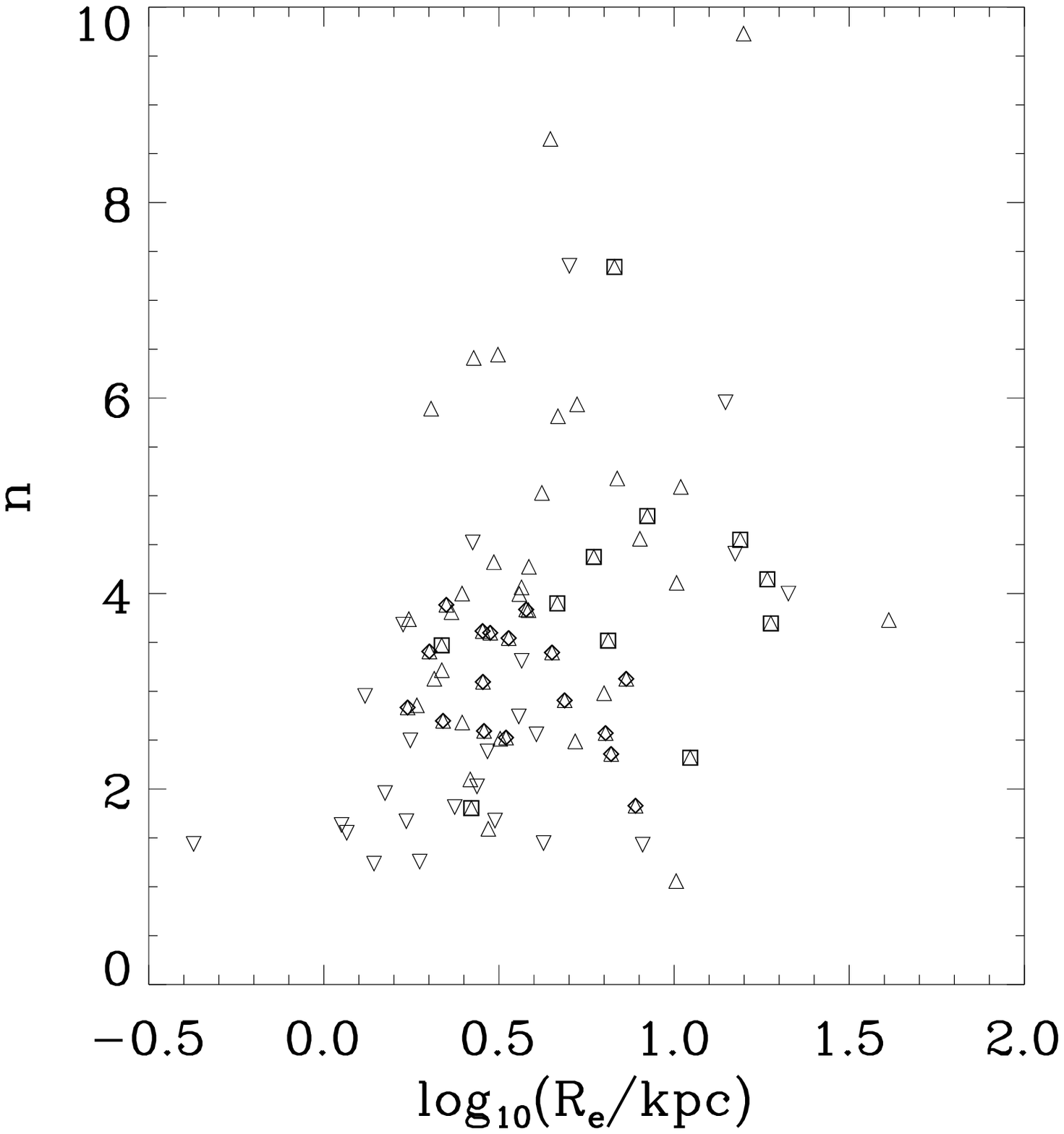}
\caption{\label{fig-n} Weak correlations are observed between the \sersic index, $n$, and the host galaxies' luminosity and scale length. Powerful radio galaxies can only be hosted by disk-dominated galaxies if the galaxy is exceptionally massive. Symbols are the same as in Fig.~1.}
\end{figure*}

We find a broad distribution of \sersic indices in our sample, spanning the full range from $n=1$ (exponential disks) to $n=4$ (de Vaucouleurs ellipticals), and higher (see Fig.~\ref{fig-n-hist}). 
There is a moderately strong correlation (Spearman's rank $\rho=0.42, p=1.7E-5$) between scale-length and \sersic $n$ (see Fig.~\ref{fig-n}). 
Interestingly, neither the FR~I's nor the FR~II's considered alone show such a significant correlation (FR~I: $\rho=0.56, p=0.0055$; FR~II: $\rho=0.35, p=0.0078E-3$).
There is a somewhat weaker anti-correlation between the host galaxy absolute magnitude and the \sersic index, $n$ ($\rho=-0.32, p=0.001$) -- once again somewhat weaker in the individual subsamples. Thus the larger, more luminous galaxies tend to be more ``bulgy''. The \sersic index is also found to correlate weakly with redshift ($\rho=0.28;~p=0.006$). Thus the disky sources tend to be lower redshift, and lower IR luminosity, but interestingly, no strong relationship is found between \sersic index and radio power (either core or total). However, there are very few sources that have \sersic indices $n<2$, thus this is not an indication that true ``disk'' galaxies are capable of hosting powerful radio sources. Rather, it indicates that within the radio source population, there is a strong requirement for an elliptical, or bulge-dominated host galaxy, and a strong dependence of radio luminosity on host galaxy size and luminosity, but not on \sersic index within the range $2<n<10$. A more ``bulgy'' bulge does not allow for a more powerful radio source, while a more luminous bulge does.

There are a number of exceptions to the general rule; Seventeen of the host galaxies have $n<2$, and would thus be classed as disky.
%3C~29; 3C~31 (NGC~0383); %3C~71???, 3C~84 (Per~A, NGC~1275); 3C~129; 3C~129.1; 3C~173.1; 3C~234; 3C~264 (NGC~3862); 3C~270 (NGC 4261); 3C~272.1; 3C~274 (M~87); 3C~288; 3C~323.1 (PG~1545+210); 3C~332 (B2~1615+32); 3C~348 (PKS~1648+05, Her~A); 3C~388; 3C~405 (Cyg~A); 3C~442 (Arp~169); and 3C~449.
Ten of these sources are low-redshift ($z<0.05$), nine of those being FR~I low-power radio sources. However, there are eight FR~II sources, and five of these are at $z>0.1$: 3C~173.1, 3C~234, 3C~288, 3C~323.1 and 3C~332. All five of these FR~II sources have close companion objects, and are either merging, or are likely to be post-merger systems. They generally show evidence of dusty environments from their optical and IR images, except for 3C~323.1 which has a quasar-like nucleus in both the optical and IR. 

\subsubsection{Ellipticity}
\begin{figure}%[htbp]
\plotone{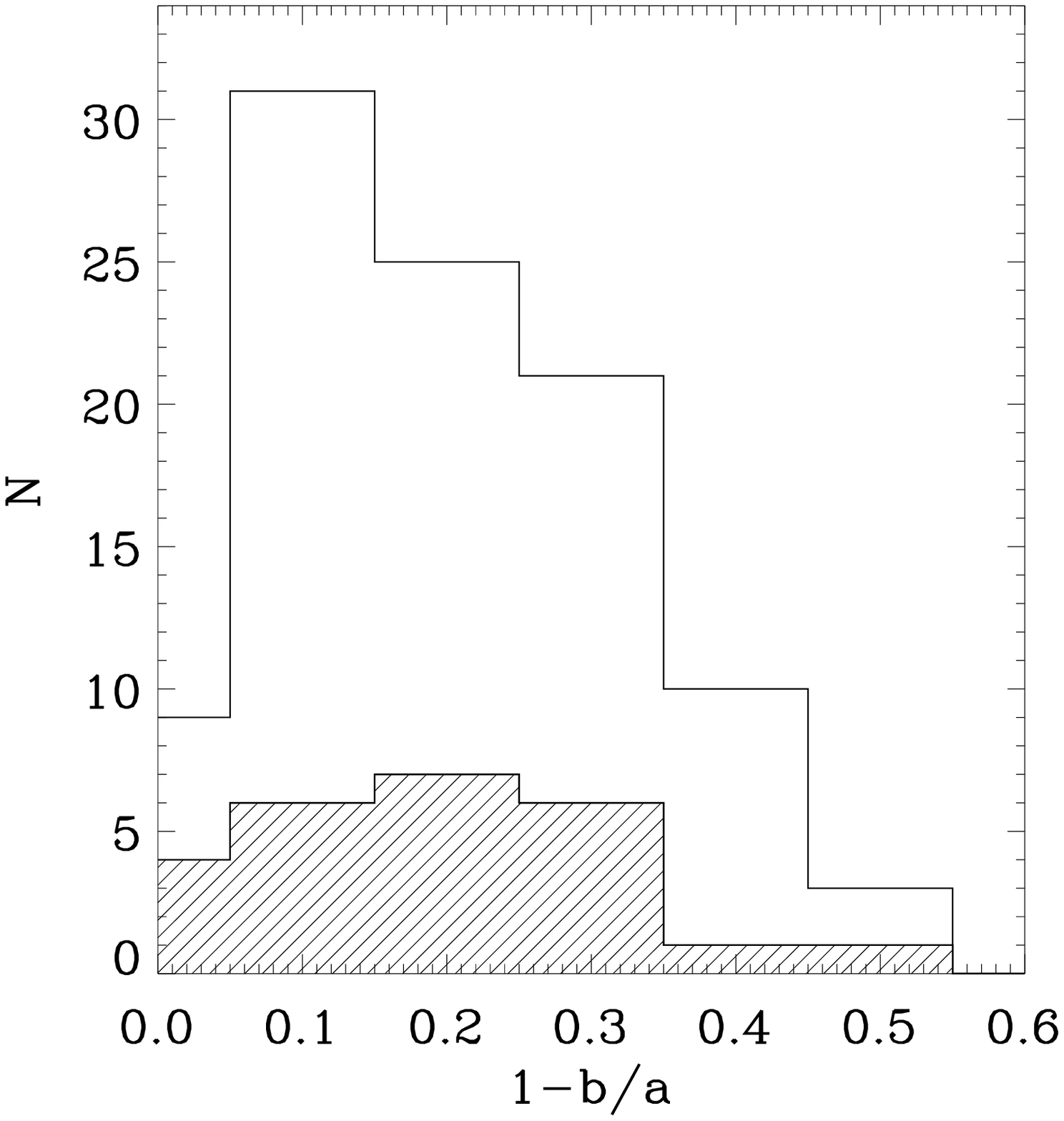}
\caption{\label{fig-ellip-hist} Ellipticity distribution for the full sample (unshaded) and for the FR~I's (shaded).}
\end{figure}

%Ellipticities - look up what old papers did with ellipse data... Milvang-Jensen et al...
The 3CR host galaxies are slightly rounder, on average, than the general elliptical galaxy population, with a sharp peak at E1, and no objects more eccentric than E5. See Fig.~\ref{fig-ellip-hist}.

\subsubsection{Isophotal twists}
\begin{figure*}%[htbp]
\centering
{\includegraphics[width=5.0cm]{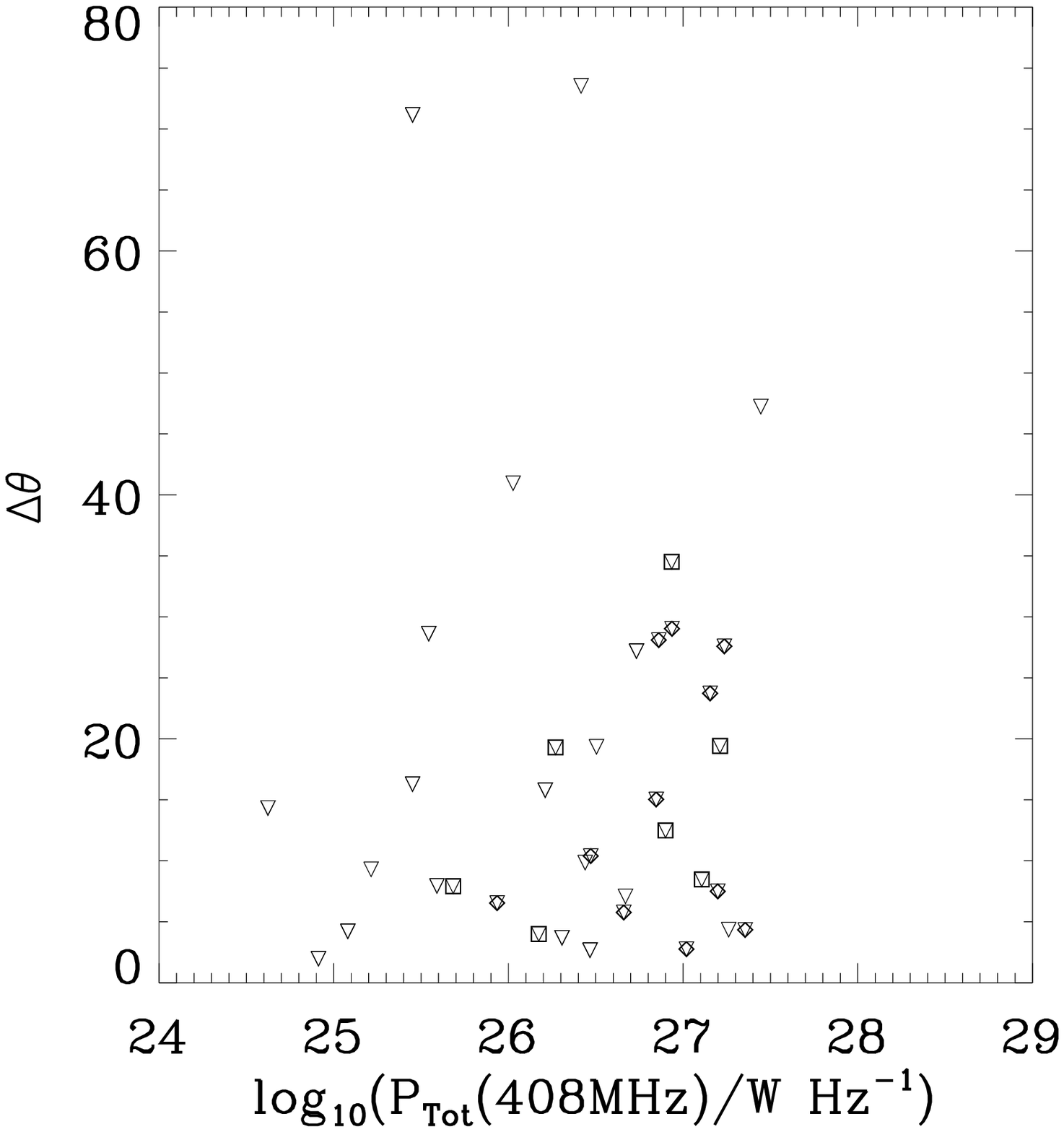}}%{/Users/floyd/3C/Work/plots/DPA_408.eps}}
{\includegraphics[width=5.0cm]{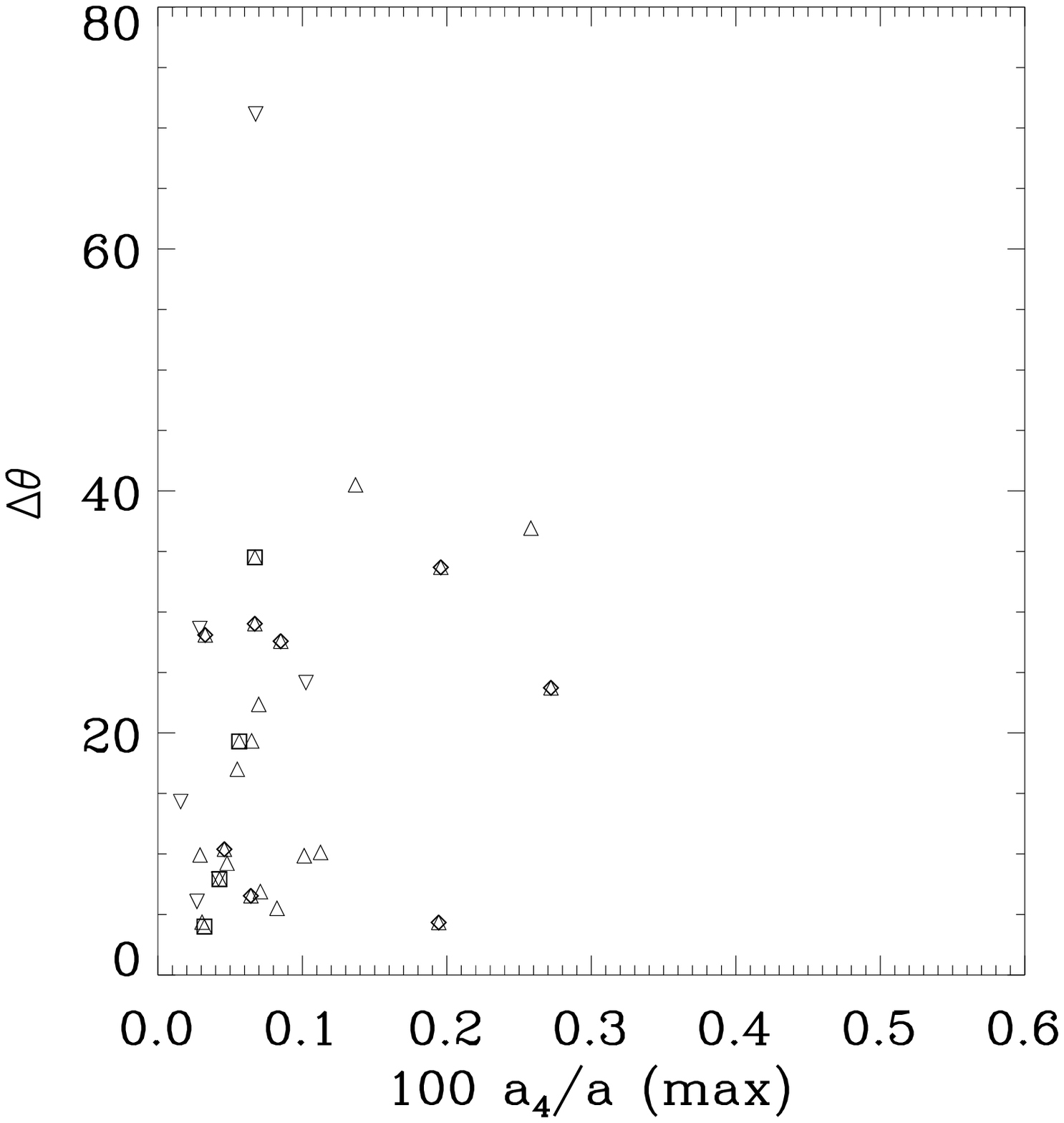}}%{/Users/floyd/3C/Work/plots/DPA_A4.eps}}
{\includegraphics[width=5.0cm]{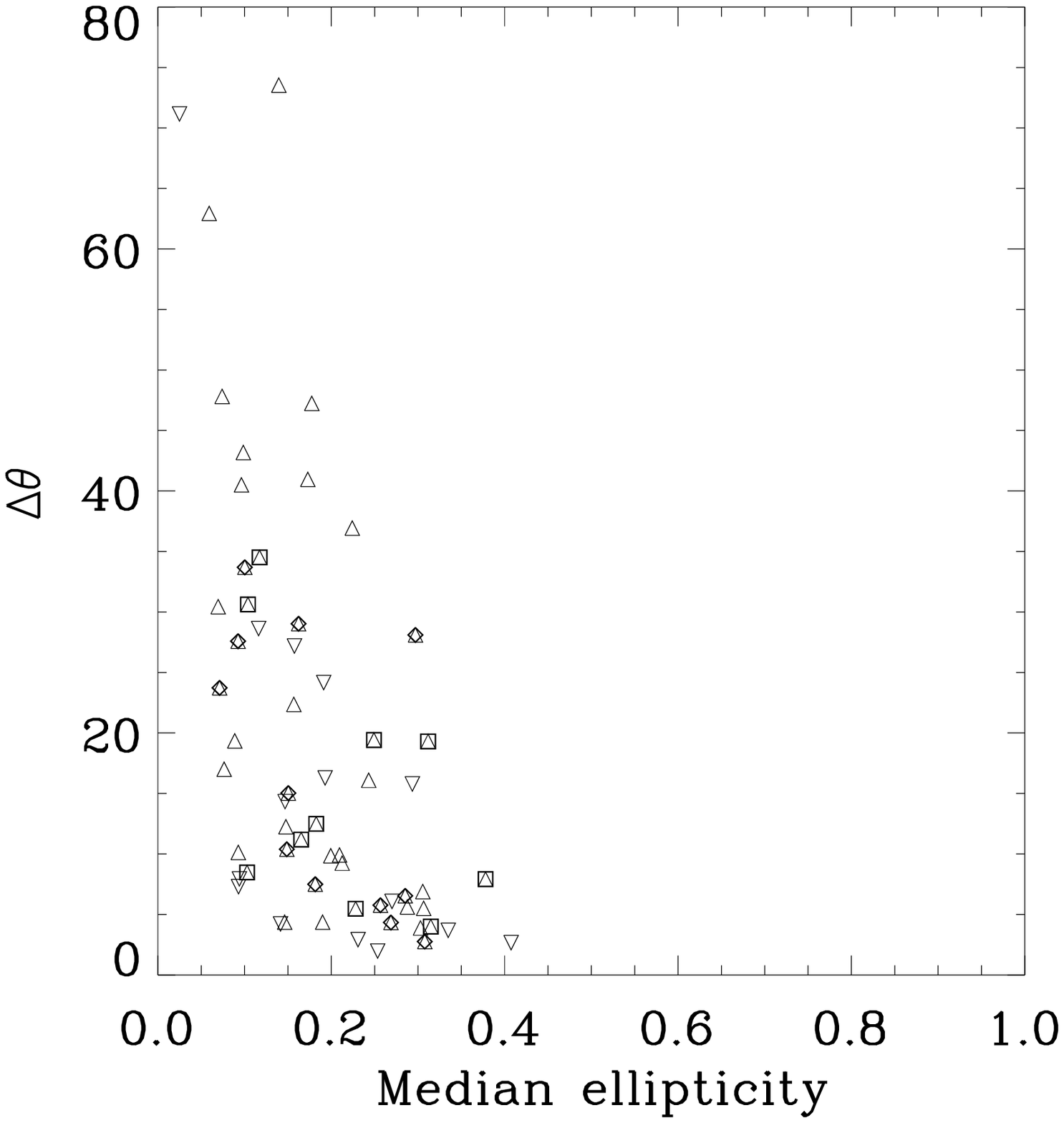}}%{/Users/floyd/3C/Work/plots/DPA_E.eps}}
\caption{\label{fig-DPA} Isophotal twist angle, $\Delta\theta$ against (left-to-right): total radio power, boxiness ($a_4$), and median ellipticity. Rounder galaxies tend to have larger isophotal twists, and as found by~\citep{gopalkrishna+03}, radio galaxies exhibit a correlation between twist angle and total radio power. Symbols are the same as in Fig.~1.}
\end{figure*}

The total isophotal twist angle, $\Delta\theta$ was computed for each galaxy from the position angle profile. We neglect the inner 0\farcs5, and neglect $>1\sigma$ outliers in the $\theta$ profile. We then quantify the twist in terms of the total position angle variation of the major axis over the entire range in surface brightness (after~\citealt{gopalkrishna+03}). Objects that are less elliptical than $1-b/a<0.2$ over their entire range are also discounted, as for these extremely round objects, a spuriously large twist angle can arise due to the circular symmetry of the isophotes. The resulting twist angles are plotted in Fig.~\ref{fig-DPA}. We confirm that the trend between total radio power and twist angle first noted by~\citet{gopalkrishna+03} for the radio galaxy sample of~\citet{govoni+00} is also seen in the present sample.

\subsection{Galaxy luminosities and scale lengths}
\begin{figure}
\plotone{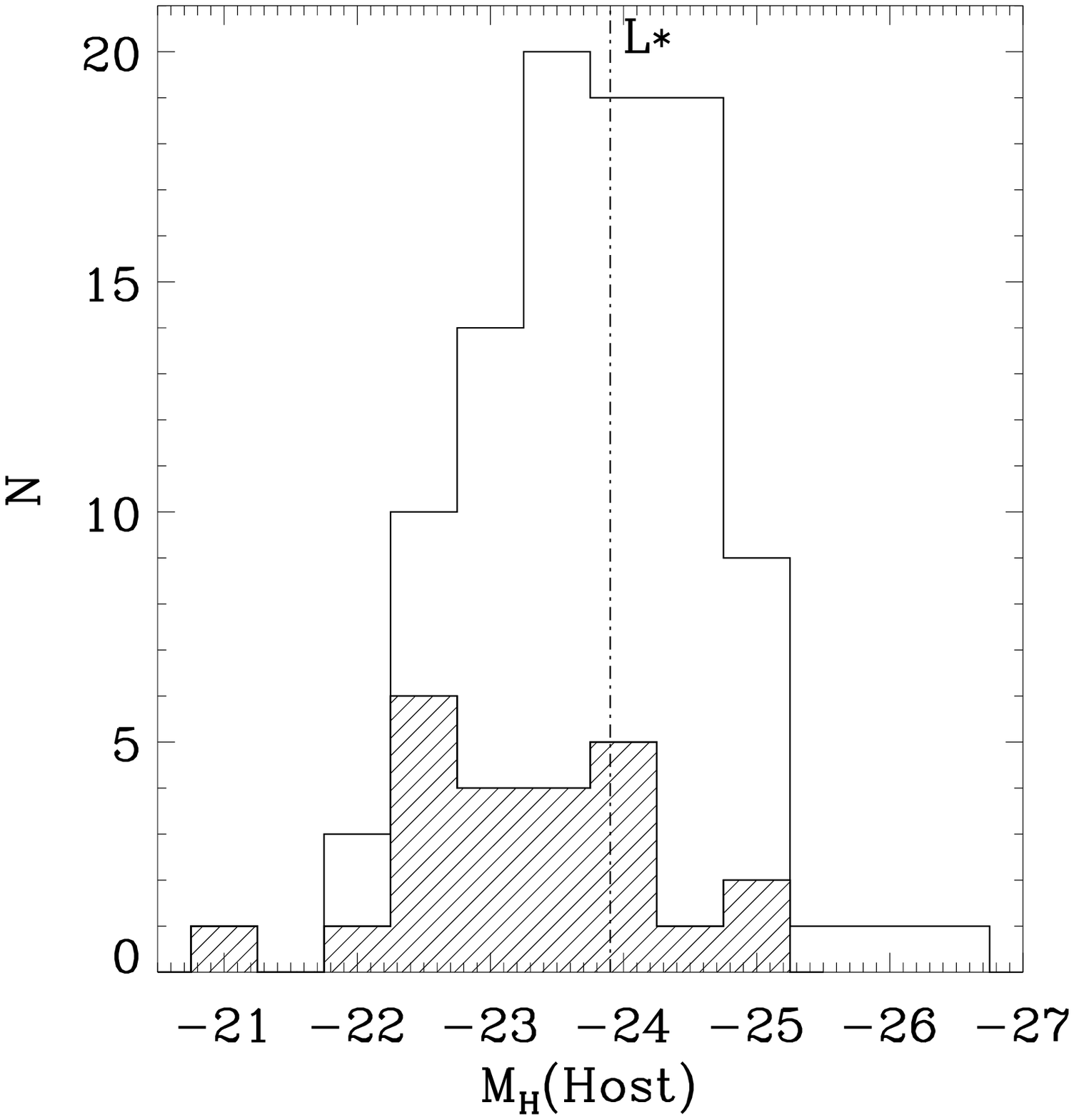}
\caption{\label{fig-MH-hist} Galaxy luminosity distribution for the full sample (unshaded) and for the FR~I's (shaded): 3CR sources generally hosted by super-$L^\star$ objects. The low-luminosity tail is dominated by low-redshift ($z<0.09$) and FR~I targets. }
\end{figure}

\begin{figure}
\plotone{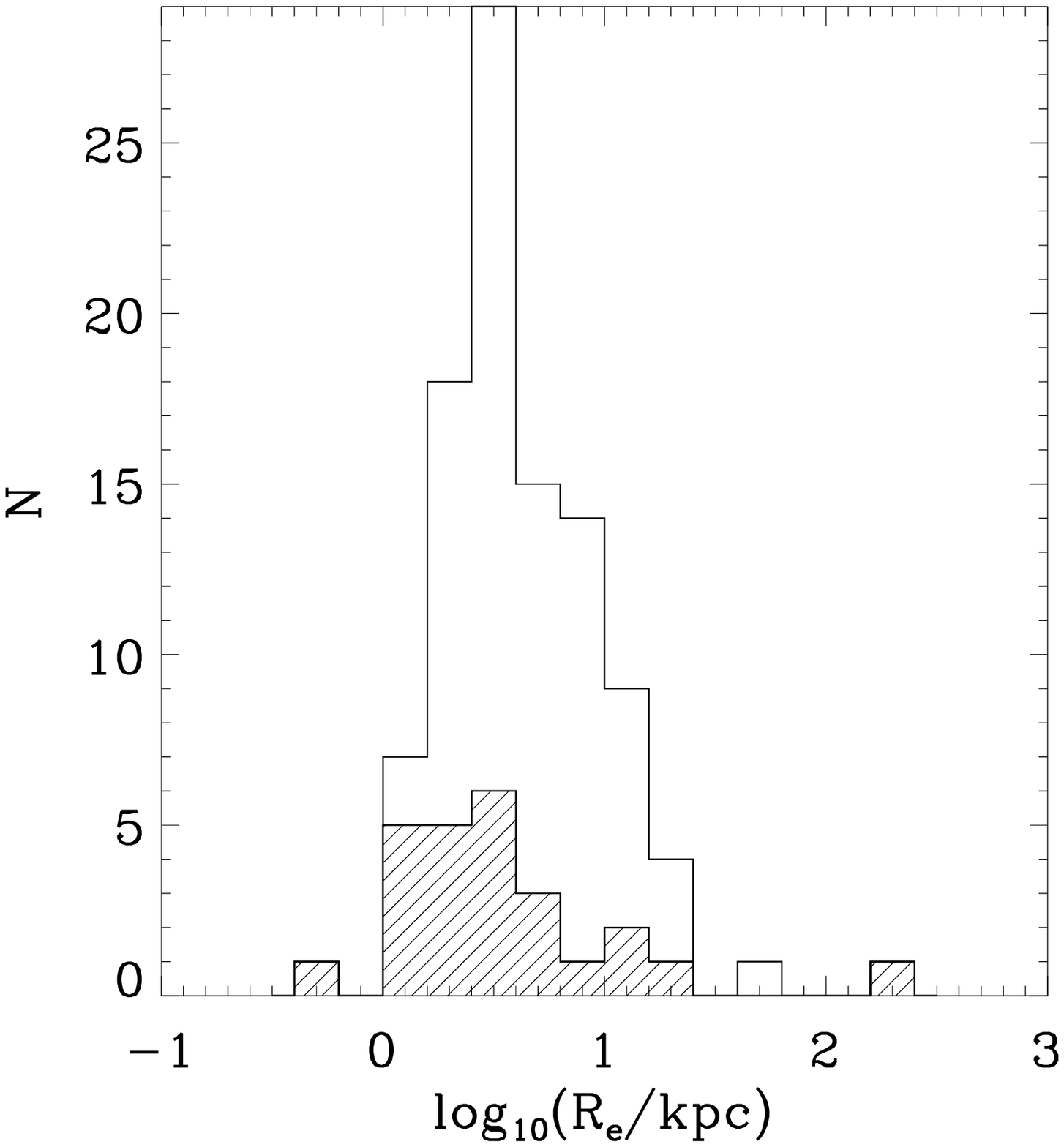}
\caption{\label{fig-R-hist} Scale-length distribution for the full sample (unshaded) and for the FR~I's (shaded).}
\end{figure}

The host galaxy luminosity distribution is shown in Fig.~\ref{fig-MH-hist}, with their scale length distribution shown in Fig.~\ref{fig-R-hist}.
All of the 3CR sources studied, except for 3C~258, 3C~270, and 3C~296, are hosted by galaxies with luminosities above $M_H=-22$, and scale lengths $>1$~kpc. The highly peculiar and disturbed 3C~258 (see earlier comment in section~\ref{sec-3c258}) is found to have an extremely compact and low-luminosity host galaxy, consistent with it being at a much higher redshift. 

Half of the sources have host galaxies brighter than $L^\star$, and larger than $3$~kpc, with the peak in the luminosity and scale-length distributions just below $L^\star$ and $3$~kpc respectively. This is as to be expected based on our knowledge of radio galaxies (e.g.~\citealt{zibetti+02}), quasar host galaxies~\citep{dunlop+03,floyd+04} and unification schemes~\citep{barthel89}: Our understanding of the AGN phenomenon hinges upon the presence of a supermassive black-hole as the ultimate energy source, and our growing demographic understanding of galaxies is that a massive black-hole requires a massive galaxy bulge or spheroid to host it.

The low-luminosity tail of the sample is dominated by low-redshift and low-radio-power FR~I's:
Of the twenty-two sources with luminosities $M_H > -23$, nineteen are at $z<0.09$, with only 3C~61.1, 3C~258 (once again), and 3C~314.1 breaking the general rule. 

3C~61.1 at $z=0.186$ has an unusual radio morphology, and appears to be hosted by an unusually faint, small galaxy at the  center of a small group. The optical image shows complex structure~\citep{dekoff+96}, with tails of emission that resemble spiral arms, and which do not correlate with the radio emission features~\citep{leahyperley91}. However, the published 20~cm VLA radio map is of low resolution, and it is difficult to confirm the location of the optical-IR counterpart accurately. 

3C~314.1 has an unusually low-surface-brightness host galaxy at $z=0.119$, with a slightly dusty appearance from the optical-IR images.

\subsection{Infrared nuclei}
\begin{figure}
\plotone{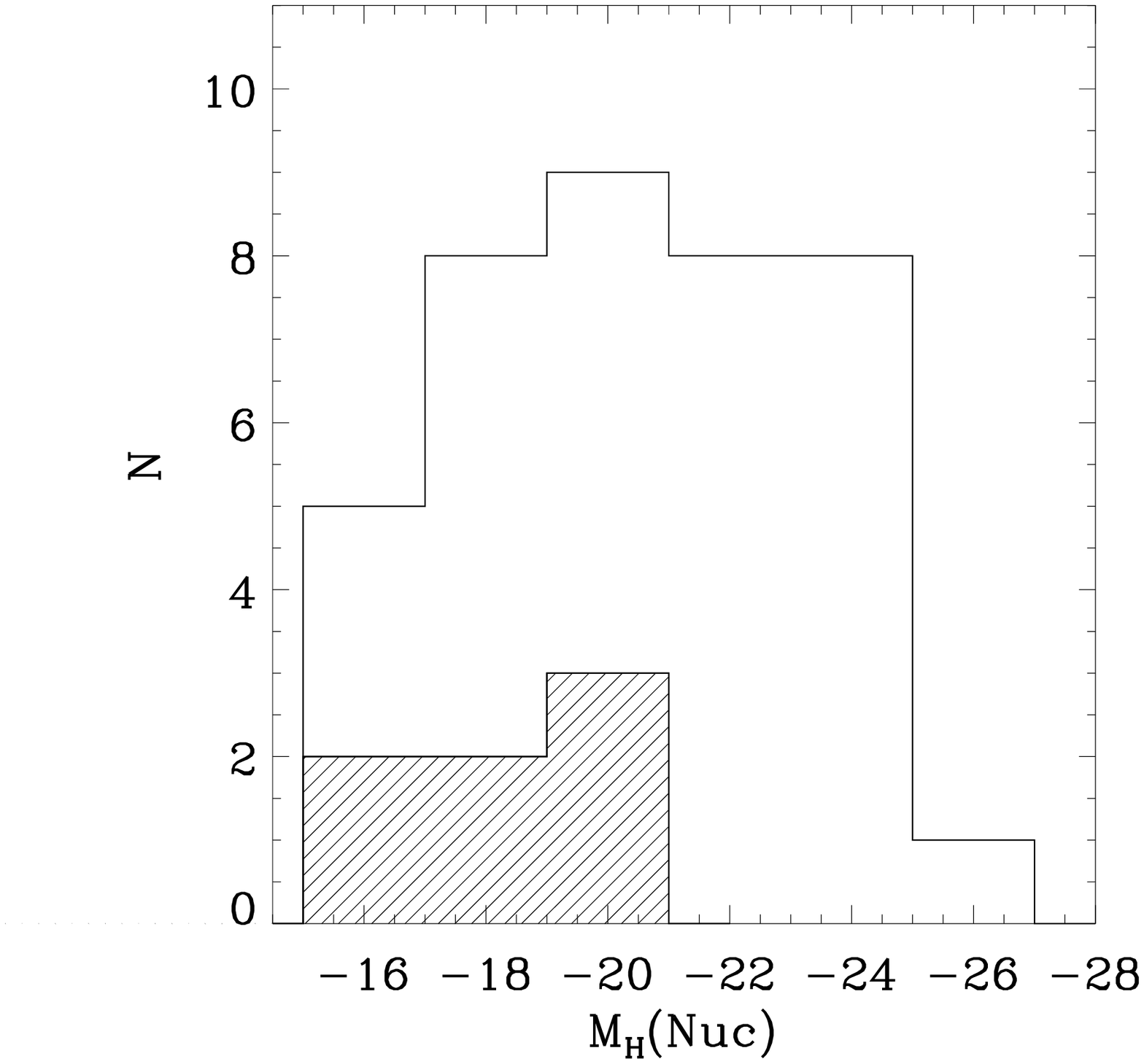}
\caption{\label{fig-MN-hist} Nuclear luminosity distribution for the full sample (unshaded) and for the FR~I's (shaded): We have detected unresolved IR nuclei in 39 objects, down to an absolute H-band magnitude of $-15$.}
\end{figure}

\begin{figure*}
\centering
{\includegraphics[width=5.0cm]{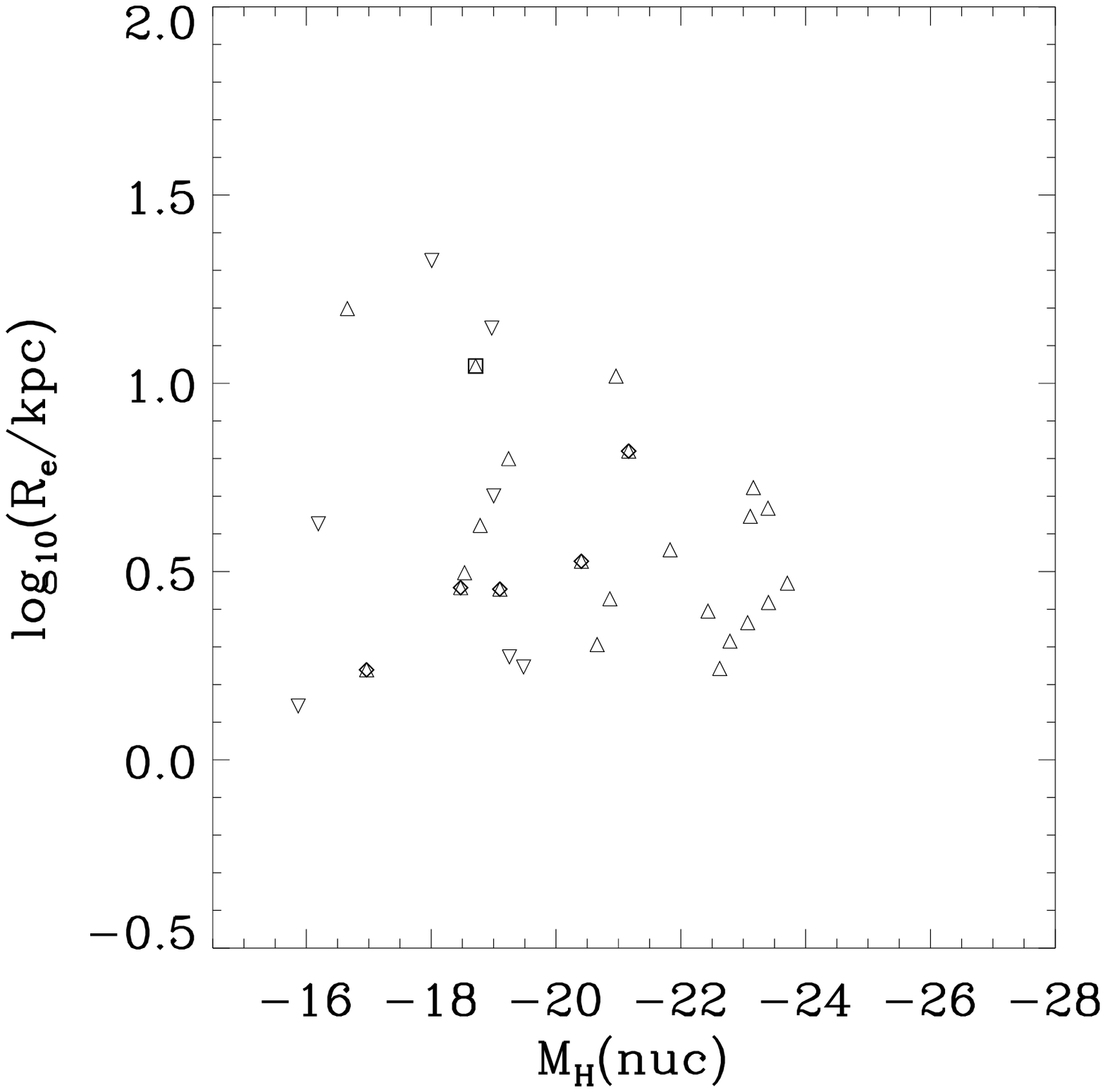}}%{/Users/floyd/3C/Work/plots/MN_R.eps}}
{\includegraphics[width=5.0cm]{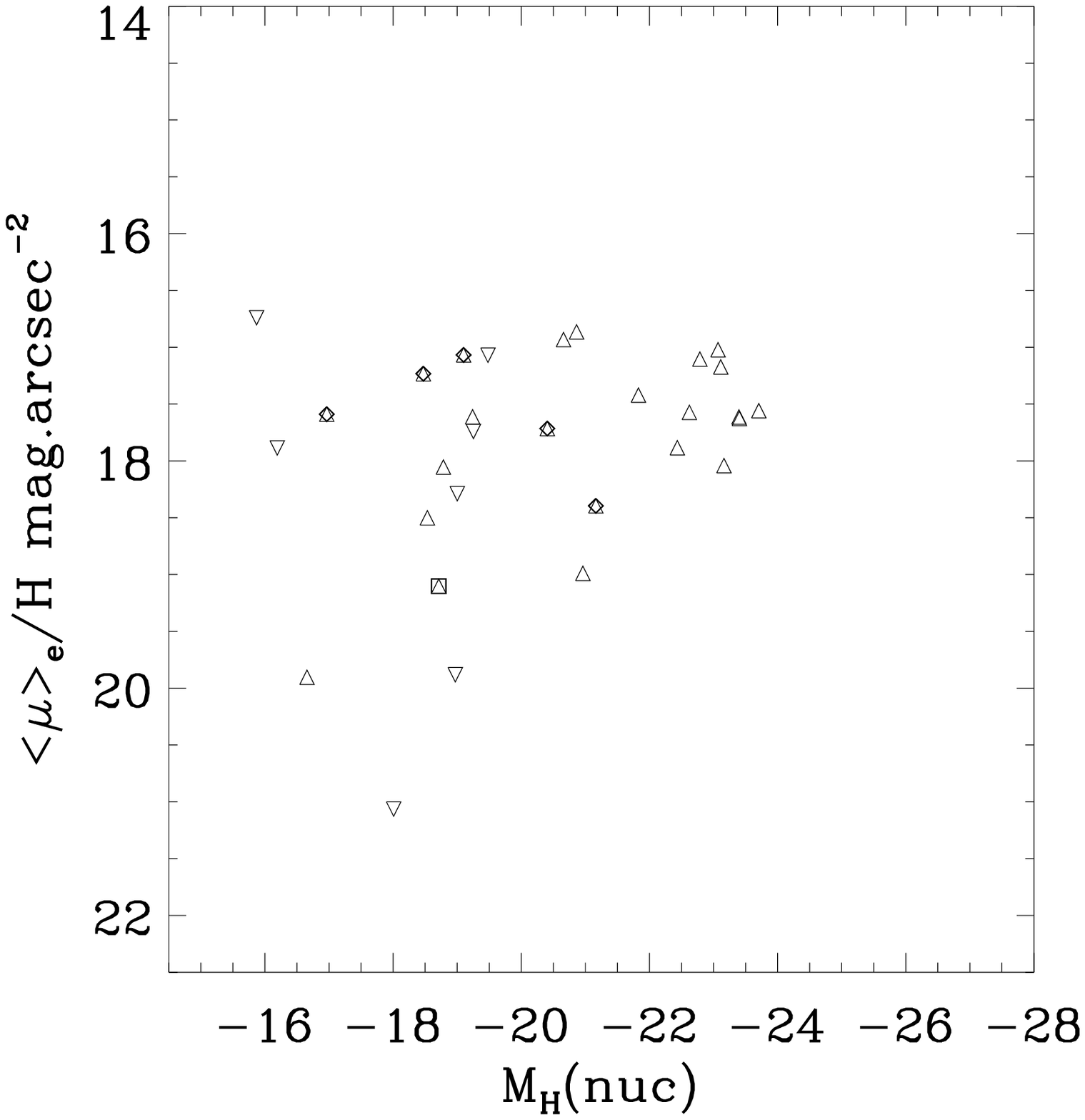}}%{/Users/floyd/3C/Work/plots/MN_mu.eps}}
{\includegraphics[width=5.0cm]{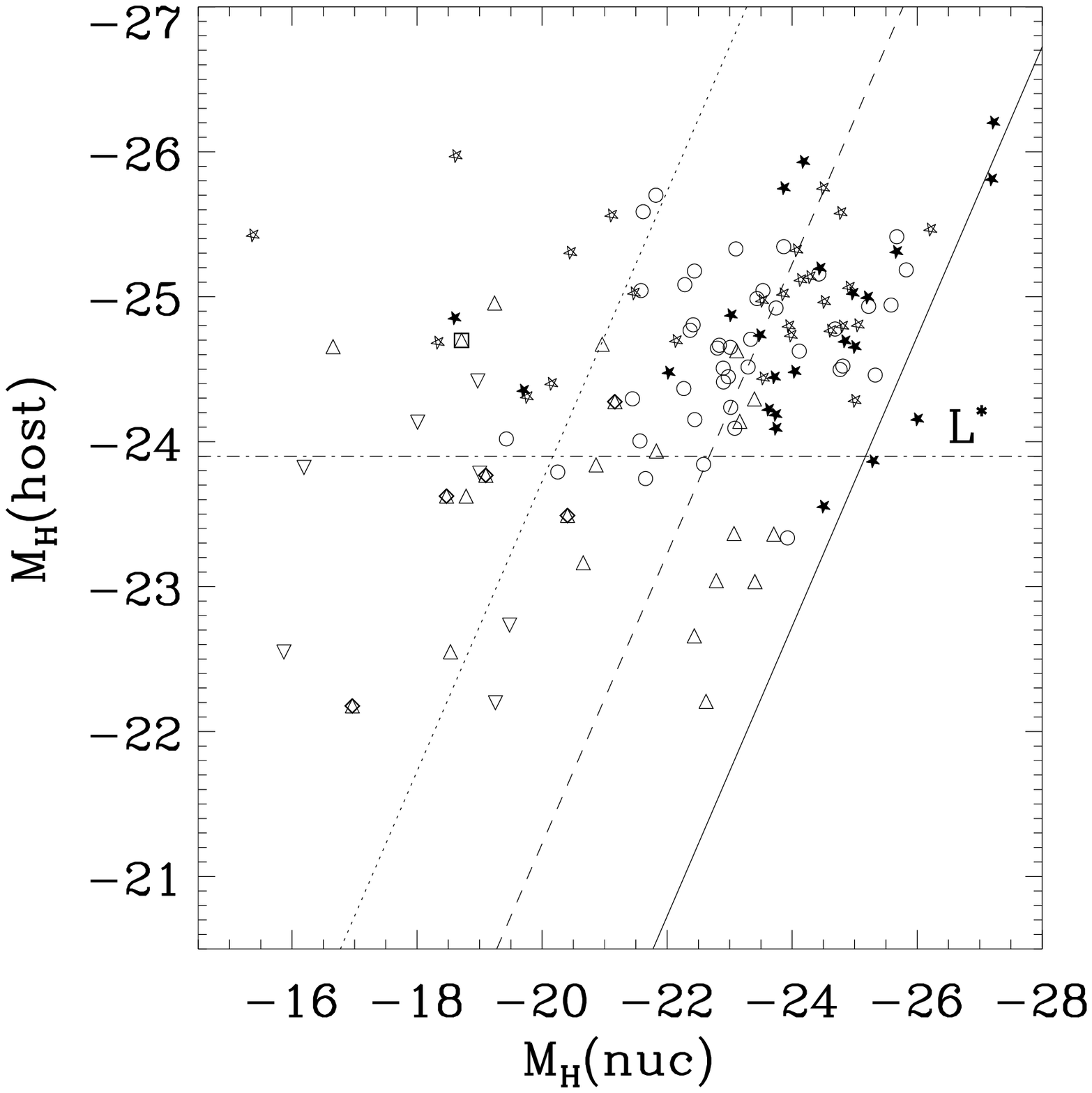}}%{/Users/floyd/3C/Work/plots/MN_MH.eps}}
\caption{\label{fig-MN} Nuclear IR luminosity vs. host galaxy properties, left-to-right: Scale-length, surface brightness, and host galaxy luminosity. There is a slight tendency for stronger nuclear sources to be located in higher surface-brightness galaxies. No trend is observed between nuclear luminosity and host luminosity (right) or scale-length (left).}
\end{figure*}

A number of our sources (40\%) have clearly detectable infrared nuclear activity. We were able to detect nuclear point sources down to an absolute magnitude of $M_H(Nuc)=-15$ (table~\ref{tab-sers}, Fig.~\ref{fig-MN-hist}). A more exhaustive search for faint nuclei, and a discussion of their origin will be the subject of another paper (Chiaberge et al. {\em in preparation}).
Firstly, we note that nuclear point sources are detected in {\em both} FR~I's and FR~II's, with roughly equal probability within our sample (38\% of FR~I's, 41\% of FR~II's).
No significant correlation is found between the NIR nuclear luminosities of the sources in which we have identified unresolved nuclear point sources, and the host luminosity, nor scale-length, although a weak correlation is seen with host galaxy surface brightness ($\rho=0.34$; $p=0.035$) -- see Fig.~\ref{fig-MN}.
%A correlation is observed between the infrared nuclear absolute magnitude and the core 5~GHz radio power ($\rho=-0.46$, $p=0.087$) of the full sample. The weakness of the correlation may in part be due to the small number of sources for which we have both core radio powers and detected IR nuclei. 
%However, we only present luminosities for the most obvious nuclei here, which one would suppose are dominated by thermal emission from the circumnuclear environment.
%This relationship  is studied in greater depth, in particular for the FR~I's, by Chiaberge et al. ({\em in preparation}), who detect IR nuclei to far lower flux levels in $\sim 80$\% of the present sample of FR~I's.
The difference in the luminosity range of FR~I and FR~II sources is very clearly seen from Fig.~\ref{fig-MN-hist}: FR~I's although they can possess nuclei, do not contain luminous quasar-like nuclei that are seen in the FR~II population.

\subsection{Kormendy relation (R-$\mu$)}
The strong correlation between scale length and surface brightness is a well-known feature of elliptical galaxies~\citep{kormendy77}. The relation followed by the present low-redshift 3CR sample is somewhat lower than that of $3.33\pm0.09$ established for early type galaxies using the Sloan Digital Sky Survey~\citep{bernardi+03}, following a slope of $3.0\pm0.1$.

\[ \mu_{\mathrm 3CR}=16.0\pm0.1+(3.0\pm0.1)\log_{10}(R_{e}) \]

\begin{figure}%[htbp]
\plotone{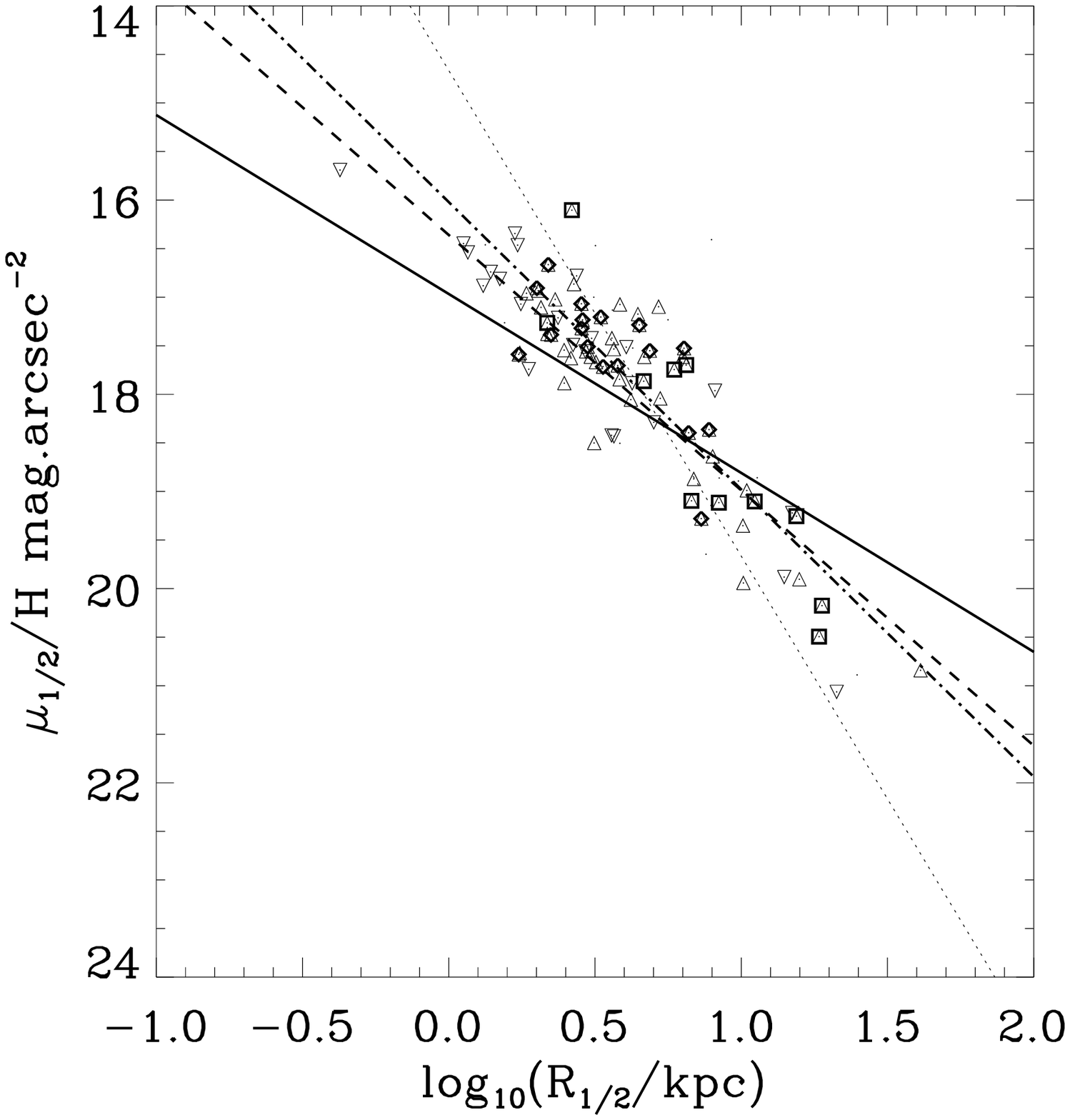}
\caption{\label{fig-korm} Best fitting scale-length--surface brightness relation for the full sample (solid line), the FR~I's (dashed) and the FR~II's (dot-dash). The locus of an $L^\star$ galaxy is shown by the dotted line.}
\end{figure}

Kormendy fit to the FR~I's and FR~II's yield somewhat different slopes of $3.46\pm0.1$ and $2.67\pm0.1$ respectively (Fig.~\ref{fig-korm}).
\[ \mu_{\mathrm FRI}=15.9\pm0.1+(3.46\pm0.1)\log_{10}(R_{e}) \]
\[ \mu_{\mathrm FRII}=16.1\pm0.1+(2.67\pm0.1)\log_{10}(R_{e}) \]

\subsection{Fundamental plane}
\begin{figure}
\plottwo{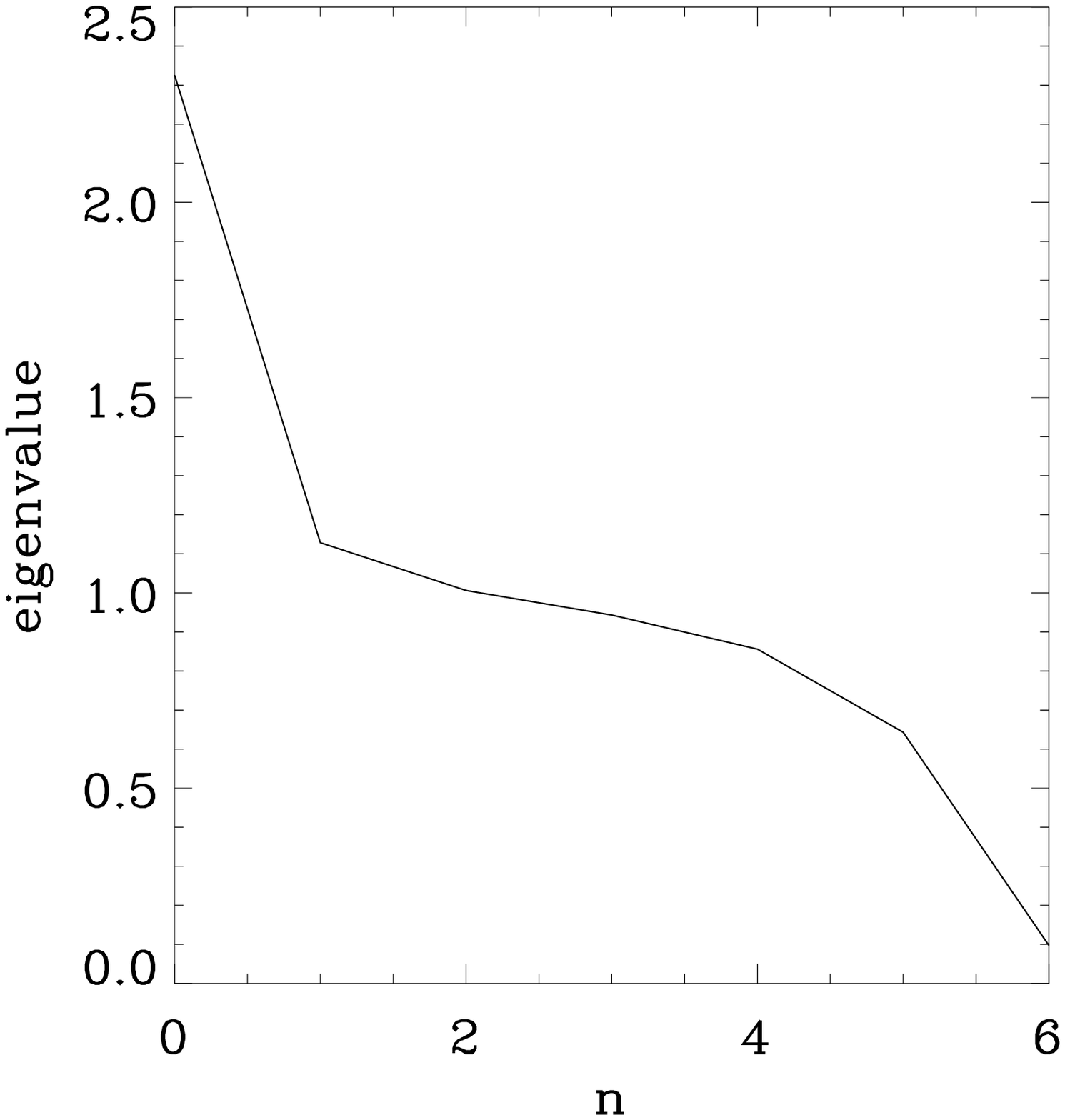}{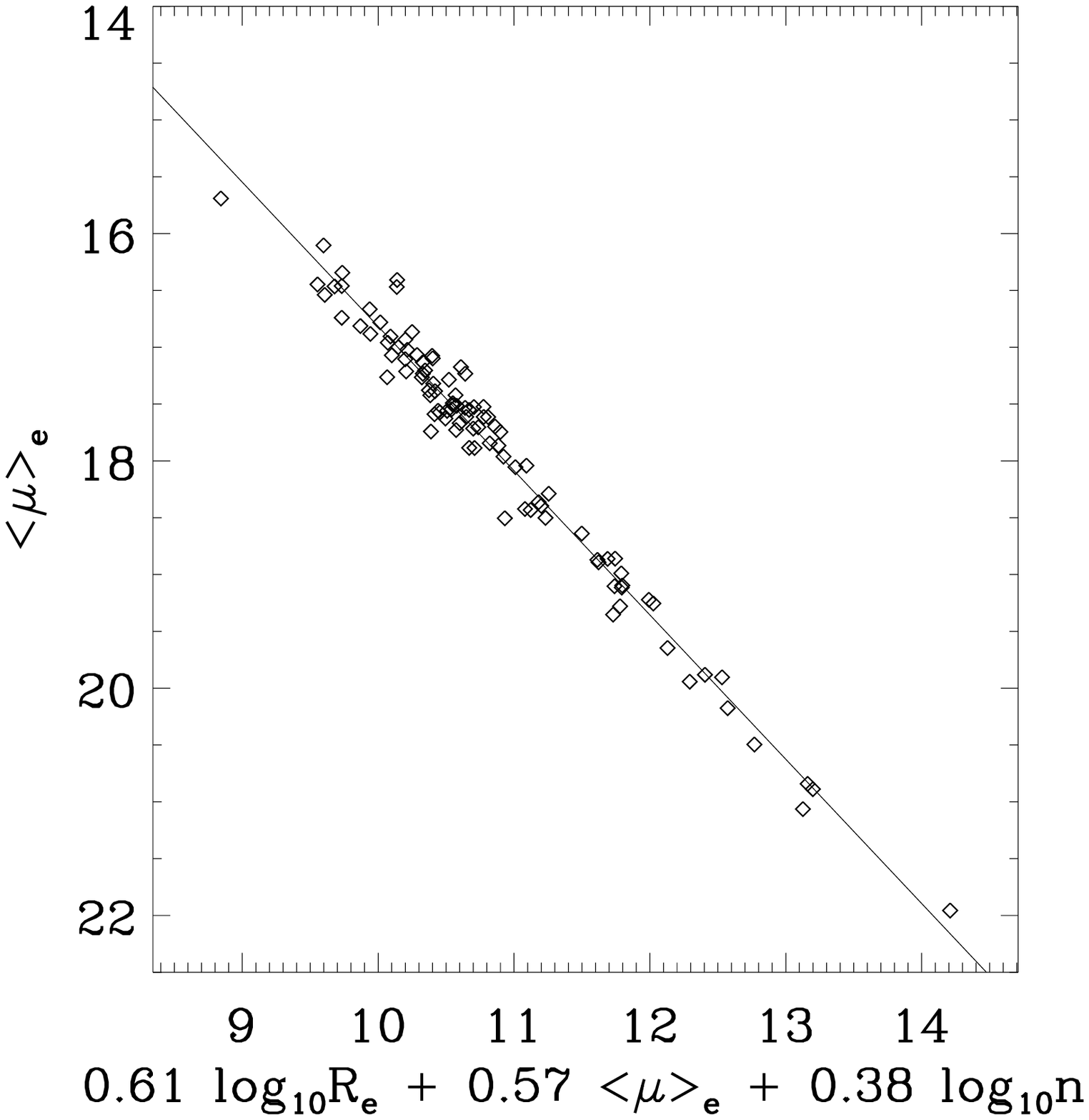}
\caption{\label{fig-FP} Left: Eigenvalues from the Principal Component Analysis on the fitted parameters from the 2-dimensional (Galfit) modeling. Only the first eigenvector accounts for a significant proportion of the scatter in the entire dataset. Right: Photometric fundamental plane (eigenvector 1).}
\end{figure}

We used Principal Component Analysis (PCA) to explore the model parameter space for any 3-dimensional correlations, like the photometric ``fundamental plane'' reported for normal elliptical galaxies by~\citep{khosroshahi+00a, khosroshahi+00b}. PCA looks for correlations among a set of multi-dimensional data and rotates the coordinate axes from being aligned with the input parameters to new directions that are aligned with the directions of greatest  variance of the data--the principal axes. In order to prevent the choice of units from artificially weighting some parameters more than others, each parameter is first normalized by subtracting its mean and dividing by its variance.  The PCA is then performed on these normalized variables.  In the output, each of the eigenvectors -- the principal axes-- is written as a linear combination of the original (but normalized) parameters.  The eigenvalues are scaled so that the sum of all eigenvalues equals the total number of eigenvectors (and therefore the total number of parameters as well). 
In Fig.~\ref{fig-FP} we show the ``scree plot'' of eigenvalues associated with the seven eigenvectors fit, along with a plot of the data over the first eigenvector. Only the first eigenvector accounts for a significant quantity of the scatter in the parameter space, and is made up almost entirely of $R_e$, $\mu_e$, and \sersic $n$. The recovered form of the relationship is: \[ \langle\mu\rangle_{e}= 2.77 \log_{10} R_{e} + 1.73\log_{10} n + 14.78\]

\subsection{Host galaxy colors}
\begin{figure}
\plotone{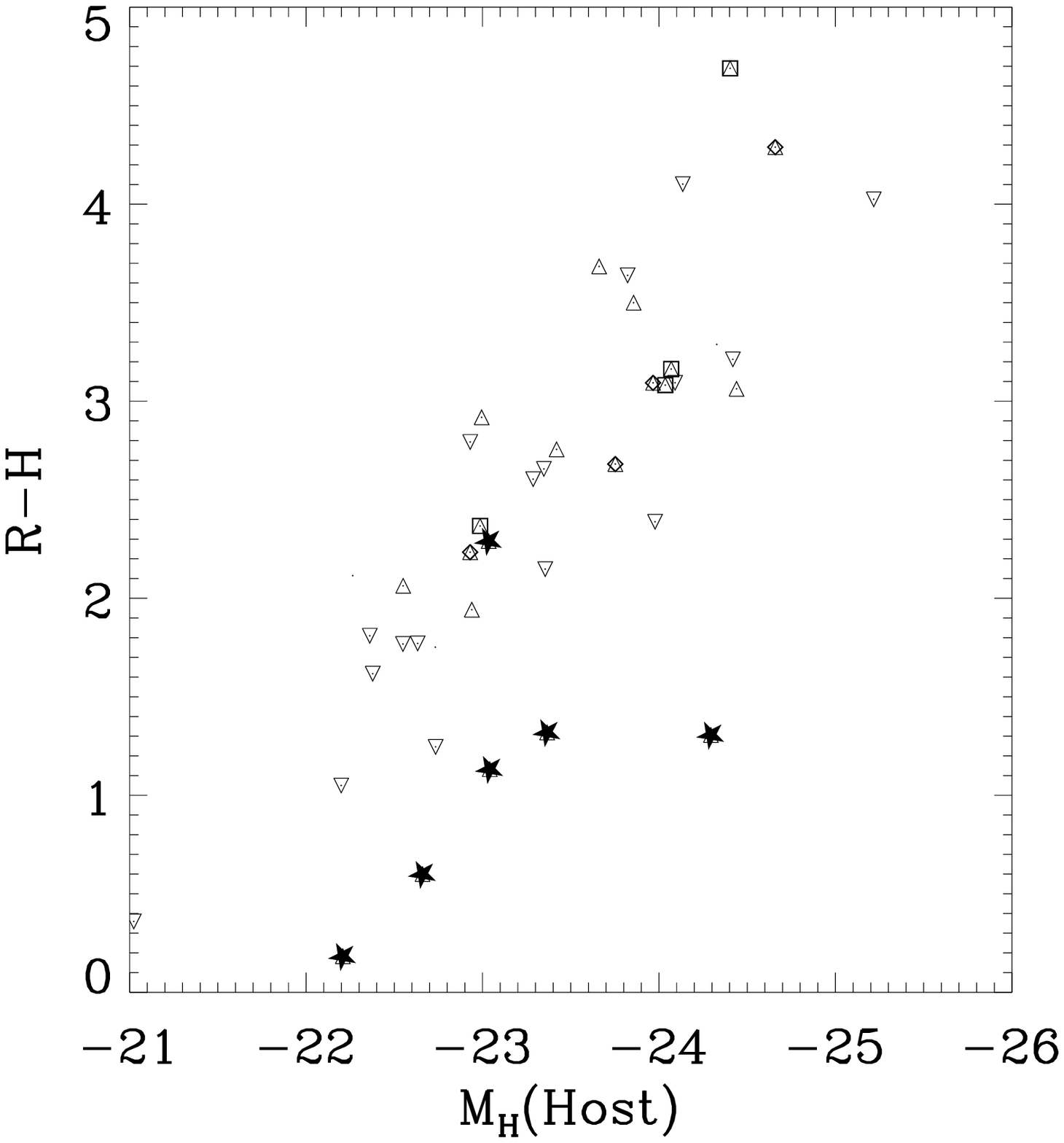}
\caption{\label{fig-CM} Color -- absolute magnitude diagram for the sample. More luminous hosts are significantly redder. Nucleated sources (with nuclei brighter than $M_H=-21$) are shown by a filled star, and dominate the blue end off the distribution.}
\end{figure}

For the cross-sample of this paper and Martel et al. (1999) we can produce accurate $R-H$ colors for the host galaxy, shown in table~\ref{tab-col}. A color -- absolute magnitude diagram for this subsample is presented in Fig.~\ref{fig-CM}. An extremely steep trend is seen between color of the host galaxy ($R-H$) and absolute $H$-band magnitude of the host galaxy, with a slope of roughly 1. %~mag.~mag$^{-1}$.
The extremely red colors at the top end of the sequence are likely due to dust absorption in the optical images of sources like 3C~293,  3C~296, 3C~388, and 3C~403. The blue outliers are all nucleated sources, in which the Martel et al. fluxes are likely to be higher than the true galaxy flux, as no attempt was made by those authors to separate host and nucleus. %3C~227, 3C~371, 3C~382, 3C~390.3, 3C~445

\subsection{Host galaxy masses}
Based on the apparent similarity between the present sample, and the normal quiescent elliptical galaxy population, we have applied the finding of~\citet{zibetti+02} that $H$-band luminosity provides a first order estimate of the dynamical mass of an elliptical galaxy. They find a mean mass-to-light ratio of $\log (M/L_H) = 0.09$ (in solar units), that is independent of total luminosity. There is an approximately 0.2 dex 1$\sigma$ scatter on this relationship.
%In Table~\ref{tab-mass} we show the approximate masses of each galaxy, with the 1$\sigma$ upper and lower limits indicated. 
The mean mass of a 3CR host galaxy from the present sample is $2\times10^{11} ~M_\odot$, but with very large scatter, in particular due to the 2 enormous outliers, 3C~130 ($10^{12}~M_{\odot}$) and 3C~338 ($5\times10^{12}~M_{\odot}$).
Clearly the assumption of constancy of the mass-to-light ratio for our sample needs verification by dynamical means, and we would argue that a significant hole in the current 3C literature is a fundamental plane relationship ($R$--$\mu$--$\sigma$) study to test whether these objects truly are dynamically representative of normal elliptical galaxies. In the absence of long-slit spectra in the public domain such a work is currently non-existent.

%%%%%%%%%%%%%%%%%%%%%%%%%%%%%%%%%%%%%%%%%%%%%%%%%%
\section{Summary and conclusions}
\label{sec-conc}

Ellipse and Galfit models have been presented for 101 $z<0.3$ 3CR radio source host galaxies. The two fitting methods (1D and 2D) are found to be statistically consistent for non-nucleated sources, although individual sources can provide significantly differing results using the two methods. Simulations have revealed that, with the addition of an artificial nuclear point source to a real source, we are still able to obtain accurate results using the 2-D technique, after the 1-D technique has failed. 

Unresolved nuclear point sources are detected with equal likelihood down to $M_H=-15$ in both FR~I and FR~II galaxies. FR~I's can have faint unresolved nuclei (all dimmer than $M_H=-20$), but these mostly exist in low-redshift sources in this radio-flux limited sample. 
%At the present level, no correlation is seen between the IR and radio luminosities of the nuclei of FR~I sources, but a weak correlation is observed in the FR~II's. 
We find no significant difference in properties of the host galaxies between nucleated and non-nucleated sources.

The host galaxies of the 3CR are generally consistent with the elliptical galaxy population: They exhibit a Kormendy relation that is similar to that of quiescent ellipticals and quasar host galaxies; the peak in the host galaxy luminosity distribution is close to $L^\star$, with steeper drop-off to high luminosity than low; there is a strong correlation between host $H$-band luminosity and ($R-H$) color. 
However, the 3CR exhibit a large spread in \sersic index to low values, and includes a number of quite disky sources. \sersic $n$ is not found to correlate with the radio luminosity, and yet the disky (low-$n$) wing of the sample is dominated by low-$z$, low radio-power sources, merging sources, and sources with luminous companions. This suggests that disky galaxies are capable of hosting powerful radio sources, providing that they are massive enough, or are undergoing a major merger.
%A far weaker correlation exists between total radio power and host galaxy color. 
%Sizes and luminosities of the host galaxies correlate with each other and with radio power.%More massive galaxies are well-known to be redder, but it seems that AGN-powered radio luminosities do not correlate as well. 

Approximately 50\% of our sample exhibit compact, often unresolved, companion sources that are too red and faint to have been detected in previous optical snapshot programs. Spectroscopy is required to determine the nature and redshifts of these sources. Their typical colors (where this can be determined) are around $R-H=2-3$, consistent with mature stellar populations at the redshift of the primary source. These sources are found predominantly in elliptical host galaxies, and have 3 likely origins:
Foreground stars - likely in cases with low galactic latitude;
Synchrotron Hotspots - particularly in a few peculiar radio sources;
Merger remnants - compact cores from cannibalized small galaxies (e.g. \citealt{canalizo+03}); Or molecular gas clouds infalling into the galaxy~\citep{bellamytadhunter04}.
By comparison, only $\sim$10\% of our sample show signs of an ongoing or recent major merger, typical of the elliptical galaxy population in general (though many more show other signs of disturbance). If many of these sources do turn out to be galactic nuclei, it would suggest that a minor merger is sufficient to fuel or re-fuel a quiescent black hole in an elliptical galaxy into radio-loud AGN activity, while disky galaxies appear to require a more major disturbance.
There is a correlation between radio power, and the isophotal twist angle within our sample, with all sources exhibiting significant isophotal twists.
We thus conclude that while bulgy elliptical galaxies host the majority of powerful radio sources, it is possible to trigger such activity in diskier objects through mergers. The diskiness may simply be a morphological distortion resulting from the merger, or the merger may be able to provide sufficient fuel to a black-hole in a disky galaxy to trigger behavior only normally seen in the more massive black holes at the centers of ellipticals.

A detailed spectroscopic study of the 3CR host galaxies is overdue. It is important to explore the dynamical states of these objects to place them on the ``fundamental plane'' ($R-\mu-\sigma$) of galaxies. The nature of the companion sources needs to be explored, ideally using imaging spectroscopy to determine their redshifts and compositions at the same time as exploring the makeup of the 3CR host galaxy itself.

%%%%%%%%%%%%%%%%%%%%%%%%%%%%%%%%%%%%%%%%%%%%%%%%%%
\section{Acknowledgments}
Based on observations with the NASA/ESA Hubble Space Telescope, obtained at the Space Telescope Science Institute, which is operated by the Association of Universities for Research in Astronomy, Inc. (AURA), under NASA contract NAS5-26555. We gratefully acknowledge support from HST grant STGO-10173.
We wish to thank the anonymous referee for constructive comments and suggestions which significantly improved the reading of this paper.

\bibliographystyle{astron}
\bibliography{floyd_3Chosts}

%%%%%%%%%%%%%%%%%%%%%%%%%%%%%%%%%%%%%%%%%%%%%%%%%%
% TABLES: 
% 1: NEWLY OBSERVED TARGETS:
\input{tab1}

%%%%%%%%%%%%%%%%%%%%%%%%%%%%%%%%%%%%%%%%%%%%%%%%%
% 2: ARCHIVAL OBJECTS:
\input{tab2}

%%%%%%%%%%%%%%%%%%%%%%%%%%%%%%%%%%%%%%%%%%%%%%%%%
% 3: FULL SAMPLE PROPERTIES
\begin{deluxetable}{lllrllrrrcrr}  
  \tabletypesize{\tiny}
  \tablecolumns{12}
  \tablewidth{0pc} 
  \renewcommand{\arraystretch}{.6}
  \tablecaption{\label{tab-props} Radio Properties of the NICMOS Snapshot Survey Sample}
  \tablehead{
    \colhead{Source} & \colhead{$z$} & \colhead{$V$} &
    \colhead{$S(178~\mathrm{MHz})^{a}$/Jy} & \colhead{log$_{10}$ P$_{178}$} &  
    \colhead{$\alpha^{a}$} & \colhead{LAS$^{b,c}$} & \colhead{PA$^{b,c}$} & \colhead{FR$^{a}$} &
    \colhead{Class$^{d}$} & \colhead{$L_{OII}^{d}$} & \colhead{$L_{OIII}^{d}$}}
  \startdata
3C~15 & 0.073 & 15.3 & 15.8 & 26.3 & 0.64 & ... & ... & ... & ... & ... & ... \\ 
3C~17 & 0.219 & 18.0 & 20.0 & 27.4 & 0.52 & ... & ... & ... & ... & ... & ... \\ 
3C~20 & 0.174 & 19.0 & 42.9 & 27.6 & 0.67 & 51 & 101 & 2 & HEG & 33.99 & 34.47 \\ 
3C~28 & 0.195 & 17.6 & 16.3 & 27.2 & 1.06 & 30 & 166 & 2 & LEG & 35.07 & ... \\ 
3C~29 & 0.044 & 14.1 & 15.1 & 25.8 & 0.50 & 139 & 160 & 1 & ... & ... & ... \\ 
3C~31 & 0.016 & 12.1 & 16.8 & 25.0 & 0.57 & 1833 & 159 & 1 & FRI & ... & ... \\ 
3C~33 & 0.059 & 15.2 & 54.4 & 26.7 & 0.76 & ... & ... & 2 & HEG & 34.72 & 35.37 \\ 
3C~33.1 & 0.181 & 19.5 & 13.0 & 27.1 & 0.62 & 216 & 45 & 2 & WQ & ... & 35.05 \\ 
3C~35 & 0.067 & 15.6 & 10.5 & 26.1 & 0.77 & 704 & 12 & 2 & LEG & ... & ... \\ 
3C~52 & 0.285 & 18.5 & 13.5 & 27.5 & 0.62 & 51 & 20 & 2 & ... & ... & ... \\ 
3C~61.1 & 0.186 & 19.0 & 31.2 & 27.5 & 0.77 & 186 & 2 & 2 & HEG & 34.70 & 35.48 \\ 
3C~66B & 0.021 & 12.9 & 24.6 & 25.4 & 0.62 & 330 & 54 & 1 & FRI & 33.16 & 33.01 \\ 
3C~71 & 0.004 & 8.9 & 16.1 & 23.7 & 0.55 & ... & ... & ... & ... & ... & ... \\ 
3C~75N & 0.024 & 13.6 & 25.8 & 25.5 & 0.71 & 692 & 111 & 1 & ... & ... & ... \\ 
3C~76.1 & 0.032 & 14.9 & 12.2 & 25.5 & 0.77 & 200 & ... & 1 & FRI & ... & ... \\ 
  \enddata
  \tablecomments{Main radio and optical properties for each source:
    redshift; 
    V-band magnitude; 
    radio flux density in Janskys at 178~MHz; 
    radio  power at  178~MHz  in W~Hz$^{-1}$; 
    radio spectral index; 
    radio source largest angular size; 
    radio structure position angle; 
    Faranoff-Riley class;
    Ionization Class;
    [OII]~3727 emission line luminosity in W;
    [OIII]~5007 emission line luminosity in W.
    [The complete version of this table is in the electronic edition of
the Journal.  The printed edition contains only a sample.]}
  \tablerefs{$^{a}$~\citet{spinrad+85} - updated values were taken from the 
    NASA Extragalactic Database (NED);
    $^{b}$~\citet{dekoff+96} and references therein; 
    $^{c}$~\citet{martel+99} and references therein;
    $^{d}$~\citet{jacksonrawlings97}.}
\end{deluxetable}

%%%%%%%%%%%%%%%%%%%%%%%%%%%%%%%%%%%%%%%%%%%%%%%%%
% 4: ELLIPSE RESULTS
\begin{deluxetable}{llrrllrrrr}
  \tabletypesize{\tiny}
  \tablecolumns{10}
  \tablewidth{0pc} 
  \renewcommand{\arraystretch}{.6}
  \tablecaption{\label{tab-ellipse} Ellipse and 1-D \sersic model fits}
  \tablehead{
    \colhead{Source} & 
    \colhead{$(1-b/a)_{\mathrm med}$} & 
    \colhead{$\theta_{\mathrm med}$} &
    \colhead{$a_3/a_{\mathrm min}$} &
    \colhead{$a_3/a_{\mathrm max}$} &
    \colhead{$a_4/a_{\mathrm min}$} & 
    \colhead{$a_4/a_{\mathrm max}$} &
    \colhead{$r_e$/kpc} & 
    \colhead{$\langle\mu\rangle_e$} & 
    \colhead{$n$}}
  \startdata
3C~15 & 0.06 & 21.32 & -0.08 & 0.02 & -0.04 & 0.02 &$4.96\pm 0.45$&$19.76\pm 0.13$&$4.17\pm 0.08$\\ 
3C~17 & 0.17 & 21.32 & -0.17 & 0.06 & -0.08 & 0.14 &$5.84\pm 1.53$&$20.69\pm 0.52$&$10.00\pm 0.00$\\ 
3C~20 & 0.10 & -63.77 & -0.06 & 0.10 & -0.03 & 0.20 &$2.72\pm 1.45$&$19.17\pm 1.05$&$3.23\pm 0.54$\\ 
3C~28 & 0.18 & -26.43 & -0.03 & 0.11 & -0.04 & 0.03 &$6.74\pm 0.63$&$20.26\pm 0.11$&$2.01\pm 0.08$\\ 
3C~29 & 0.03 & -76.28 & -0.05 & 0.02 & -0.05 & 0.03 &$4.44\pm 0.15$&$19.05\pm 0.02$&$1.50\pm 0.02$\\ 
3C~31 & 0.12 & -32.23 & -0.02 & 0.08 & -0.03 & 0.01 &$1.46\pm 0.05$&$17.20\pm 0.02$&$1.43\pm 0.02$\\ 
3C~33 & 0.07 & -43.84 & -0.09 & 0.01 & -0.04 & 0.06 &$2.75\pm 0.21$&$18.97\pm 0.11$&$3.19\pm 0.07$\\ 
3C~33.1 & 0.06 & 13.96 & -0.02 & 0.03 & -0.05 & 0.02 &$2.19\pm 0.65$&$18.66\pm 0.65$&$4.49\pm 0.33$\\ 
3C~35 & 0.23 & -65.28 & -0.18 & 0.02 & -0.07 & 0.03 &$4.59\pm 0.67$&$19.89\pm 0.20$&$3.60\pm 0.14$\\ 
3C~52 & 0.31 & 10.96 & -0.03 & 0.06 & -0.02 & 0.08 &$3.55\pm 0.36$&$18.74\pm 0.18$&$1.73\pm 0.09$\\ 
3C~61.1 & 0.09 & 0.95 & -0.17 & 0.22 & -0.07 & 0.08 &$1.87\pm 0.16$&$19.65\pm 0.20$&$2.07\pm 0.11$\\ 
3C~66B & 0.06 & -50.32 & -0.08 & 0.04 & -0.03 & 0.06 &$2.42\pm 0.15$&$18.42\pm 0.04$&$1.97\pm 0.05$\\ 
3C~71 & 0.18 & 36.54 & -0.04 & 0.03 & -0.06 & 0.02 &$2.21\pm 0.51$&$19.51\pm 0.35$&$10.00\pm 0.00$\\ 
3C~75N & 0.19 & -80.47 & -0.01 & 0.05 & -0.02 & 0.04 &$1.64\pm 0.22$&$17.75\pm 0.16$&$3.09\pm 0.12$\\ 
3C~76.1 & 0.14 & -47.79 & -0.02 & 0.04 & -0.05 & 0.02 &$1.12\pm 0.07$&$17.95\pm 0.09$&$2.26\pm 0.06$\\ 
  \enddata
  \tablecomments{Properties of the elliptical isophotal fit to each source, including 1-D radial-profile \sersic fit:
    median ellipticity;
    median position angle (East of North);
    minimum value of $a_3/a$;
    maximum value of $a_3/a$;
    minimum value of $a_4/a$;
    maximum value of $a_4/a$.
    scale length (half-light radius) of best-fit model to elliptical isophotes;
    mean surface brightness of model within scale-length, $\langle\mu\rangle_e$;
    \sersic index of model.
    [The complete version of this table is in the electronic edition of
the Journal.  The printed edition contains only a sample.]}
\end{deluxetable}

\clearpage

%%%%%%%%%%%%%%%%%%%%%%%%%%%%%%%%%%%%%%%%%%%%%%%%%
% 5: SERSIC RESULTS
\begin{deluxetable}{lrrrrrrrr}
  \tabletypesize{\tiny}
  \tablecolumns{9}
  \tablewidth{0pc}
  \renewcommand{\arraystretch}{.6}
  \tablecaption{\label{tab-sers} 2-D \sersic model fits}
  \tablehead{
  \colhead{Source} &
  \colhead{$M_{H}$(Nuc)} &
  \colhead{$M_{H}$(Host)} &
  \colhead{$R_e$/kpc} &
  \colhead{$\mu_H$} &
  \colhead{$n$} &
  \colhead{$1-\frac{b}{a}$} &
  \colhead{$\theta$} &
  \colhead{$d$}
  }
  \startdata
3C~15 &$...$&$-24.33\pm 0.01$&$8.41\pm 0.05$&$18.86\pm 0.01$&$6.10\pm 0.02$&$0.07\pm 0.00$&$19.33\pm 0.38$&$0.11\pm 0.01$\\
3C~17 &$-21.51$&$-24.19\pm 0.01$&$4.25\pm 0.05$&$17.52\pm 0.00$&$7.49\pm 0.11$&$0.23\pm 0.00$&$20.94\pm 0.31$&$-0.21\pm 0.02$\\
3C~20 &$...$&$-23.61\pm 0.00$&$2.19\pm 0.01$&$16.66\pm 0.00$&$2.70\pm 0.01$&$0.10\pm 0.00$&$-65.53\pm 0.62$&$-0.15\pm 0.01$\\
3C~28 &$-18.71$&$-24.70\pm 0.01$&$11.12\pm 0.07$&$19.10\pm 0.01$&$2.32\pm 0.02$&$0.24\pm 0.00$&$-33.50\pm 0.23$&$-0.26\pm 0.01$\\
3C~29 &$-16.19$&$-23.82\pm 0.01$&$4.24\pm 0.01$&$17.89\pm 0.01$&$1.45\pm 0.00$&$0.04\pm 0.00$&$-81.55\pm 0.00$&$0.08\pm 0.00$\\
3C~31 &$...$&$-23.29\pm 0.00$&$1.72\pm 0.00$&$16.46\pm 0.00$&$1.67\pm 0.00$&$0.12\pm 0.00$&$-32.98\pm 0.08$&$-0.01\pm 0.00$\\
3C~33 &$...$&$-23.43\pm 0.00$&$2.98\pm 0.01$&$17.51\pm 0.00$&$3.60\pm 0.01$&$0.08\pm 0.00$&$-35.60\pm 0.26$&$-0.13\pm 0.01$\\
3C~33.1 &$-20.86$&$-23.84\pm 0.02$&$2.68\pm 0.02$&$16.86\pm 0.02$&$6.41\pm 0.00$&$0.01\pm 0.01$&$54.91\pm 2.56$&$0.00\pm 0.00$\\
3C~35 &$...$&$-24.07\pm 0.00$&$8.38\pm 0.04$&$19.11\pm 0.00$&$4.79\pm 0.01$&$0.28\pm 0.00$&$-66.34\pm 0.07$&$-0.02\pm 0.00$\\
3C~52 &$...$&$-25.06\pm 0.00$&$5.22\pm 0.02$&$17.10\pm 0.00$&$2.49\pm 0.01$&$0.35\pm 0.00$&$10.16\pm 0.18$&$-0.17\pm 0.02$\\
3C~61.1 &$-16.96$&$-22.18\pm 0.11$&$1.73\pm 1.56$&$17.59\pm 0.00$&$2.83\pm 0.00$&$0.13\pm 0.01$&$-2.02\pm 1.22$&$0.41\pm 0.00$\\
3C~66B &$...$&$-23.35\pm 0.01$&$2.93\pm 0.01$&$17.56\pm 0.01$&$2.39\pm 0.01$&$0.07\pm 0.00$&$-59.24\pm 0.34$&$0.00\pm 0.00$\\
3C~71 &$...$&$0.86\pm 0.01$&$1.17\pm 0.01$&$39.77\pm 0.00$&$3.12\pm 0.02$&$0.38\pm 0.00$&$44.99\pm 0.09$&$0.00\pm 0.00$\\
3C~75N &$...$&$-23.35\pm 0.00$&$1.68\pm 0.00$&$16.34\pm 0.00$&$3.68\pm 0.01$&$0.14\pm 0.00$&$-78.88\pm 0.15$&$0.22\pm 0.01$\\
3C~76.1 &$...$&$-22.28\pm 0.00$&$1.31\pm 0.00$&$16.88\pm 0.00$&$2.95\pm 0.00$&$0.16\pm 0.00$&$-48.12\pm 0.12$&$-0.08\pm 0.00$\\
  \enddata
\tablecomments{Properties of the best-fit 2-D \sersic model to each source:
Absolute magnitudes of the nucleus and host galaxy; 
scale-length (half-light radius) in kpc $R_e$; 
surface brightness at the scale-length $\mu$; 
\sersic index $n$; 
ellipticity $1-\frac{b}{a}$; 
position angle $\theta$ (relative to North); 
diskiness, $d$.
diskiness, $d$.
Notes: The following objects did not admit a complete or normal fit, as discussed in the main text: (a) 3C~71 is too extended to be properly constrained on either the NIC2 or WFPC2 chips; (b) 3C~258 -- values are presented as if the target lay at $z=0.165$, however it is likely that the source is actually at higher redshift.
[The complete version of this table is in the electronic edition of the Journal.  The printed edition contains only a sample.]}
\end{deluxetable}

%%%%%%%%%%%%%%%%%%%%%%%%%%%%%%%%%%%%%%%%%%%%%%%%%
% Table 6: Optical data 
\begin{deluxetable}{llll}
  \tabletypesize{\tiny}
  \tablecolumns{4}
  \tablewidth{0pc}
  \renewcommand{\arraystretch}{.6}
  \tablecaption{\label{tab-opt} Optical data from the WFPC2 archive}
  \tablehead{
  \colhead{Source} &
  \colhead{PROPOSID} & 
  \colhead{Filter} & 
  \colhead{exptime}}
\startdata
3C~15	&	6348,5476	&	F555W,F702W 	&	600,280 \\
3C~17	&	6967,5476	&	F555W,F702W	&	600,280 \\
3C~20	&	5476		&	F702W		&	300 \\
3C~28	&	5476		&	F702W		&	280 \\
3C~29	&	6967,5476	&	F555W,F702W	&	600,280 \\
3C~31	&	6673,5476,6673&	F555W,F702W,F814W&460,280,460 \\
3C~33	&	5156		&	F702W		&	1700 \\
3C~33.1	&	6967,5476	&	F555W,F702W	&	600,300 \\
3C~35	&	6967,5476	&	F555W,F702W	&	600,280 \\
3C~52	&	6967,5476	&	F555W,F702W	&	600,280 \\
3C~61.1	&	6348,5476	&	F555W,F702W	&	600,300 \\
3C~66B	&	6673,5476,6673&	F555W,F702W,F814W&460,280,460 \\
3C~71	&	5754,5754,5754,5479&	F218W,F336W,F791W,F606W&2400,900,440,500 \\
3C~75	&	5476,5927	&	F702W,F791W	&	280,750 \\
3C~76.1	&	6967		&	F555W		&	600 \\
\enddata
\tablecomments{Optical (WFPC2) images for each sample were obtained from the MAST (or from http://acs.pha.jhu.edu/$\sim$martel/ ). The table indicates the proposal ID for each image used, along with the filter for the observation, and the integration time.
[The complete version of this table is in the electronic edition of
the Journal.  The printed edition contains only a sample.]}
\end{deluxetable}

%%%%%%%%%%%%%%%%%%%%%%%%%%%%%%%%%%%%%%%%%%%%%%%%%
% 7: COMPANIONS
\begin{deluxetable}{llllrlrrr|ccccccccc}  
  \tabletypesize{\tiny}
  \tablecolumns{18}
  \tablewidth{0pc} 
  \renewcommand{\arraystretch}{.6}
  \tablecaption{\label{tab-companions} Companion Sources} 
  \tablehead{
    \colhead{Source} & \colhead{$N$} & 
    \colhead{$\alpha_{\mathrm J2000}$} & \colhead{$\delta_{\mathrm J2000}$} & 
    \colhead{$D$/\arcsec} & \colhead{$H$} & \colhead{FWHM} &
    \colhead{Elong.} & \colhead{Ellip.} &
    \colhead{Sh} & \colhead{TT} & \colhead{M} & \colhead{PM} & 
    \colhead{Maj} & \colhead{Min} & \colhead{Pt} & \colhead{Jet} & \colhead{DD}}
  \startdata
3C~15&      3& 00 37 04.05 & -01 09 05.5&16.57&19.21& 8.60&1.26&0.21 & ...&...&...&...&...& Y &...&...& Y   \\
3C~17&      3& 00 38 20.75 & -02 07 41.8& 6.73&22.43& 7.15&2.55&0.61 & ...&...&...& Y &...& Y &...&...&...  \\
3C~20&   $>9$& 00 43 09.67 & +52 03 30.5&18.31&17.80& 2.20&1.04&0.04 &  ? &...&...&...&...& Y & Y &...&...  \\
3C~28&   $>9$& 00 55 50.24 & +26 24 45.5&22.65&18.84& 4.31&1.04&0.04 & ...&...&...&...&...& Y &...& Y &...  \\
3C~29&      1& 00 57 35.33 & -01 23 37.4&23.63&24.04& 1.98&1.47&0.32 & ...&...&...&...&...&...&...&...&...  \\
3C~31&      3& 01 07 24.44 & +32 24 44.0&15.67&21.20& 2.18&1.28&0.22 & ...&...&...& Y & Y &...&...& Y &...  \\
3C~33&      0& ...         & ...        & ... & ... & ... & ...& ... & ...& Y &...&...&...&...&...& Y &...  \\
3C~33.1&    1& 01 09 44.11 & +73 11 54.4& 4.49&17.51& 4.87&1.99&0.50 & ...&...&...& Y & Y &...&...&...&...  \\
3C~35&      8& 01 12 01.93 & +49 28 44.2&15.93&17.81& 2.08&1.06&0.06 & ...&...&...& Y &...& Y & Y &...&...  \\
3C~52&   $>9$& 01 48 28.08 & +53 32 44.8& 6.60&18.66& 5.12&1.40&0.29 & ...&...&...&...&...& Y & Y & Y &...  \\
3C~61.1& $>9$& 02 22 36.07 & +86 19 07.7&11.73&18.10& 5.47&1.12&0.10 & ...& Y &...& Y & Y & Y & Y & Y &...  \\
3C~66B&     4& 02 23 11.54 & +42 59 23.1&16.55&19.52& 2.60&1.07&0.06 & ...&...&...& Y & Y &...& Y &...& Y   \\
3C~71&   $>9$& 02 42 41.38 & -00 00 49.4&17.70&18.06& 3.23&1.13&0.11 & ...&...&...&...&...& Y & Y &...&...  \\
3C~75N&     4& 02 57 43.63 & +06 02 31.5&18.89&18.74&12.54&2.69&0.63 & ...&...&...& Y & Y &...& Y &...&...  \\
3C~76.1&    5& 03 03 15.19 & +16 26 24.2& 9.65&19.99&17.36&2.24&0.55 & ...& Y &...&...&...& Y & Y & Y &...  \\
%3C~15&      3& 00 37 04.05 & -01 09 05.5&16.57&19.21&8.60&1.26&0.21\\
%3C~17&      3& 00 38 20.75 & -02 07 41.8&6.73&22.43&7.15&2.55&0.61\\
%3C~20&   $>9$& 00 43 09.67 & +52 03 30.5&18.31&17.80&2.20&1.04&0.04\\
%3C~28&   $>9$& 00 55 50.24 & +26 24 45.5&22.65&18.84&4.31&1.04&0.04\\
%3C~29&      1& 00 57 35.33 & -01 23 37.4&23.63&24.04&1.98&1.47&0.32\\
%3C~31&      3& 01 07 24.44 & +32 24 44.0&15.67&21.20&2.18&1.28&0.22\\
%3C~33&      0& ... & ... & ... & ... & ... & ... & ... \\
%3C~33.1&    1& 01 09 44.11 & +73 11 54.4&4.49&17.51&4.87&1.99&0.50\\
%3C~35&      8& 01 12 01.93 & +49 28 44.2&15.93&17.81&2.08&1.06&0.06\\
%3C~52&   $>9$& 01 48 28.08 & +53 32 44.8&6.60&18.66&5.12&1.40&0.29\\
%3C~61.1& $>9$& 02 22 36.07 & +86 19 07.7&11.73&18.10&5.47&1.12&0.10\\
%3C~66B&     4& 02 23 11.54 & +42 59 23.1&16.55&19.52&2.60&1.07&0.06\\
%3C~71&   $>9$& 02 42 41.38 & -00 00 49.4&17.70&18.06&3.23&1.13&0.11\\
%3C~75N&     4& 02 57 43.63 & +06 02 31.5&18.89&18.74&12.54&2.69&0.63\\
%3C~76.1&    5& 03 03 15.19 & +16 26 24.2&9.65&19.99&17.36&2.24&0.55\\
\enddata
\tablecomments{Number and properties of companion sources for each 3CR target:
3CR number; 
Number of companions on NIC2 image;
Coordinates of brightest companion source;
Distance (arcsec) of brightest companion from radio source;
H-band absolute magnitude of brightest secondary; 
Diameter (FWHM) of secondary;
Elongation (if resolved);
Ellipticity (if resolved). 
The right-hand side of the table lists the presence of various types of artefact ('Y'): 
Shells; tidal tails; mergers; pre-mergers; major companion sources; minor companion sources; unresolved point-like companion sources; jets; dust disks. See the main test for a fuller description of each category.
A question mark '?' implies the presence of the artifact is suggested but cannot be confirmed categorically.
[The complete version of this table is in the electronic edition of
the Journal.  The printed edition contains only a sample.]}
\end{deluxetable}

 %%%%%%%%%%%%%%%%%%%%%%%%%%%%%%%%%%%%%%%%%%%%%%%%%
% 8: colorS
\begin{deluxetable}{lrrr}
  \tabletypesize{\tiny}
  \tablecolumns{4}
  \tablewidth{0pc}
  \renewcommand{\arraystretch}{.6}
  \tablecaption{\label{tab-col} Host galaxy $R-H$ colors}
  \tablehead{
  \colhead{Source} &
  \colhead{$R^{a}$} & 
  \colhead{$H$} & 
  \colhead{$R-H$}}
\startdata
3c15 &	16.84 &	13.55 &	3.29 \\
3c29 &	16.53 &	12.89 &	3.64 \\
3c31 &	13.78 &	11.18 &	2.60 \\
3c35 &	16.84 &	13.68 &	3.16 \\
3c66B &	14.38 &	11.72 &	2.66 \\
3c75N &	14.22 &	12.07 &	2.15 \\
3c83.1 &	13.92 &	11.53 &	2.39 \\
3c84 &	13.50 &	12.45 &	1.05 \\
3c88 &	15.92 &	12.23 &	3.68 \\
3c98 &	15.26 &	13.03 &	2.23 \\
3c111 &	17.12 &	14.83 &	2.29 \\
3c198 &	17.64 &	15.58 &	2.06 \\
3c227 &	16.21 &	15.61 &	0.60 \\
3c236 &	17.62 &	14.54 &	3.08 \\
3c264 &	14.17 &	12.55 &	1.62 \\
\enddata
\tablecomments{Approximate $R-H$ colors for the sources based on $R$-band fluxes at a 20~kpc aperture.}
\tablerefs{$^{a}$~\citet{martel+99}.
[The complete version of this table is in the electronic edition of
the Journal.  The printed edition contains only a sample.]}
\end{deluxetable}

%%%%%%%%%%%%%%%%%%%%%%%%%%%%%%%%%%%%%%%%%%%%%%%%%
 %%%%%%%%%%%%%%%%%%%%%%%%%%%%%%%%%%%%%%%%%%%%%%%%%
 %%%%%%%%%%%%%%%%%%%%%%%%%%%%%%%%%%%%%%%%%%%%%%%%%

\appendix

\section{Consistency of Galfit and 2DM}
\label{sec-test}

In this paper we have used Galfit~\citep{galfit} to perform 2-dimensional modeling of the host galaxies of the 3CR. But we have adopted the error treatment of 2DM~\citep{mclure+99,floyd+04} in cases where the nucleus dominates (i.e. provides $\gtsim 1/2$ of the total flux). In this section we explore the consistency of the two approaches in a series of test cases using both real and simulated galaxies.

2DM was designed for use on QSO host galaxies, and is a simpler fitting code than Galfit. The major difference is in assignment of errors, as for the PSF-dominated cores of QSO images it is the PSF sampling error that dominates any fitted model, and the Poissonian error on the flux in a given pixel is an insufficient determination of the error in the model fit~\citep{floyd+04}. In its standard form it fits a single \sersic host galaxy component and a nuclear component, centered at the same fixed position.

Galfit and 2DM were used to model the host galaxy flux of three radio galaxies from the sample: 3C~66B, 3C~111, and 3C~449. These three were specifically chosen to span the range of infrared nuclear activity seen across the sample: 
3C~449 shows no sign of an unresolved nuclear point source, while 3C~66B shows a weak point-like nucleus, and 3C~111 is a quasar.
All data has been reduced following the technique described in section~\ref{sec-sampDR}. We have modelled each object using a single \sersic component model in Galfit, as described in section~\ref{sec-mod}, and using 2DM as described in~\citet{floyd+04}. The resulting models are presented in table~\ref{tab-comp}.
%Model fits are compared, and then 
Next, synthetic galaxies were constructed {\em based on the parameters recovered from these three objects}, using Galfit. Random noise was added to each synthesized test case at incrementing levels, and these noisy synthetic targets are then remodeled with Galfit and 2DM to explore the stability of fit to noise, and ability to recover the ``true'' properties of the host galaxy. The results of this test are explored in section~\ref{sec-synth} below.

\subsection{Models of the three real test cases}
\begin{figure*}[htb]
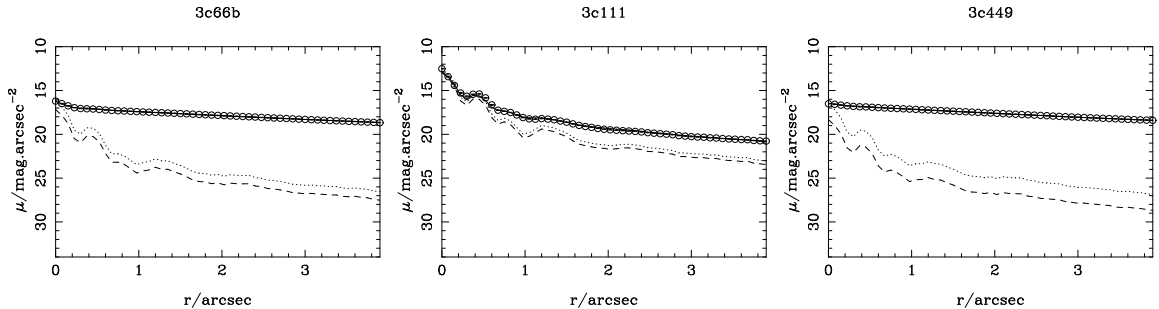

  \begin{center}
	{\includegraphics[height=5.0cm,angle=270]{f17a.eps}}
	{\includegraphics[height=5.0cm,angle=270]{f17b.eps}}
	{\includegraphics[height=5.0cm,angle=270]{f17c.eps}}
	\caption{\label{fig-2dmprof}2DM radial profiles for the three test cases. Solid lines indicate best-fit 2DM model. Dashed lines indicate the best-fit nuclear component, with the dotted lines indicating the PSF scaled to match the nuclear flux of the source.}
  \end{center}
\end{figure*}

The best-fit model parameters for the three objects, 3C~66B, 3C~111, and 3C~449, using each of the modeling techniques, are presented side-by-side for comparison in table~\ref{tab-comp}. Radial profiles are shown in Fig.~\ref{fig-2dmprof}.
2DM and Galfit are clearly seen from the table to be in good agreement. The host fluxes generally agree well (to within $\sim2\%$) and the scale-lengths within $\sim10\%$. However, in the case of the quasar, 3C~111, there is a larger discrepancy, and greater dependence on the treatment of errors in the central region. There is also a significant discrepancy in the value of the nuclear flux in all three sources. This is once again, very dependent on the error treatment. 

\input{tab9}

\subsection{\label{sec-synth} Synthetic galaxy tests}
We constructed simple synthetic galaxy models using Galfit (in non-optimization mode), and the parameters recovered from the three real test cases (3C~66B; 3C~111; 3C449). To these idealized galaxy models, varying amounts of noise were added, and different PSF's convolved with the data, in order to produce a synthetic NICMOS observation of an idealized galaxy. These synthetic observations were then modelled using 2DM.
We began with the simplest case of the original models convolved with the PSF's, with no noise or background added.
2DM was then run just once to see how reliably it could recover the original parameter set used to define the synthetic observation. Results are presented for this ``quick'' run in table~\ref{tab-test}. 2DM is found to recover the properties of the Galfit models to within $\sim10\%$. Note that in normal use, 2DM is run repeatedly to refine these fitting results. However, we adopt 10\% as a reasonable estimate for the systematic errors inherent in fitting non-ideal galaxies with idealized forms.

\input{tab10}

Next we added in an additional random ``readnoise'' component to the
simulated galaxies, and reran the Galfit minimization, starting from a
generic starting point, as described in~\citet{floyd+04}. 
Recovered parameters are presented in table~\ref{tab-noise}.
The parameters are recovered almost perfectly from each galaxy with
read-noise of $1-10e^{-}$, plus Poisson.

\input{tab11}

\subsection{Working Conclusions}
The results from 2DM and Galfit agree typically to within
$\sim10\%$ on morphological parameters, $\sim2\%$ on host
flux. However, there is a degree of variation, depending 
in particular on the treatment of errors in the central regions. 
This is in part due to the different convolution algorithms applied. 
For highly luminous objects this leads to a disagreement in some host
properties, while for all sources, there is still a large margin of
error in the nuclear luminosity. 

 %%%%%%%%%%%%%%%%%%%%%%%%%%%%%%%%%%%%%%%%%%%%%%%%%
 %%%%%%%%%%%%%%%%%%%%%%%%%%%%%%%%%%%%%%%%%%%%%%%%%
 %%%%%%%%%%%%%%%%%%%%%%%%%%%%%%%%%%%%%%%%%%%%%%%%%

\end{document}

%% file: tab1.tex
% Table 1: NEWLY OBSERVED TARGETS:

\begin{deluxetable}{lllrrl}
  \tabletypesize{\tiny}
  \tablecolumns{6}
  \tablewidth{0pc}
  \renewcommand{\arraystretch}{.6}
  \tablecaption{\label{tab-new} Observing log}
  \tablehead{
    \colhead{Source} &
    \colhead{$\alpha_{2000}$} &
    \colhead{$\delta_{2000}$} &
    \colhead{$z$} &
    \colhead{$\lambda$} &
    \colhead{Obs. Date}}
  \startdata
  3C~15	 & 00:37:04.1 & -01:09:08 & 0.073 & -64 & 2006-06-24\\
  3C~17	 & 00:38:20.5 & -02:07:40 & 0.219 & -65 & 2006-07-02\\
  3C~33	 & 01:08:53.3 & +13:20:25 & 0.059 & -49 & 2006-06-29\\
  3C~98  & 03:58:54.4 & +10:26:03 & 0.030 & -31 & 2005-11-25\\
  3C~132 & 04:56:43.0 & +22:49:22 & 0.214 & -13 & 2005-11-24\\
  3C~153 & 06:09:32.5 & +48:04:15 & 0.277 &  13 & 2005-11-25\\
  3C~166 & 06:45:24.1 & +21:21:51 & 0.245 &   8 & 2005-11-04\\
  3C~234 & 10:01:49.5 & +28:47:09 & 0.185 &  53 & 2005-11-03\\
  3C~258 & 11:24:43.5 & +19:19:12 & 0.165 &  69 & 2006-01-25\\
  3C~284 & 13:11:04.7 & +27:28:08 & 0.239 &  86 & 2006-03-04\\
  3C~296 & 14:16:52.9 & +10:48:26 & 0.025 &  64 & 2006-04-21\\
  3C~300 & 14:23:01.0 & +19:35:17 & 0.270 &  68 & 2006-03-04\\
  3C~323.1&15:47:43.5 & +20:52:17 & 0.264 &  49 & 2006-04-21\\
  3C~326 & 15:52:09.1 & +20:05:24 & 0.090 &  48 & 2006-04-21\\
  3C~332 & 14:23:01.0 & +19:35:17 & 0.270 &  45 & 2006-01-12\\
  3C~357 & 17:28:18.5 & +31:46:14 & 0.166 &  31 & 2006-03-25\\
  3C~403.1&19:52:30.4 & -01:17:20 & 0.055 & -14 & 2006-06-23\\
  3C~410 & 20:20:06.5 & +29:42:14 & 0.248 &  -4 & 2006-06-30\\
  3C~424 & 20:48:12.1 & +07:01:18 & 0.127 & -22 & 2006-06-22\\
  3C~442 & 22:14:46.9 & +13:50:27 & 0.026 & -34 & 2006-06-27\\
  3C~459 & 23:16:35.1 & +04:05:19 & 0.219 & -51 & 2006-06-24\\
  \enddata
  \tablecomments{Observation dates and positions of the 21 newly observed sources in the sample. We present the source name, J2000 Equatorial coordinates, redshift, galactic latitude, and observing date. Two other targets, 3C~18 and 3C~63 were observed as part of the program but were missed by the NIC2 chip.}
\end{deluxetable}

%% file: tab2.tex
% Table 2: ARCHIVAL OBJECTS:
\begin{deluxetable}{lcllrrll}
  \tabletypesize{\tiny}
  \tablecolumns{8}
  \tablewidth{0pc}
  \renewcommand{\arraystretch}{.6}
  \tablecaption{\label{tab-arch} Archival Objects: Observation Dates}
  \tablehead{
    \colhead{Source} &
    \colhead{Alt. names} &
    \colhead{$\alpha_{2000}$} &
    \colhead{$\delta_{2000}$} &
    \colhead{$z$} &
    \colhead{$\lambda$} &
    \colhead{Obs. Date} & 
    \colhead{PROPOSID}}
  \startdata
  3C~71  & M~77, NGC~1068          & 02:42:40.7 & -00:00:48 & 0.003793 &-52 & 1998-02-21 & 7215$^{(a)}$\\
  3C~84  & Per~A, NGC~1275, A~0426 & 03:19:48.1 & +41:30:42 & 0.017559 &-13 & 1998-03-16 & 7330$^{(b)}$\\
  3C~264 & NGC~3862, A~1367        & 11:45:05.0 & +19:36:23 & 0.021718 & 73 & 1998-05-12 & 7862$^{(c)}$\\
  3C~270 & NGC~4261                & 12:19:23.2 & +05:49:31 & 0.007465 & 67 & 1998-04-23 & 7868$^{(d)}$\\
  3C~272.1&M~84, NGC~4374          & 12:25:03.7 & +12:53:13 & 0.003536 & 74 & 1998-07-13 & 7868$^{(e)}$\\
  3C~274 & M~87, NGC~4486          & 12:30:49.4 & +12:23:28 & 0.004360 & 75 & 1997-11-20 & 7171$^{(f)}$\\
  3C~293 & UGC~08782               & 13:52:17.8 & +31:26:46 & 0.045034 & 76 & 1998-08-19 & 7853$^{(g)}$\\
  3C~305 & UGC~09553, IC~1065      & 14:49:21.6 & +63:16:14 & 0.041639 & 49 & 1998-07-19 & 7853$^{(h)}$\\
  3C~317 & UGC~09799, A~2052       & 15:16:44.5 & +07:01:17 & 0.034457 & 50 & 1998-08-26 & 7886$^{(d)}$\\
  3C~338 & NGC~6166, A~2199        & 16:28:38.5 & +39:33:06 & 0.030354 & 44 & 1997-12-18 & 7453$^{(i)}$\\
  3C~405 & Cyg~A                   & 19:59:28.3 & +40:44:02 & 0.056075 &  6 & 1997-12-16 & 7258$^{(j)}$\\
  \enddata
  \tablecomments{Observation dates and positions of the 11 archival objects in the sample.  
    All observations are on NIC2 through the F160W filter. We list HST proposal ID's and 
    references for the original publication of the data. 
    3C~273 also has deep NICMOS F160W imaging of its jet, but is excluded from this paper as the on-galaxy
    integration time is too short for an accurate characterisation of the host galaxy.}
  \tablerefs{
    (a)~\citet{thompsoncorbin99}; 
    (b)~\citet{martini+03};
    (c)~\citet{capetti+00}; 
    (d)~\citet{quillen+00}; 
    (e)~\citet{bower+00}; 
    (f)~\citet{corbin+02}; 
    (g)~\citet{floyd+06a}; 
    (h)~\citet{jackson+03}; 
    (i)~\citet{jensen+01};
    (j)~\citet{tadhunter+99}.}
\end{deluxetable}

%% file: tab9.tex
\begin{deluxetable}{lrrrrrr}
  \tabletypesize{\small}
  \tablecolumns{7}
  \tablewidth{0pc}
  \tablecaption{\label{tab-comp}Galfit/2DM results: Real test-case galaxies}
  \tablehead{
    \colhead{Code} &
    \colhead{Mag(host)} & \colhead{$R_{e}/kpc$} &  \colhead{$n(=1/\beta)$} &
    \colhead{$b/a$}&\colhead{$\theta$}&\colhead{Mag(nuc)}}
  \startdata
  \cutinhead{\bf 3C~66b}
  Galfit     &12.47&2.93&1.64&0.92&-79.4&19.40\\
  2DM(nosamp)&12.52&2.67&1.55&0.91&-86.5&19.29\\
  2DM(samp)  &12.47&2.81&1.67&0.91&-85.9&18.64\\
  \cutinhead{\bf 3C~111}
  Galfit         &14.40&2.26&6.98&0.72&16.8&15.05\\
  2DM(nosamp)    &14.58&2.24&4.28&0.63&17.4&15.04\\  
  2DM(samp)      &14.71&2.95&3.75&0.63&17.2&14.71\\  
  \cutinhead{\bf 3C~449}
  Galfit     &12.48&1.90&1.36&0.84&-6.4&21.34\\
  2DM(nosamp)&12.51&1.81&1.30&0.86&-7.7&20.46\\
  2DM(samp)  &12.49&1.84&1.37&0.85&-7.2&20.52\\

  \enddata
  \tablecomments{Comparison of results of fitting the three test case
  objects with Galfit and 2DM. Position angles are in degrees
  anti-clockwise from North. All magnitudes are uncorrected AB mag's
  from the PHOTFNU header keyword. Scale-lengths are half-light radii
  on the semi-major axis. The centre of the galaxy and
  nucleus are fixed at the same position in galfit as in 2DM, in order
  to force a comparison based on the same number of parameters. For
  3C~111 this posed problems (see text) so we present the galfit models
  both with and without a fixed centre.}
\end{deluxetable}

%% file: tab10.tex
\begin{deluxetable}{lrrrrrr}
  \tabletypesize{\small}
  \tablecolumns{7}
  \tablewidth{0pc}
  \tablecaption{\label{tab-test}Galfit/2DM results: Synthetic Galaxies}
  \tablehead{
    \colhead{Code} &
    \colhead{Mag(host)} & \colhead{$R_{e}/kpc$} &  \colhead{$n(=1/\beta)$} &
    \colhead{$b/a$}&\colhead{$\theta$}&\colhead{Mag(nuc)}}
  \startdata
  \cutinhead{\bf A (Simulated 3C~66B)}
  Galfit&12.47&2.93&1.64&0.92&-79.4&19.40\\
  2DM   &12.38&3.15&1.89&0.90&-72.4&19.87\\
  \cutinhead{\bf B (Simulated 3C~111)}
  Galfit&14.40&2.26&6.98&0.72&16.8&15.05\\
  2DM   &14.56&2.15&3.64&0.74&13.5&15.34\\
  \cutinhead{\bf C (Simulated 3C~449)}
  Galfit&12.48&1.90&1.36&0.84&-6.4&21.34\\
  2DM   &12.49&1.87&1.34&0.84&-6.3&21.22\\
  \enddata
  \tablecomments{Comparison of results of fitting the three synthetic
  galaxies, constructed with galfit using the parameters deduced for
  the three real test cases. Galfit recovers the parameter values to within 10\% in a single run (no added noise).}
\end{deluxetable}

%% file: tab11.tex
\begin{deluxetable}{lrrrrrr}
  \tabletypesize{\small}
  \tablecolumns{7}
  \tablewidth{0pc}
  \tablecaption{\label{tab-noise}Response of Galfit to synthetic
  galaxies plus noise}
  \tablehead{
    \colhead{Galaxy} &
    \colhead{Mag(host)} & \colhead{$R_{e}/kpc$} &  \colhead{$n(=1/\beta)$} &
    \colhead{$b/a$}&\colhead{$\theta$}&\colhead{Mag(nuc)}}
  \startdata
  \cutinhead{\bf A (Simulated 3C~66b)}
  Original                   &12.47&2.93&1.64&0.92&-79.4&19.40\\
  Recovered($1e^{-}$Poisson) &12.47&2.93&1.64&0.92&-79.4&19.40\\
  Recovered($10e^{-}$Poisson)&12.47&2.93&1.64&0.92&-79.4&19.41\\
  \cutinhead{\bf B (Simulated 3C~111)}
  Original                     &14.40&2.26&6.98&0.72&16.8&15.05\\
  Recovered ($1e^{-}+$Poisson) &14.40&2.26&6.93&0.72&16.8&15.05\\
  Recovered ($10e^{-}+$Poisson)&14.39&2.26&7.01&0.72&16.8&15.06\\
  \cutinhead{\bf C (Simulated 3C~449)}
  Original                     &12.48&1.90&1.36&0.84&-6.4&21.34\\
  Recovered ($1e^{-}+$Poisson) &12.48&1.89&1.36&0.84&-6.4&21.32\\
  Recovered ($10e^{-}+$Poisson)&12.48&1.89&1.36&0.84&-6.4&21.33\\
  \enddata
  \tablecomments{Galfit results for fits to three synthetic test case
  galaxies constructed with varying degrees of readnoise (1 and 10
  $e^{-}$) plus Poisson noise. The three test cases are based on the
  Galfit models for 3C~66B, 3C~111, and 3C~449. The top line of each row
  gives the true parameters for the synthetic galaxy. The subsequent
  rows show the results recovered by running galfit. Each run was
  initiated at a generic starting point, not tailored to the
  individual cases.}
\end{deluxetable}